\documentclass[11pt, reqno]{amsart}
\pdfoutput=1

\usepackage{amsmath, amsthm, amssymb, mathtools, mathrsfs}
\usepackage{amsaddr}
\usepackage{bbm}
\usepackage{pgffor}
\usepackage{graphicx}
\usepackage[breakable]{tcolorbox}

\usepackage{enumitem}
\usepackage{fullpage}
\usepackage{hyperref}
\definecolor{rufous}{HTML}{A81C07}
\definecolor{orangish}{HTML}{C55C58}
\hypersetup{
    colorlinks=true,                            
    linkcolor=rufous,                          
    citecolor=rufous,                          
    filecolor=rufous,                           
    urlcolor=rufous                             
}

\newcommand{\makeheading}[1]%
        {\hspace*{-\marginparsep minus \marginparwidth}%
         \begin{minipage}[t]{\textwidth}%
                {\large \bfseries #1}\\[-0.15\baselineskip]%
                 \rule{\columnwidth}{1pt}%
         \end{minipage}}

\theoremstyle{definition}
\newtheorem{defn}[equation]{Definition}
\newtheorem{rem}[equation]{Remark}
\newtheorem{ex}[equation]{Example}

\theoremstyle{plain}
\newtheorem*{lem*}{Lemma}
\newtheorem{lem}[equation]{Lemma}
\newtheorem{thm}[equation]{Theorem}
\newtheorem{thmalpha}{Theorem}

\newtheorem*{cor*}{Corollary}
\newtheorem{cor}[equation]{Corollary}
\newtheorem{prop}[equation]{Proposition}

\numberwithin{equation}{section}

\def\semicolon{;}
\def\applytolist#1{
    \expandafter\def\csname multi#1\endcsname##1{
        \def\multiack{##1}\ifx\multiack\semicolon
            \def\next{\relax}
        \else
            \csname #1\endcsname{##1}
            \def\next{\csname multi#1\endcsname}
        \fi
        \next}
    \csname multi#1\endcsname}

\def\calc#1{\expandafter\def\csname c#1\endcsname{{\mathcal #1}}}
\applytolist{calc}QWERTYUIOPLKJHGFDSAZXCVBNM;
\def\bbc#1{\expandafter\def\csname bb#1\endcsname{{\mathbb #1}}}
\applytolist{bbc}QWERTYUIOPLKJHGFDSAZXCVBNM;
\def\bfc#1{\expandafter\def\csname bf#1\endcsname{{\mathbf #1}}}
\applytolist{bfc}QWERTYUIOPLKJHGFDSAZXCVBNM;
\def\sfc#1{\expandafter\def\csname s#1\endcsname{{\sf #1}}}
\applytolist{sfc}QWERTYUIOPLKJHGFDSAZXCVBNM;
\def\fc#1{\expandafter\def\csname f#1\endcsname{{\mathfrak #1}}}
\applytolist{fc}QWERTYUIOPLKJHGFDSAZXCVBNM;
\def\rmc#1{\expandafter\def\csname rm#1\endcsname{{\mathrm #1}}}
\applytolist{rmc}QWERTYUIOPLKJHGFDSAZXCVBNM;
\def\scrc#1{\expandafter\def\csname scr#1\endcsname{{\mathscr #1}}}
\applytolist{scrc}QWERTYUIOPLKJHGFDSAZXCVBNM;

\newcommand{\set}[2]{\left\{ #1 : #2\right\}}

\DeclareMathOperator{\Span}{span}

\DeclareMathOperator{\supp}{supp}
\newcommand{\loc}{\operatorname{loc}}

\newcommand{\End}{\operatorname{End}}
\newcommand{\id}{\operatorname{id}}
\newcommand{\Ad}{\operatorname{Ad}}
\newcommand{\Star}{\operatorname{star}}
\newcommand{\plaq}{\operatorname{plaq}}
\newcommand{\sa}{\operatorname{sa}}

\usepackage{xcolor}

\usepackage{tikz}
\usetikzlibrary{cd}
\usetikzlibrary{calc}
\usetikzlibrary{decorations,decorations.pathreplacing,decorations.markings}

\tikzstyle{mid>}=[decoration={markings, mark=at position 0.5 with {\arrow{>}}}, postaction={decorate}]
\tikzstyle{mid<}=[decoration={markings, mark=at position 0.5 with {\arrow{<}}}, postaction={decorate}]
\definecolor{violet}{RGB}{148,0,211}
\definecolor{rufous}{HTML}{A81C07}

\begin{document}
\title{
An algebraic quantum field theoretic approach to toric code with gapped boundary
}
\author{Daniel Wallick}
\address{Department of Mathematics, The Ohio State University, Columbus, OH 43210, USA}
\date{\today}
\maketitle
\begin{abstract}
Topologically ordered quantum spin systems have become an area of great interest, as they may provide a fault-tolerant means of quantum computation.  One of the simplest examples of such a spin system is Kitaev's toric code.  Naaijkens made mathematically rigorous the treatment of toric code on an infinite planar lattice (the thermodynamic limit), using an operator algebraic approach via algebraic quantum field theory.  We adapt his methods to study the case of toric code with gapped boundary.  In particular, we recover the condensation results described in Kitaev and Kong and show that the boundary theory is a module tensor category over the bulk, as expected.
\end{abstract}


\section{Introduction}


Kitaev's quantum double model is a quantum spin system exhibiting topological order, and it is a useful model to study since it exhibits non-abelian anyons \cite{MR1951039}.  
These non-abelian anyons allow for fault-tolerant quantum computation, which is of value in quantum information \cite{MR2443722}.  
The simplest example of Kitaev's quantum double model is toric code.  
While the toric code model exhibits abelian anyons, and is therefore less useful for computational purposes, it is nonetheless well studied due to its simplicity \cite{MR1951039, MR2942952, MR2345476}.  
Like in other topologically ordered quantum spin systems, the fusion and braiding of the excitations in toric code are modeled by a unitary modular tensor category, specifically $\cZ(\mathsf{Hilb_{fd}}(\bbZ/2\bbZ))$ \cite{MR1951039}.  
We refer the reader to \cite[Appendix E]{MR2200691} for more details on how topologically ordered spin systems are modeled by unitary modular tensor categories.  

Recently, Naaijkens used techniques from algebraic quantum field theory to study the case of toric code on an infinite planar lattice \cite{MR2804555, MR2956822, MR3135456}.  
In particular, he used these techniques to rigorously analyze the thermodynamic limit.  
The more general case of Kitaev's quantum double model for abelian groups has been studied in analogous fashion \cite{MR3426207}.  
Using these approaches, Naaijkens \cite{MR2804555} was able to recover the fusion and braiding statistics described in \cite{MR1951039}.  

In algebraic quantum field theory \cite{MR297259, MR334742},
one has a quasi-local $\rmC^*$-algebra $\fA$ that is the $\rmC^*$-inductive limit of a net $\fA_i$ of von Neumann algebras corresponding to local regions $i$.  
Here we are using an unspecified choice of local regions, with $i$ an arbitrary index, as in this paragraph we are sketching an abstract description of AQFT.  
We will later adapt this to our specific setting.
We will assume that $\fA$ is faithfully represented on some Hilbert space $\cH$ by means of a vacuum representation $\pi_0$.  
One then considers \emph{superselection sectors}, which are representations of $\fA$ satisfying that for any region $i$, 
\[
\pi|_{\fA_j} 
\cong
\pi_0|_{\fA_j}
\]
for all $j$ disjoint from $i$.  
An additional assumption that is often necessary is that of \emph{Haag duality}, which is that for all regions $i$,
\[\fA_i
=
\left(\bigcup_{j \cap i = \emptyset} \fA_j\right)'.
\]
(Here we identify the $\fA_i$ with their images under $\pi_0$, a practice we will generally avoid in the remainder in the text.)
Under the assumption of Haag duality, the superselection sectors form a braided $\rmC^*$-tensor category, as described in detail in \cite{Halvorson_2006}.

Naaijkens' treatment of toric code used the following blueprint.  
He first constructed the superselection sectors corresponding to the known excitations in toric code \cite{MR2804555}.  
He showed that these sectors and their intertwiners formed the unitary modular tensor category $\cZ(\mathsf{Hilb_{fd}}(\bbZ/2\bbZ))$, as expected from previous work \cite{MR1951039}, even though he had not proven Haag duality at this point.  
He later showed Haag duality \cite{MR2956822} and proved that the known excitations exhausted all of the superselection sectors \cite{MR3135456}.  

A natural next step is to examine the case of toric code with boundary; see \cite{MR2942952} for an in-depth discussion of gapped boundaries in topologically ordered quantum spin models.  
We specifically consider the case of a boundary where there is toric code on one side and vacuum on the other.  
There are two types of gapped boundaries, namely the \emph{rough} boundary and the \emph{smooth} boundary; see Figure \ref{fig:SmoothAndRoughBoundary} below.  
\begin{figure}[h]
\centering
\includegraphics{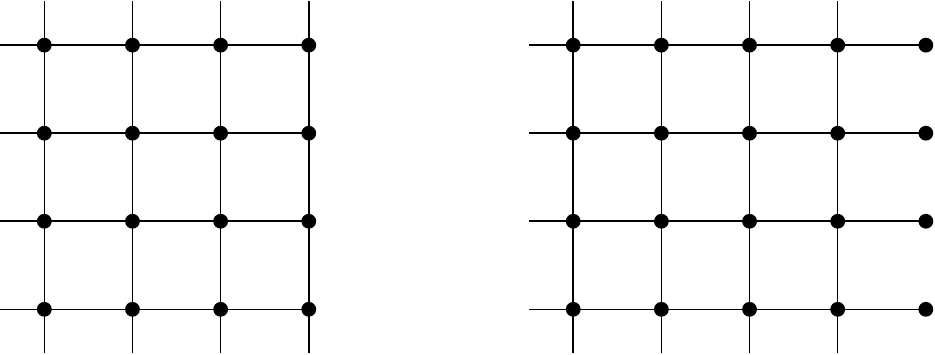}
\caption{Lattices illustrating toric code with smooth boundary (left) and rough boundary (right).}
\label{fig:SmoothAndRoughBoundary}
\end{figure}
Due to the simplicity of the toric code model, these two types of boundary are in fact equivalent in some sense.  
In particular, one can pass from one boundary to the other by taking the dual lattice and changing bases.  
We therefore focus our attention on the smooth boundary for convenience.
According to \cite{MR2942952}, the boundary excitations for a gapped boundary system exhibiting topological order are given by a module tensor category over the unitary modular tensor category of bulk excitations.
Here, a \emph{module tensor category} over a braided tensor category $\cC$ is a tensor category $\cM$ equipped with a braided tensor functor $F \colon \cC \to \cZ(\cM)$, where $\cZ(\cM)$ is the Drinfeld center of $\cM$ \cite{MR3578212}.
The module tensor category structure described in \cite{MR2942952} is given by bringing a bulk excitation to the boundary and mapping it to a half-braiding (an object in the Drinfeld center).  
For toric code specifically, the boundary excitations should be described by the fusion category $\mathsf{Hilb_{fd}}(\bbZ/2\bbZ)$, as certain bulk excitations condense at the boundary.  
This category of boundary excitations is then a module tensor category over the braided category of bulk excitations in the way just described.

We adapt the work in \cite{MR2804555, MR2956822, MR3135456} to the case of toric code with smooth boundary, recovering the previously known description of how the excitations behave \cite{MR2942952}.

\begin{thmalpha}
\label{thm:SectorCatModuleTensor}
The fusion category of boundary excitations---more precisely, the fusion category of superselection sectors localized in a fixed cone along the boundary---is a module tensor category over the category of sectors for the bulk toric code.  
\end{thmalpha}

We prove this theorem in many parts.
We begin by briefly reviewing some categorical notions in \textsection \ref{sec:BriefCatOverview}.
In \textsection\ref{sec:ToricCodeWithBoundaryModel}, we present the model for toric code with smooth boundary.  
We construct a canonical ground state in \textsection\ref{sec:GroundState}, and in \textsection\ref{sec:ExcitationsAndSuperselectionSectors} we construct the superselection sectors for the known excitations.  
We show that these superselection sectors are localized and transportable along the boundary, and we also show that the condensation results described in \cite{MR2942952} hold. 
In \textsection\ref{sec:Intertwiners}, we present descriptions of the intertwiners between these superselection sectors analogous to the one found in \cite{MR2804555}, and we define the tensor product of superselection sectors and intertwiners between them.  
In \textsection\ref{sec:ConeAlgebras}, we prove that the cone regions we consider give rise to infinite factors, allowing us to construct a fusion category in \textsection\ref{sec:FusionCategoryForBoundaryExcitations} whose objects correspond to the known excitations.  
In \textsection\ref{sec:BulkToBoundary}, we construct a braided tensor functor from the bulk toric code to the Drinfeld center of the fusion category constructed in \textsection\ref{sec:FusionCategoryForBoundaryExcitations}, which equips this category with the structure of a module tensor category.  
Finally, in \textsection\ref{sec:HaagDuality} and \textsection\ref{sec:DistalSplit}, we prove Haag duality and a property called the \emph{distal split property} for the state $\omega_0$, allowing us to show in \textsection\ref{sec:BoundingExcitations} that we have accounted for all of the excitations in our model.  


\section{A brief overview of categorical definitions}
\label{sec:BriefCatOverview}


In this section, we present a brief overview of category theory definitions that will be used.  
For more detail see \cite[\textsection 2]{MR3578212}.  
We use the term \emph{tensor category} to refer to a linear monoidal category, as done in \cite{MR3578212}.
We say that a tensor category is \emph{rigid} if every object has a dual and a predual, and we said that a tensor category $\cC$ is \emph{braided} if it is equipped with a collection of isomorphisms $\beta_{a, b} \colon a \otimes b \to b \otimes a$ for each $a, b \in \cC$ that is natural in both inputs and satisfies the following \emph{braid equations}: 
\begin{align*}
\beta_{a \otimes b, c}
&=
(\beta_{a,c} \otimes \id_b)(\id_a \otimes a_{b,c}),
\\
\beta_{a, b \otimes c}
&=
(\id_b \otimes \beta_{a,c})(\beta_{a,b} \otimes \id_c).
\end{align*}
Note that in the above equations we have suppressed the associator isomorphisms that are part of the data of being a monoidal category; however, all of the tensor categories we consider are \emph{strict}, meaning that the associator isomorphisms are all identity morphisms.  
A \emph{fusion category} is a semisimple rigid tensor category with finitely many isomorphism classes of simple objects and with simple tensor unit.  
Most of the categories we consider are \emph{unitary fusion categories}.  
These are fusion categories equipped with a \emph{dagger structure}, that is, for each morphism $f \colon a \to b$, there exists a morphism $f^\dag \colon b \to a$, and the map $f \mapsto f^\dag$ is an anti-linear involution.  
We further require that with this choice of dagger structure, the endomorphism algebra for each object in the category is a finite-dimensional $\rmC^*$-algebra.  
In our examples, the dagger structure will generally correspond to the adjoint in the $\rmC^*$-algebraic setting.  

We remark that the braided unitary fusion categories we consider satisfy a nondegeneracy condition making them \emph{unitary modular tensor categories}.  
However, we will not make further mention of this fact in what follows.  
For more information about modular tensor categories see \cite[\textsection 8.13-8.14]{MR3242743}.

\begin{ex}
One example of a fusion category we will use in the text is that of $\mathsf{Hilb_{fd}}(\bbZ/2\bbZ)$.  
This is the category of finite-dimensional Hilbert spaces graded by elements of the group $\bbZ/2\bbZ$.  
The tensor product is given by the group structure in $\bbZ/2\bbZ$.  
That is to say, if $\bbZ/2\bbZ = \{1, g\}$, then $(V_1 \oplus V_g) \otimes (W_1 \oplus W_g)$ has $1$-graded component $(V_1 \otimes W_1) \oplus (V_g \otimes W_g)$ and $g$-graded component $(V_1 \otimes W_g) \oplus (V_g \otimes W_1)$.
The associator for this category is the standard ``move parentheses map"; that is to say, $\mathsf{Hilb_{fd}}(\bbZ/2\bbZ)$ has trivial associator.
\end{ex}

Given a tensor $\cC$, one can build a braided tensor category $\cZ(\cC)$ called the \emph{Drinfeld center}. 
We remark that if $\cC$ is a fusion category, then $\cZ(\cC)$ is a modular tensor category.
The objects in $\cZ(\cC)$ are \emph{half-braidings}, which are pairs $(z, \sigma_{-, z})$, where $z \in \cC$ and $\sigma_{-, z} \colon - \otimes z \to z \otimes -$ is a natural isomorphism satisfying that for all $a, b \in \cC$, 
\[
\sigma_{a \otimes b, z}
=
(\sigma_{a, z} \otimes \id_b)(\id_a \otimes \sigma_{b, z}).
\]
As before, we suppress associator isomorphisms (which will be identities in our examples).
A morphism $f \colon (z, \sigma_{-, z}) \to (w, \sigma_{-, w})$ in $\cZ(\cC)$ is a morphism $f \colon z \to w$ in $\cC$ satisfying that for all $a \in \cC$, 
\[
(f \otimes \id_a) \sigma_{a, z}
=
\sigma_{a, w}(\id_a \otimes f).
\]
For $(z, \sigma_{-, z}), (w, \sigma_{-, w}) \in \cZ(\cC)$, we have that $(z, \sigma_{-, z}) \otimes (w, \sigma_{-, w}) = (z \otimes w, \sigma_{-, z \otimes w})$, where $\sigma_{-, z \otimes w}$ is given by the following formula: 
\[
\sigma_{a, z \otimes w}
\coloneqq
(\id_z \otimes \sigma_{a, w})(\sigma_{a, z} \otimes \id_w).
\]
We then have that $\cZ(\cC)$ is a braided tensor category, with braiding given by $\beta_{(z, \sigma_{-, z}), (w, \sigma_{-, w})} \coloneqq \sigma_{z, w}$.

The goal of this paper will be to show that a specific fusion category $\cM$ is a module tensor category over a braided fusion category $\cC$ in the manner described in \cite{MR2942952}.
\begin{defn}
Let $\cM$ be a tensor category, and let $\cC$ be a braided tensor category.  
We say that $\cM$ is a \emph{module tensor category} over $\cC$ if it equipped with a braided tensor functor $F \colon \cC \to \cZ(\cM)$.  
\end{defn}
Here a functor $F \colon \cC \to \cD$ between tensor categories is a \emph{tensor functor} if it is equipped with natural isomorphisms $F^2_{a, b} \colon F(a \otimes b) \to F(a) \otimes F(b)$ and $F^1 \colon F(1_\cC) \to 1_\cD$ that satisfy certain coherence relations.  
In the example we will construct, these isomorphisms are idenities.  
We say that a tensor functor $F \colon \cC \to \cD$ between braided tensor categories is \emph{braided} if $F^2_{a, b} F(\beta^\cC_{a, b}) = \beta^\cD_{F(a), F(b)} F^2_{a, b}$ for all $a, b \in \cC$.
Note that if $F^2_{a, b}$ is the identity for all $a, b \in \cC$ (as it will be in the example we construct), the braided condition becomes simply $F(\beta^\cC_{a, b}) = \beta^\cD_{F(a), F(b)}$.  

Finally, occasionally in the course of the text, we refer to the notion of a $\rmC^*$-tensor category.  
We will typically use this term when we are referring to a tensor category with a dagger structure that is not a unitary fusion category.  
In the literature, $\rmC^*$-tensor category is often a stronger notion than simply a tensor category with a dagger structure; see \cite{MR4362722} for one paper using a stronger definition.  
However, further discussion about this topic would take us too far afield.


\section{Toric code with smooth boundary}
\label{sec:ToricCodeWithBoundaryModel}

\begin{figure}[h]
\centering
\includegraphics{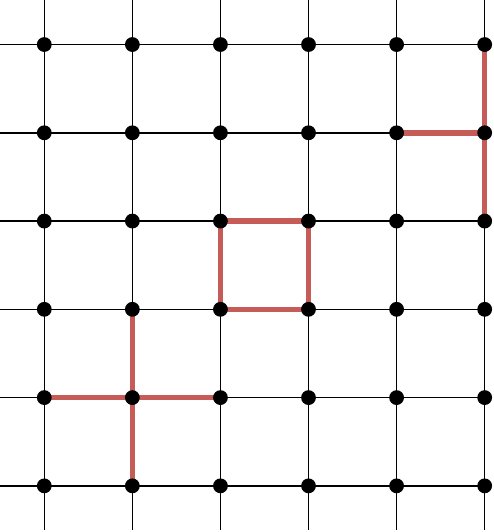}
\caption{The lattice for toric code with smooth boundary.
A star at a vertex in the boundary, a plaquette, and a star at a vertex not in the boundary are illustrated going from top right to bottom left.}
\label{fig:ToricCodeSmoothBoundaryLattice}
\end{figure}

We consider an infinite lattice with a smooth boundary, as shown in Figure \ref{fig:ToricCodeSmoothBoundaryLattice}.
We associate with each bond in the lattice a copy of $\bbC^2$.  
We let $\bfB$ denote the collection of all bonds in the lattice.  
As in \cite{MR2804555}, for any finite subset $\Lambda \subseteq \bfB$, we let $\fA(\Lambda)$ be the finite-dimensional $\rmC^*$-algebra corresponding to the tensor product of $M_2(\bbC) = \scrB(\bbC^2)$ over the bonds in $\Lambda$.  
Note that if $\Lambda_1 \subseteq \Lambda_2$, then we have a canonical inclusion $\fA(\Lambda_1) \subseteq \fA(\Lambda_2)$.  
We define the algebra of \emph{local operators} $\fA_{\loc}$ to be 
\[
\fA_{\loc}
\coloneqq
\bigcup_{\Lambda \subseteq \bfB \text{ finite}}
\fA(\Lambda),
\]
and we define the \emph{quasi-local algebra} $\fA$ to be the completion of $\fA_{\loc}$ in norm.  
If $\Lambda \subseteq \bfB$ is any subset, we define the algebra $\fA(\Lambda)$ of operators \emph{localized in $\Lambda$} to be the norm-completion of the algebra 
\[
\fA(\Lambda)_{\loc}
\coloneqq
\bigcup_{\Lambda_0 \subseteq \Lambda \text{ finite}} \fA(\Lambda_0).
\]
For an operator $A \in \fA_{\loc}$, we define the \emph{support of $A$} to be the collection of bonds $\supp(A) \subseteq \bfB$ on which $A$ does not act as the identity.  

We now describe the local Hamiltonians for the toric code model with smooth boundary.  
Note that we have the following Pauli $X$, $Y$, and $Z$ matrices in $M_2(\bbC)$: 
\[
\sigma^X 
\coloneqq
\begin{pmatrix}
0 & 1
\\
1 & 0
\end{pmatrix},
\qquad
\sigma^Y
\coloneqq
\begin{pmatrix}
0 & -i
\\
i & 0
\end{pmatrix},
\qquad
\sigma^Z
\coloneqq
\begin{pmatrix}
1 & 0
\\
0 & - 1
\end{pmatrix}.
\]
For a vertex $s$ in the lattice, we let $\Star(s)$ be the subset of $\bfB$ consisting of all bonds adjacent to $s$ (illustrated in Figure \ref{fig:ToricCodeSmoothBoundaryLattice}).  
Note that if $s$ is on the boundary, $\Star(s)$ only consists of three bonds, while otherwise $\Star(s)$ consists of four bonds.
Similarly, for face (or plaquette) $p$ in the lattice, we let $\plaq(p)$ be the subset of $\bfB$ consisting of all bonds adjacent to $p$ (illustrated in Figure \ref{fig:ToricCodeSmoothBoundaryLattice}).  
For a vertex $s$, we let $A_s$ be the tensor product of Pauli $X$ matrices over each of the bonds in $\Star(s)$.
Similarly, for a plaquette $p$, we let $B_p$ be the tensor product of Pauli $Z$ matrices over each of the bonds in $\plaq(p)$.
Note that the matrices $A_s$ and $B_p$ commute for all vertices $s$ and plaquettes $p$.  
For a finite subset $\Lambda \subseteq \bfB$, the local Hamiltonian has the form 
\begin{equation}
\label{eq:ToricCodeLocalHamiltonian}
H_\Lambda
=
-\sum_{\operatorname{star}(s) \subseteq \Lambda} A_s - \sum_{\operatorname{plaq}(p) \subseteq \Lambda} B_p.
\end{equation}

We now want to frame the local Hamiltonians in \eqref{eq:ToricCodeLocalHamiltonian} using the concept of interactions.  
An \emph{interaction} \cite[p.~241]{MR1441540} is a map $\Phi$ from the set of finite subsets of $\bfB$ to $\fA$ satisfying that for any finite subset $\Lambda \subseteq \bfB$, 
\begin{itemize}
\item $\Phi(\Lambda) \in \fA(\Lambda)$,
\item $\Phi(\Lambda)^* = \Phi(\Lambda)$.
\end{itemize}
Observe that if we define the interaction $\Phi$ to be 
\begin{equation}
\label{eq:ToricCodeInteractions}
\Phi(\Lambda)
=
\begin{cases}
-A_s & \text{if } \Lambda = \Star(s), 
\\
-B_p & \text{if } \Lambda = \plaq(p),
\\
0 & \text{otherwise,}
\end{cases}
\end{equation}
then for any finite subset $\Lambda \subseteq \bfB$, 
\[
H_\Lambda
=
\sum_{\Lambda_0 \subseteq \Lambda} \Phi(\Lambda_0).
\]
We thus say that the local Hamiltonians $H_\Lambda$ are given by the interaction $\Phi$.  
Note that the interactions in \eqref{eq:ToricCodeInteractions} are invariant under the action of $\bbZ$ on the lattice by vertical translations (i.e., translations parallel to the boundary).  
More precisely, if we let $\tau_n \colon \fA \to \fA$ denote translation by $n$ in the direction parallel to the boundary, then we have that for all finite $\Lambda \subseteq \bfB$, 
\[
\tau_n(\Phi(\Lambda))
=
\Phi(n + \Lambda),
\]
where $n + \Lambda$ denotes the translation of $\Lambda$ by $n$ in the direction parallel to the boundary.


\section{Ground state}
\label{sec:GroundState}


We proceed as in \cite{MR2804555}, which is based on the treatment in \cite{MR2345476}.
Note that for every finite subset $\Lambda \subseteq \bfB$, we have an action $\alpha_t^{\Lambda}$ of $\bbR$ on $\fA(\Lambda)$ given by 
\[
\alpha_t^{\Lambda}(A)
\coloneqq
e^{itH_\Lambda} A e^{-itH_\Lambda}
\qquad
\text{for all }
A \in \fA(\Lambda).
\]
We wish to take a limit over the net of finite subsets of $\bfB$ to obtain an action $\alpha_t$ of $\bbR$ on $\fA$, and we also want a closed operator $\overline{\delta}$ defined on a dense subspace of $\fA$ generating the dymanics, meaning that $\alpha_t = e^{t\overline{\delta}}$.  
To do so, we will invoke \cite[Thm.~6.2.4]{MR1441540}.  
Observe that if $\Lambda \subseteq \bfB$ is a finite subset with $|\Lambda| > 4$, then $\Phi(\Lambda) = 0$, where $\Phi$ is the interaction described in \eqref{eq:ToricCodeInteractions}.  
Furthermore, $\|\Phi(\Lambda)\| = 1$ if $\Phi(\Lambda) \neq 0$, and if $j \in \bfB$, then there are at most four finite subsets $\Lambda \subseteq \bfB$ such that $j \in \Lambda$ and $\Phi(\Lambda) \neq 0$.  
Hence, we have that 
\[
\sum_{n \geq 0} e^n \left(\sup_{j \in \bfB} \sum_{\substack{\Lambda \ni j \\ |\Lambda| = n + 1}} \|\Phi(\Lambda)\| \right)
<
\infty.
\]
Thus, the hypothesis of \cite[Thm.~6.2.4]{MR1441540} holds.  
We define a derivation $\delta$ with domain $D(\delta) = \fA_{\loc}$, given on $A \in \fA(\Lambda)$ (where $\Lambda \subseteq \bfB$ is finite) by 
\[
\delta(A)
\coloneqq
i \sum_{\Lambda_0 \cap \Lambda \neq \emptyset} [\Phi(\Lambda_0), A].  
\]
By \cite[Thm.~6.2.4]{MR1441540}, $\delta$ is norm-closable, and the closure $\overline{\delta}$ generates a strongly continuous one-parameter family of $*$-automorphisms $\alpha_t$ of $\fA$, i.e., $\alpha_t = e^{t\overline{\delta}}$.  
Furthermore, we have that for all $A \in \fA$, 
\[
\lim_{\Lambda \subseteq \bfB \text{ finite}} \|\alpha_t(A) - \alpha^{\Lambda}_t(A)\|
=
0,
\]
uniformly for $t$ in compact subsets of $\bbR$.  

Now, suppose $A \in \fA_{\loc}$, and let $\Lambda \subseteq \bfB$ be a finite subset such that $\supp(A) \subseteq \Lambda$ and such that any star or plaquette intersecting $\supp(A)$ is contained in $\Lambda$.  
Then we have that 
\[
\delta(A)
=
i \sum_{\Lambda_0 \cap \Lambda \neq \emptyset} [\Phi(\Lambda_0), A]
=
i [H_\Lambda, A].
\]

A \emph{ground state} for this system \cite[Thm.~6.2.52]{MR1441540} is a state $\omega_0$ on $\fA$ satisfying that for all $X \in \fA_{\loc}$, 
\[
-i\omega_0(X^*\delta(X)) \geq 0.
\]
We now construct such a state on $\fA$.  
To do so, we will make use of the following lemma, which is a simple application of Cauchy-Schwarz.

\begin{lem}[{\cite[\textsection 2.1.1]{MR2345476}}]
\label{lem:AFHStateLemma}
Suppose $\cA$ is a unital $\rmC^*$-algebra, and let $\omega$ be a state on $\cA$.  
Suppose $X \leq I$ in $\cA$ such that $\omega(X) = 1$.  
(Here, $I$ is the unit of $\cA$.)  
Then for all $Y \in \cA$, we have that 
\[
\omega(XY)
=
\omega(YX)
=
\omega(Y).
\]
\end{lem}

The existence of a ground state for the dynamics described above, which is also energy-minimizing, is given by the following result.  

\begin{thm}
\label{thm:GroundState}
There exists a ground state $\omega_0 \colon \fA \to \bbC$ for the dynamics given by the interactions in \eqref{eq:ToricCodeInteractions}, which satisfies that $\omega_0(A_s) = \omega_0(B_p) = 1$ for all stars $s$ and plaquettes $p$.  
Furthermore, $\omega_0$ is the unique state on $\fA$ satisfying this property, and $\omega_0$ is pure.  
\end{thm}

\begin{proof}
We proceed as in \cite[\textsection 2.2.1]{MR2345476}, using arguments from \cite[\textsection 2]{MR2804555} for details omitted in that paper.
Let $\fA_{XZ}$ be the abelian, unital $*$-algebra generated by the star operators $A_s$ and the plaquette operators $B_p$.  
We define a state $\omega$ on $\fA_{XZ}$ by $\omega(A_s) = \omega(B_p) = 1$ for all star operators $A_s$ and plaquette operators $B_p$.  
Such a state exists since it exists in the boundary-less setting \cite{MR2804555}.
Indeed, if we consider the collection of star and plaquette operators in the boundary-less setting for which the centering vertex or face is on or left of a vertical line, then these operators satisfy the same relations as the star and plaquette operators for the toric code with smooth boundary.  
Hence, restricting the translation-invariant ground state for toric code without boundary (which takes the value 1 on all star and plaquette operators) to the algebra generated by the star and plaquette operators on or left of some vertical line gives a state on $\fA_{XZ}$ for the case with boundary that has the desired properties.  
By Lemma \ref{lem:AFHStateLemma}, the equations $\omega(A_s) = \omega(B_p) = 1$ determine $\omega$ on $\fA_{XZ}$.  

We let $\omega_0$ be a Hahn-Banach extension of $\omega$ to $\fA$.  
Then $\omega_0$ is a ground state by the argument used in \cite{MR2804555}.  
In particular, by Lemma \ref{lem:AFHStateLemma}, we have that for all $X, Y \in \fA_{\loc}$, 
\begin{align*}
-i\omega_0(X^*\delta(Y))
&=
\omega_0(X^*[H_\Lambda, Y])
\\&=
\sum_{\Star(s) \subseteq \Lambda} (\omega_0(X^*Y) - \omega_0(X^*A_sY))
+
\sum_{\plaq(p) \subseteq \Lambda} (\omega_0(X^*Y) - \omega_0(X^*B_pY)),
\end{align*}
where $\Lambda \subseteq \bfB$ is any finite subset satisfying that $\supp(Y) \subseteq \Lambda$ and every star and plaquette intersecting $\supp(Y)$ is contained in $\Lambda$.  
Taking $X = Y$ and using that $A_s \leq I$ and $B_p \leq I$ shows that $-i\omega_0(X^*\delta(X)) \geq 0$ for all $X \in \fA_{\loc}$.  
Hence $\omega_0$ is a ground state.  

We now show that there is only one Hahn-Banach extension $\omega_0$ of $\omega$ to $\fA$.  
We present an argument inspired by the one in \cite[\textsection 2.2.1]{MR2345476}.
It suffices to show that $\omega_0$ is determined on all tensor products of $\sigma^Z$, $\sigma^X$, and $\sigma^Z \sigma^X$, as these matrices along with the identity form a basis for $M_2(\bbC)$. 
Let $X$ be such an operator.  
First, observe that if $X$ anti-commutes with a star or plaquette operator $Y$, then by Lemma \ref{lem:AFHStateLemma}, we have that 
\[
\omega_0(X)
=
\omega_0(YXY)
=
-\omega_0(X)
\]
and hence $\omega_0(X) = 0$.  
We may therefore assume that $X$ does not anti-commute with any star or plaquette operator.  
In addition, if $A_s$ is a star operator and $B_p$ is a plaquette operator, then $XA_s$ and $B_pX$ are also tensor products of $\sigma^Z$, $\sigma^X$, and $\sigma^Z \sigma^X$, as $A_s$ is a tensor product of $\sigma^X$ and $B_p$ is a tensor product of $\sigma^Z$.  
Suppose $j \in \supp(X)$ is a bond in the north-most row of $\supp(X)$.  
If $j$ is a vertical bond, then $X$ must act as $\sigma^X$ on $j$, as otherwise $X$ anti-commutes with the star operator $A_{s_0}$ at the vertex $s_0$ on the north end of $j$.  
In this case, letting $s_1$ denote the vertex at the south end of $j$, we may assume that $X$ acts as the identity on $j$ by replacing $X$ with $XA_{s_1}$, since $\omega_0(XA_{s_1}) = \omega_0(X)$ by Lemma \ref{lem:AFHStateLemma}.
Note that the only bonds other than $j$ affected by replacing $X$ with $XA_{s_1}$ reside in the two rows directly south of $j$.  
Analogously, if $j$ is a horizontal bond, we may assume $X$ acts as the identity on $j$ by replacing $X$ with $B_pX$, where here $B_p$ is the plaquette operator with north-most bond $j$.  
As before, the only other bonds affected by replacing $X$ with $B_pX$ reside in the two rows directly south of $j$.  

Proceeding in this manner, we may assume that $\supp(X)$ is contained in two consecutive rows in the lattice (one with horizontal bonds and one with vertical bonds).  
We claim that at this point, $X$ must be the identity, and hence $\omega_0(X) = 1$.  
Suppose $X \neq I$, so that $\supp(X) \neq \emptyset$.  
Let $j_0 \in \supp(X)$ be the bond in the north-most row of $\supp(X)$ that is furthest to the west.  
Suppose that $j_0$ is a vertical bond.  
(The situation where $j_0$ is horizontal is handled analogously.)
We consider the bond $j_1$ directly to the southwest of $j_0$ (see Figure \ref{fig:AFHGroundStateArgument}).
\begin{figure}[h]
\centering
\includegraphics{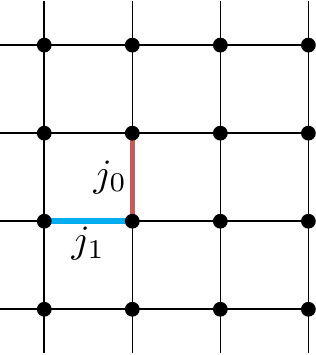}
\caption{The north-east most bond $j_0$ (bold, orange) of an operator $X$ that is the tensor product of $\sigma^Z$, $\sigma^X$, and $\sigma^Z \sigma^X$ and only acts nontrivially on two consecutive rows (here the rows containing $j_0$ and $j_1$).  
The bond $j_1$ (bold, cyan) is the bond directly southwest of $j_0$.}
\label{fig:AFHGroundStateArgument}
\end{figure}
Since $j_0$ is in the north-most row of $\supp(X)$, we have by the argument from before that $X$ acts as $\sigma^X$ on $j_0$.  
Moreover, since $X$ acts as the identity on all bonds south of $j_1$, we must have that $X$ acts as $\sigma^Z$ or the identity on $j_1$, as otherwise $X$ anti-commutes with the plaquette operator whose north-most bond is $j_1$.  
But then $X$ must anticommute with the plaquette operator $B_{p_0}$ containing $j_0$ and $j_1$, since by hypothesis $X$ acts as the identity on the other bonds of this plaquette.  
Hence $\supp(X)$ must be empty, so $X$ is the identity.

Lastly, we have that $\omega_0$ is pure.  
Indeed, suppose $\phi \colon \fA \to \bbC$ is a functional satisfying that $0 \leq \phi \leq \omega_0$.  
Then for all star operators $A_s$ and plaquette operators $B_p$, we have that 
\[
0
\leq
\phi(I - A_s)
\leq
\omega(I - A_s)
=
0,
\qquad\qquad
0
\leq
\phi(I - B_p)
\leq
\omega(I - B_p)
=
0,
\]
so $\phi(A_s) = \phi(I) = \phi(B_p)$.  
Thus by the uniqueness condition for $\omega_0$, $\phi = \phi(I) \cdot \omega_0$, so $\omega_0$ is pure.  
\end{proof}

\begin{rem}
We have that $\omega_0$ is invariant under the action of $\bbZ$ by translations parallel to the boundary, since such translations map star operators to star operators and plaquette operators to plaquette operators.  
\end{rem}

For the remainder of this paper, we let $(\pi_0, \cH, \Omega)$ be the GNS representation for $\omega_0$.  
Note that this means that $\cH$ is the GNS Hilbert space corresponding to $\pi_0$ and $\Omega$ is the canonical cyclic vector.  


\section{Excitations}
\label{sec:ExcitationsAndSuperselectionSectors}


In toric code, pairs of excitations are given by string operators corresponding to finite paths, as described in the following definition.  

\begin{defn}[{\cite{MR1951039, MR2804555}}]
A \emph{finite path of type $Z$} is a path $\gamma$ in the lattice, with endpoints vertices of the lattice (see Figure \ref{fig:FiniteStringOperators}).  
The \emph{string operator} corresponding to this path is the operator $\Gamma_\gamma^Z \coloneqq \bigotimes_{j \in \gamma} \sigma^Z_j$.  
Similarly, a \emph{finite path of type $X$} is a path $\gamma$ in the dual lattice, with endpoints faces of the lattice or the boundary (see Figure \ref{fig:FiniteStringOperators}).  
The \emph{string operator} corresponding to this path is the operator $\Gamma_\gamma^X \coloneqq \bigotimes_{j \in \gamma} \sigma^X_j$.  
(Note that we say that $j \in \gamma$ if $j$ is a bond intersected transversally by $\gamma$.)
Finally, a \emph{finite path of type $Y$} is a ribbon $\gamma$, consisting of a finite path $\gamma^1$ of type $X$ and an adjacent path $\gamma^2$ of type $Z$ (see Figure \ref{fig:FiniteStringOperators}).  
The \emph{string operator} corresponding to this path is the operator $\Gamma_\gamma^Y \coloneqq \Gamma_{\gamma^1}^X \Gamma_{\gamma^2}^Z$.
\begin{figure}[h]
\centering
\includegraphics{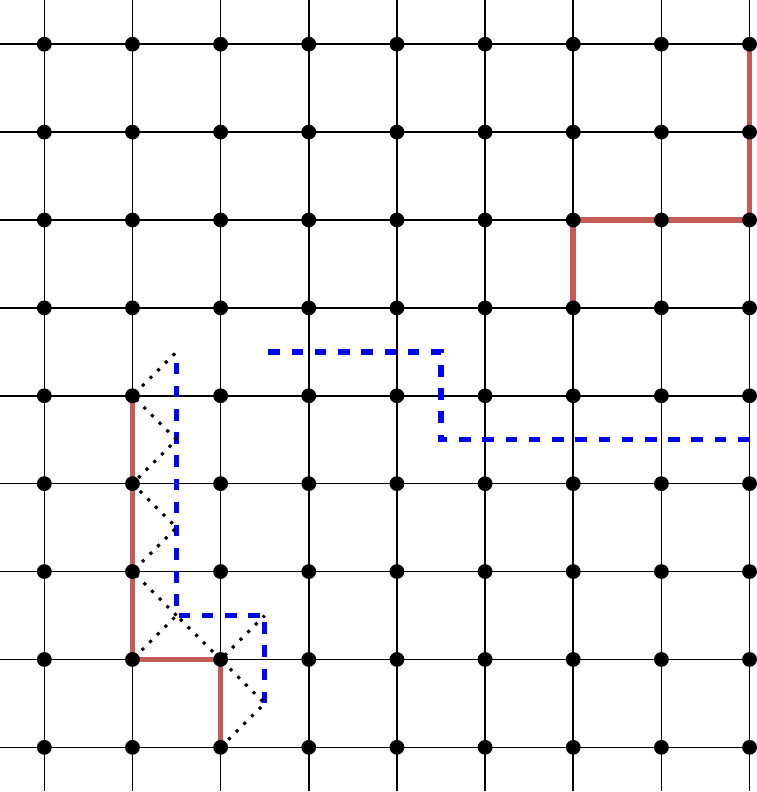}
\caption{Finite paths of types $Z$, $X$, and $Y$, shown from top right to bottom left.
The path of type $Z$ is the bold, orange path in the top right of the figure, the path of type $X$ is the dashed, blue path in the middle of the figure, and the path of type $Y$ is the path consisting of both a dashed, blue component and a bold, orange component in the bottom left of the figure.
The path of type $X$ illustrated is one with one ``endpoint" at the boundary.
The dotted lines in the path of type $Y$ illustrate the components of the ribbon, with the terminal dotted lines corresponding to the sites at the endpoints.}
\label{fig:FiniteStringOperators}
\end{figure}
\end{defn}

The endpoints of paths of type $Z$, $X$, and $Y$ are \emph{sites} of type $Z$, $X$, and $Y$ respectively.  
Note that a site of type $Z$ is a vertex, a site of type $X$ is a plaquette, and a site of type $Y$ is a vertex-plaquette pairing (illustrated in Figure \ref{fig:FiniteStringOperators}).
These sites give the location of excitations corresponding to the endpoints of a string operator.  
Note that a site of type $Z$ can be located on the boundary, but a site of type $X$ or type $Y$ must be located in the bulk of the lattice.  

We wish to create single excitations by extending one endpoint of a path to infinity.  
We will describe these excitations using the algebraic quantum field theory concept of superselection sectors.  
We first define the notion of a cone along the boundary, which will be necessary in order to define a superselection sector.  

\begin{defn}[{\cite[Def.~3.3]{MR2804555}}]
A \emph{cone along the boundary} is a subset $\Lambda \subseteq \bfB$ formed by taking all bonds of the lattice that intersect the region enclosed by two rays making an angle smaller than $\pi/2$, with one of the rays running along the boundary (see Figure \ref{fig:BoundaryCone}).
\begin{figure}[h]
\centering
\includegraphics{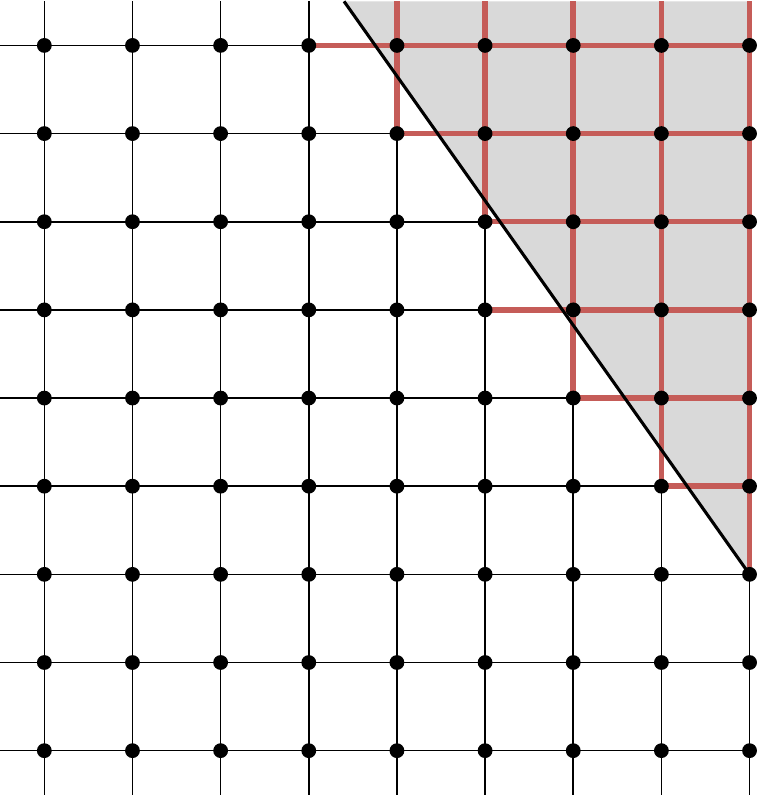}
\caption{
A cone $\Lambda$ along the boundary.  
The bolded orange bonds are those that intersect the shaded region and are the elements of the set $\Lambda \subseteq \bfB$ depicted.  
}
\label{fig:BoundaryCone}
\end{figure}
\end{defn}

\begin{defn}[{\cite{MR4362722, MR3135456}}]
\label{defn:SuperselectionSector}
A \emph{superselection sector} is a representation $\pi$ of $\fA$ satisfying that for any cone $\Lambda$ along the boundary, 
\[
\pi|_{\fA(\Lambda^c)}
\cong
\pi_0|_{\fA(\Lambda^c)},
\]
where the equivalence above is unitary equivalence.  
The above equation is called the \emph{superselection criterion}.
\end{defn}

In our work, we will need a reformulation of Definition \ref{defn:SuperselectionSector} that is easier to work with in practice.  
This will be analogous to the notion of localized and transportable endomorphism in \cite{MR2804555}, but we will rephrase the definitions of localization and transportability so that they refer not to endomorphisms but to $*$-homomorphisms $\pi \colon \fA \to \scrB(\cH)$, where $\cH$ is as usual the GNS Hilbert space corresponding to $\pi_0$.  
Not every superselection sector corresponds to an endomorphism of $\fA$, a fact noted in \cite{MR2804555}, so working with superselection sectors as $*$-homomorphisms into $\scrB(\cH)$ allows for a more comprehensive treatment.  
Furthermore, doing so allows us to avoid identifying $\fA$ with $\pi_0(\fA)$, which is done in \cite{MR2804555}.  

\begin{defn}[{\cite[Def.~3.2]{MR2804555}}]
\label{defn:Localization}
Let $\Lambda \subseteq \bfB$, and let $\pi \colon \fA \to \scrB(\cH)$ be a $*$-homomorphism.  
We say that $\pi$ is \emph{localized} in $\Lambda$ if $\pi(A) = \pi_0(A)$ for all $A \in \fA(\Lambda^c)$.  
\end{defn}

We will specifically consider the case of $*$-homomorphisms that are localized in a cone along the boundary.  
However, it will sometimes be useful for us to consider regions that are not cones, which is why Definition \ref{defn:Localization} is stated in more generality.  
Transportability, on the other hand, is only defined for cones along the boundary.  

\begin{defn}[{\cite[Def.~4.1]{MR2804555}}]
\label{defn:Transportability}
Let $\pi \colon \fA \to \scrB(\cH)$ be a $*$-homomorphism localized in a cone $\Lambda$ along the boundary.  
We say that $\pi$ is \emph{transportable} if for all cones $\widetilde \Lambda$ along the boundary, there exists a $*$-homomorphism $\pi' \colon \fA \to \scrB(\cH)$ localized in $\widetilde \Lambda$ that is unitarily equivalent to $\pi$.  
\end{defn}

\begin{rem}
\label{rem:SuperselectionSectorLocalizedTransportable}
Using arguments in \cite[\textsection 2]{MR3135456}, if $\pi$ is a superselection sector, then $\pi$ is unitarily equivalent to a $*$-homomorphism $\pi' \colon \fA \to \scrB(\cH)$ that is localized in a cone $\Lambda$ along the boundary and is transportable.
We can therefore view superselection sectors as $*$-homomorphisms that are localized and transportable.
\end{rem}

\begin{figure}[h]
\centering
\includegraphics{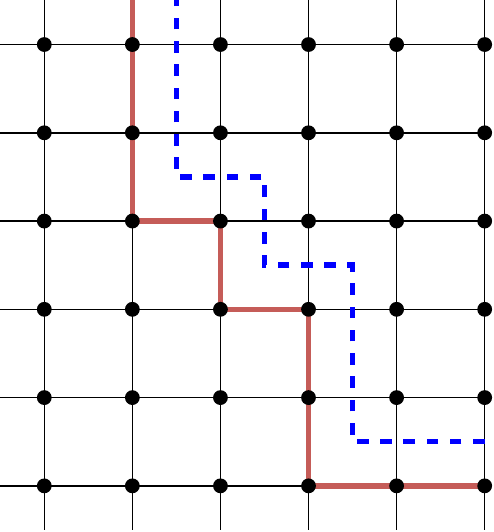}
\caption{
An infinite path $\gamma$ of type $X$ (dashed, blue) starting at the boundary, as well as an infinite path of type $Y$ (dashed, blue combined with bold, orange) starting at the boundary.
}
\label{fig:TypeXandYPathsStartingAtBoundary}
\end{figure}

Let $\Lambda$ be a cone along the boundary.  
We wish to describe superselection sectors $\pi$ localized in $\Lambda$.  
We begin by taking an infinite path $\gamma$ of type $X$, $Y$, or $Z$ starting at the boundary and contained in $\Lambda$.  
A path of type $Z$ starting at the boundary is simply a path of type $Z$ starting at a vertex site $s$ at the boundary.  
Paths of type $X$ and $Y$ starting at the boundary are shown in Figure \ref{fig:TypeXandYPathsStartingAtBoundary}.
By the argument in \cite[Prop.~3.1]{MR2804555}, we have an automorphism $\rho^k_\gamma \colon \fA \to \fA$ given on $A \in \fA_{\loc}$ by 
\[
\rho^k_\gamma(A)
=
\lim_{n \to \infty} \Gamma^k_{\gamma_n} A \Gamma^k_{\gamma_n}.
\]
Here $k$ denotes the type of $\gamma$, and $\gamma_n$ is the path consisting of the first $n$ bonds of $\gamma$.  
Note that $\pi_0 \circ \rho^k_\gamma$ is localized in any region $\Lambda_0 \subseteq \bfB$ such that $\gamma \subseteq \Lambda_0$.  
In particular, we have that $\pi_0 \circ \rho^k_\gamma$ is localized in $\Lambda$.  
In addition, we have that $\rho^k_\gamma \circ \rho^k_\gamma = \mathbbm{1}$, where $\mathbbm{1}$ is the identity automorphism.  

Note that the above discussion did not use that the path $\gamma$ started at the boundary.  
Thus, for a site $s$ in the bulk, we can define automorphisms of all types ($X$, $Y$, and $Z$) as described in \cite[Prop.~3.1]{MR2804555}.
For $k \in \{X, Y, Z\}$ and an appropriate site $s$ (in the bulk or on the boundary), we define $\omega^k_s \coloneqq \omega_0 \circ \rho^k_\gamma$, where $\gamma$ is an infinite path (of the appropriate type) starting at $s$.  
By the argument used in \cite[Lem.~3.1]{MR2804555}, $\omega^k_s$ does not depend on the path $\gamma$.  
Note that if $s$ is on the boundary, then $s$ must be of type $Z$.  

For  $k \in \{X, Y, Z\}$ and an appropriate site $s$, we let $\pi^k_s$ be the GNS representation of $\omega^k_s$.  
We then have the following proposition, which mirrors \cite[Thm.~3.1]{MR2804555}.

\begin{prop}
\label{prop:ToricCodeExcitationsBulkBehavior}
Let $\Lambda$ be a cone along the boundary.  
We then have the following: 
\begin{itemize}
\item
For any $k \in \{X, Y, Z\}$ and appropriate site $s$ in $\Lambda$, $\pi^k_s$ is localized in $\Lambda$, that is, $\pi_0|_{\fA(\Lambda^c)} \cong \pi^k_s|_{\fA(\Lambda^c)}$.
\item
(Type $Z$ particle can be moved to boundary)
For any two vertex sites $s, r$ in $\Lambda$ (bulk or boundary), $\pi_s^Z \cong \pi_r^Z$.
\end{itemize}
\end{prop}

\begin{proof}
Let $k \in \{X, Y, Z\}$ and $s$ be an appropriate site in $\Lambda$.  
The fact that $\pi_0|_{\fA(\Lambda^c)} \cong \pi^k_s|_{\fA(\Lambda^c)}$ follows by an argument given in the proof of \cite[Thm.~3.1]{MR2804555}, but we repeat it here.  
Let $\gamma$ be a path of the appropriate type starting at $s$ and entirely contained in $\Lambda$.  
Then $(\pi_0 \circ \rho^k_\gamma, \cH, \Omega)$ is also a GNS representation for $\omega_s^k$, and $\pi_0 \circ \rho^k_\gamma(A) = \pi_0(A)$ for all $A \in \fA(\Lambda^c)$.  
By uniqueness of the GNS representation, $\pi_0 \circ \rho^k_\gamma \cong \pi^k_s$, so the result follows.  

Similarly, the fact that $\pi_s^Z \cong \pi_r^Z$ for any two vertex sites $s, r$ in $\Lambda$ (bulk or boundary) also follows from an argument given in the proof of \cite[Thm.~3.1]{MR2804555}, but we repeat it here.  
Let $\gamma$ be an infinite path contained in $\Lambda$ of type $Z$ starting at $s$.  
Let $\gamma'$ be a path from $s$ to $r$ of type $Z$ entirely contained in $\Lambda$.  
Then $\omega_r^Z = \omega_0 \circ \Ad \Gamma^Z_{\gamma'} \circ \rho^Z_\gamma$.  
Hence $(\pi_0 \circ \Ad \Gamma^Z_{\gamma'} \circ \rho^Z_\gamma, \cH, \Omega)$ is a GNS representation for $\omega_r^Z$.  
But since $\Gamma_Z^{\gamma'}$ is a unitary, we have that 
\[
\pi_0 \circ \Ad \Gamma^Z_{\gamma'} \circ \rho^Z_\gamma
\cong
\pi_0 \circ \rho^Z_\gamma
\cong
\pi_s^Z.
\]
Hence by uniqueness of the GNS representation, $\pi_s^Z \cong \pi_r^Z$, as desired.
\end{proof}

Furthermore, we obtain condensation of the type $X$ excitations at the boundary, while the type $Z$ excitations do not condense at the boundary.  
This is described in the following theorem

\begin{thm}
\label{thm:ToricCodeBoundaryExcitationsCondensation}
Let $\Lambda$ be a cone along the boundary.  
We then have the following: 
\begin{itemize}
\item
(Type $X$ particle condenses at the boundary)
For any (bulk) plaquette site $s$ in $\Lambda$, $\pi_s^X \cong \pi_0$.   
\item
(Type $Y$ particle becomes type $Z$ at boundary) 
For any (bulk) combined site $s$ in $\Lambda$, $\pi_s^Y \cong \pi_r^Z$, where $r$ is a vertex site on the boundary.  
\item
(Type $Z$ particle does not condense at the boundary) $\pi_s^Z \not \cong \pi_0$ for a vertex site $s$ in $\Lambda$.  
\end{itemize}
\end{thm}

\begin{proof}
We first show show that $\pi_s^X \cong \pi_0$ for any (bulk) plaquette site $s$.
Let $s$ be a plaquette site.  
We let $\gamma$ be an infinite path contained in $\Lambda$ of type $X$ starting at $s$, with $\gamma_n$ the finite path consisting of the first $n$ segments of $\gamma$.  
Let $\gamma'$ be a path from $s$ to the boundary of type $X$ entirely contained in $\Lambda$ (see Figure \ref{fig:TypeXParticleCondensation}).  
\begin{figure}[h]
\centering
\includegraphics{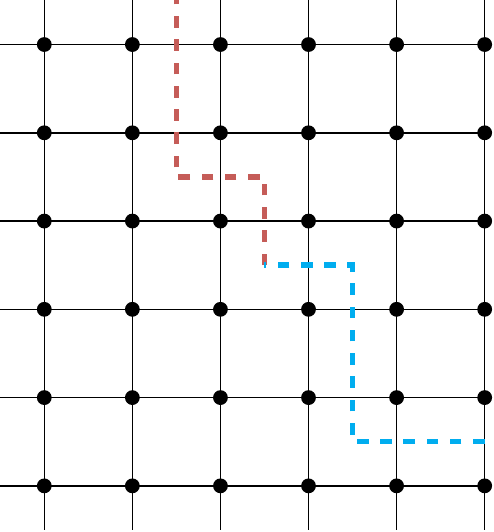}
\caption{
An infinite path $\gamma$ (dashed, orange), along with a path $\gamma'$ (dashed, cyan) from the starting site of $\gamma$ to the boundary.  
}
\label{fig:TypeXParticleCondensation}
\end{figure}
Then for any local operator $A \in \fA_{XZ}$, eventually $A$ commutes with $\Gamma_{\gamma'}^X  \Gamma_{\gamma_n}^X$, since this is true for any star or plaquette operator.  
We thus have that $\Ad \Gamma_{\gamma'}^X \circ \rho^X_\gamma(A) = A$ for all $A \in \fA_{XZ}$, and hence $\omega_0 \circ \Ad \Gamma_{\gamma'}^X \circ \rho^X_\gamma = \omega_0$ since $\omega_0$ is determined by its values on $\fA_{XZ}$.  
We then have that $(\pi_0 \circ \Ad \Gamma_{\gamma'}^X \circ \rho^X_\gamma, \cH, \Omega)$ is a GNS representation for $\omega_0$, so we have that
\[
\pi_0
\cong
\pi_0 \circ \Ad \Gamma_{\gamma'}^X \circ \rho^X_\gamma
\cong
\pi_0 \circ  \rho^X_\gamma
\cong
\pi_s^X,
\]
as desired.

Now, consider a (bulk) combined site $s$ in $\Lambda$.  
Let $\gamma$ be an infinite ribbon path contained in $\Lambda$ starting at $s$, and let $\gamma^1$ and $\gamma^2$ be the paths of type $X$ and $Z$ respectively that comprise $\gamma$.  
Let $\gamma'$ be a path from the plaquette site in $s$ to the boundary, and let $r$ denote the vertex site contained in $s$.  
Let $\gamma^1_n$ and $\gamma^2_n$ denote the finite paths consisting of the first $n$ segments of $\gamma^1$ and $\gamma^2$ respectively.  
Then for any local operator $A \in \fA_{XZ}$, eventually $A$ commutes with $\Gamma_{\gamma'}^X \Gamma_{\gamma^1_n}^X$, so for large enough $n$, 
\[
\Ad \Gamma_{\gamma'}^X \circ \Ad \Gamma_{\gamma^1_n}^X \circ \Ad \Gamma_{\gamma^2_n}^Z(A)
=
\Ad \Gamma_{\gamma^2_n}^Z \circ \Ad \Gamma_{\gamma'}^X \circ \Ad \Gamma_{\gamma^1_n}^X(A)
=
\Ad \Gamma_{\gamma^2_n}^Z(A).
\]
Hence for all local $A \in \fA_{XZ}$, $\Ad \Gamma_{\gamma'}^X \circ \rho^Y_\gamma(A) = \rho^Z_{\gamma^2}(A)$.  
It follows that $\omega_0 \circ \Ad \Gamma_{\gamma'}^X \circ \rho^Y_\gamma = \omega_r^Z$.  
Indeed, $\omega_r^Z = \omega_0 \circ \rho^Z_{\gamma^2}$ takes the value 1 on all star and plaquette operators except for the star operator $A_r$ at the site $r$, as $\omega_r^Z(A_r) = -1$.  
By the argument used in the proof of Theorem \ref{thm:GroundState}, therefore, $\omega_r^Z$ is determined by its values on $\fA_{XZ}$, so $\omega_0 \circ \Ad \Gamma_{\gamma'}^X \circ \rho^Y_\gamma = \omega_r^Z$.  
We then have that $(\pi_0 \circ \Ad \Gamma_{\gamma'}^X \circ \rho^Y_\gamma, \cH, \Omega)$ is a GNS representation for $\omega_r^Z$, so we have that \[
\pi_r^Z
\cong
\pi_0 \circ \Ad \Gamma_{\gamma'}^X \circ \rho^Y_\gamma
\cong
\pi_0 \circ \rho^Y_\gamma
\cong
\pi_s^Y.
\]
The desired result follows by Proposition \ref{prop:ToricCodeExcitationsBulkBehavior}.

Finally, we show that $\pi_s^Z \not \cong \pi_0$ for a vertex site $s$.  
Let $s$ be a vertex site.
We use an argument in the proof of \cite[Thm.~3.1]{MR2804555}, modified to fit our setting.  
For clarity, we repeat the argument in full.  
Since $\omega_0$ is a pure state, the GNS representation $\pi_0$ is irreducible.  
Since $\pi^Z_s$ can be obtained by precomposing $\pi_0$ with an automorphism of $\fA$, $\pi^Z_s$ is also irreducible.  
Hence $\omega_0$ and $\omega^Z_s$ are factor states.  
By \cite[Prop.~10.3.7]{MR1468230}, $\pi_0 \cong \pi^Z_s$ if and only if $\pi_0$ and $\pi^Z_s$ are quasi-equivalent.  
Thus, since $\omega_0$ and $\omega^Z_s$ are factor states, by \cite[Cor.~2.6.11]{MR887100}, in order to show that $\pi_s^Z \not \cong \pi_0$, it suffices to show that there exists $\varepsilon > 0$ such that for all finite sets $\widehat{\Lambda}$ of the lattice, there exists a local operator $B \in \fA(\widehat{\Lambda}^c)$ such that $|\omega_0(B) - \omega_s^Z(B)| \geq \varepsilon \|B\|$.  

Let $\varepsilon = 1$, and let $\widehat{\Lambda}$ be a finite subset of the lattice.  
Without loss of generality, we may assume that $\widehat{\Lambda}$ contains the star at $s$.  
Let $\gamma$ be a non-self-intersecting curve on the dual lattice starting and ending at the boundary such that $\widehat{\Lambda}$ is contained in the interior of $\gamma$ (see Figure \ref{fig:BoundingCurveIllustration}), and let $B \coloneqq \Gamma_\gamma^X$.  
\begin{figure}[h]
\centering
\includegraphics{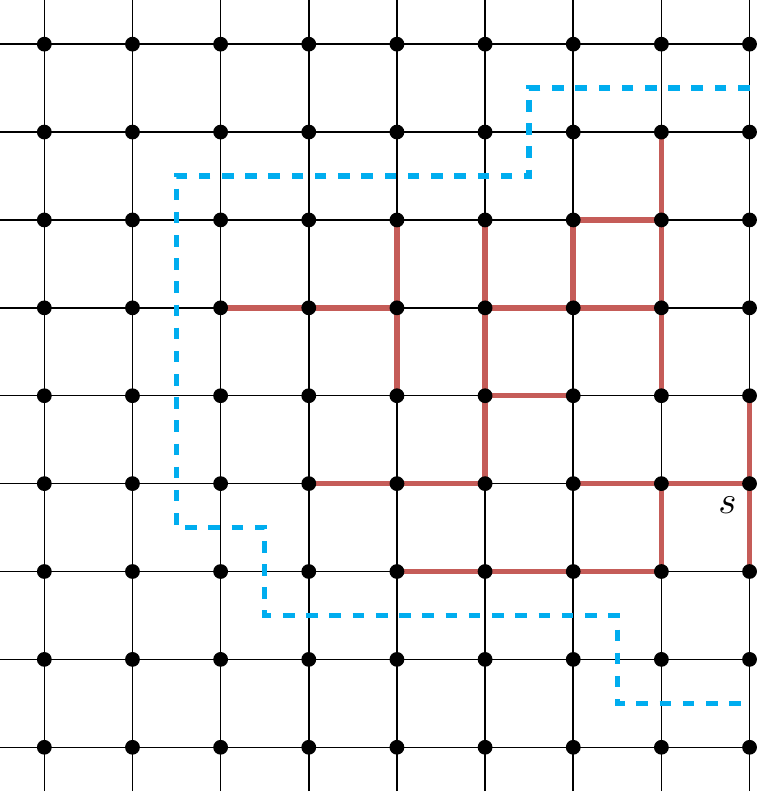}
\caption{A finite subset $\widehat{\Lambda} \subseteq \bfB$ (bolded orange bonds) containing the star at $s$, and a curve $\gamma$ (dashed cyan curve) starting and ending at the boundary with $\widehat{\Lambda}$ contained in the interior.}
\label{fig:BoundingCurveIllustration}
\end{figure}
Then $B \in \fA(\widehat{\Lambda}^c)$ is a local operator, and $B$ is the product of all star operators in the region bounded by $\gamma$, including the star operator at $s$.  
We thus have that $\omega_0(B) = 1$ and $\omega_s^Z(B) = -1$, so 
\[
|\omega_0(B) - \omega_s^Z(B)|
=
2
>
\|B\|.
\qedhere
\]
\end{proof}

Proposition \ref{prop:ToricCodeExcitationsBulkBehavior} and Theorem \ref{thm:ToricCodeBoundaryExcitationsCondensation}, along with their proofs, give the following corollary.

\begin{cor}
Let $k \in \{X, Y, Z\}$ and $s$ be an appropriate site.  
Then $\pi^k_s$ is a superselection sector.  
More specifically, if $\gamma$ is an infinite path of type $k$ localized in the cone $\Lambda$ along the boundary, then $\pi_0 \circ \rho^k_\gamma$ is transportable in addition to being localized in $\Lambda$.  
\end{cor}

For simplicity of notation, we will define $\pi^k_\gamma \coloneqq \pi_0 \circ \rho^k_\gamma$, and we will use this notation for the remainder of this paper.  


\section{Intertwiners and tensor products}
\label{sec:Intertwiners}


We wish to build a unitary fusion category in which the simple objects are given by the boundary excitations described in Theorem \ref{thm:ToricCodeBoundaryExcitationsCondensation}.  
To do so, we need that for two boundary excitations localized in a cone $\Lambda$ along the boundary, the intertwining isomorphisms are contained in $\pi_0(\fA(\Lambda))''$.  
As noted in \cite{MR2804555}, traditionally such a result is proven using Haag duality, which is the statement that for each cone $\Lambda$, the following equality holds: 
\[
\pi_0(\fA(\Lambda))''
=
\pi_0(\fA(\Lambda^c))'.
\]
We will show later that Haag duality holds in this situation (see \textsection\ref{sec:HaagDuality}).
However, there is also a direct argument showing that for toric code without boundary, the intertwining isomorphisms for excitations $\pi_s^Z$ and $\pi_r^Z$ localized in a cone $\Lambda$ are contained in $\pi_0(\fA(\Lambda))''$ \cite{MR2804555}, and this argument holds without modification in the situation with boundary.  
This direct construction will be useful for constructing a tensor functor from the bulk to the boundary (see \textsection\ref{sec:BulkToBoundary}), as well as for bounding the number of excitations (see \textsection\ref{sec:BoundingExcitations}).  

\begin{prop}[{\cite[Lem.~4.1]{MR2804555}}]
\label{prop:IntertwinersBetweenTypeZExcitations}
Let $\gamma^1$ and $\gamma^2$ be two infinite paths of type $Z$ starting at the boundary.  
Then there is a unique unitary $V$ such that $\Ad V \circ \pi^Z_{\gamma^1} = \pi^Z_{\gamma^2}$ and $V\Omega = \Gamma^Z_{\gamma'}\Omega$, where $\Omega$ is the GNS vector for $\omega_0$ and $\gamma'$ is any path of type $Z$ from the starting site of $\gamma^1$ to the starting site of $\gamma^2$.  

Furthermore, if $\widetilde \gamma_n$ is a sequence of paths of type $Z$ from the $n$th vertex of $\gamma^1$ to the $n$th vertex of $\gamma^2$ satisfying that the distances from $\widetilde \gamma_n$ to the starting sites of $\gamma^1$ and $\gamma^2$ go to infinity, then $V = \lim^{WOT} \Gamma^Z_{\gamma^1_n} \Gamma^Z_{\widetilde \gamma_n} \Gamma^Z_{\gamma^2_n}$, where $\gamma^i_n$ is the path consisting of the first $n$ bonds of $\gamma^i$.  
\end{prop}

\begin{rem}
We observe that in the above proposition, we identify the string operators $\Gamma^Z_\gamma$ with their image under the GNS representation $\pi_0$.  
This identification is common in the field of operator algebras.  
In fact, since $\fA$ is a UHF (uniformly hyperfinite) algebra, it is simple, so $\fA \cong \pi_0(\fA)$.  
(For more information on UHF algebras, see \cite{MR1468230}, specifically page 759 and section 12 of that text.)
We will generally try to distinguish $\fA$ and $\pi_0(\fA)$ for clarity, but we will often identify string operators and star and plaquette terms with their image under $\pi_0$.  
We hope that it will be clear from context when we have done so.  
\end{rem}

We also have a similar direct construction for unitaries intertwining condensed type $X$ excitations and the identity.  
We will not need to use this construction once we have shown Haag duality; however, the proof is illustrative of the techniques used in proving \cite[Lem.~4.1]{MR2804555}.  

\begin{prop}
\label{prop:IntertwinersBetweenCondensedXExcitationAndIdentity}
Let $\gamma$ be an infinite path of type $X$ starting at the boundary.  
Then there is a unique unitary $V$ such that $\Ad V \circ \pi^X_\gamma = \pi_0$ and $V\Omega = \Omega$ (where $\Omega$ is the GNS vector for $\omega_0$).  
Furthermore, if $\widetilde \gamma_n$ is a sequence of paths of type $X$ from the $n$th face of $\gamma$ to the boundary satisfying that the distance from $\widetilde \gamma_n$ to the starting bond of $\gamma$ goes to infinity, then $V = \lim^{WOT} \Gamma^X_{\gamma_n} \Gamma^X_{\widetilde \gamma_n}$, where $\gamma_n$ is the path consisting of the first $n$ bonds of $\gamma$.  
\end{prop}

\begin{proof}
First, note that as shown in the proof of \cite[Lem.~4.1]{MR2804555}, we have that $A_s\Omega = \Omega$ for all stars $s$, since 
\begin{equation}
\label{eq:StarOpsFixGNSVector}
\|A_s \Omega - \Omega\|^2
=
\omega_0((A_s - I)^*(A_s - I))
=
\omega_0(2I - 2A_s)
=
0.
\end{equation}
The existence and uniqueness of $V$ follows by the same argument used in \cite[Lem.~4.1]{MR2804555}.  
In particular, $V$ is unique by Schur's Lemma, and $V$ exists by uniqueness of the GNS representation.  

To show that $V = \lim^{WOT} \Gamma^X_{\gamma_n} \Gamma^X_{\widetilde \gamma_n}$, we follow the proof of \cite[Lem.~4.1]{MR2804555}, with modifications to fit our setting.  
For clarity, we repeat the argument in full.  
Note that $\Gamma^X_{\gamma_n} \Gamma^X_{\widetilde \gamma_n}$ is the product of the star operators in the region bounded by $\gamma_n \cup \widetilde \gamma_n$ and the boundary.  
Hence $\Gamma^X_{\gamma_n} \Gamma^X_{\widetilde \gamma_n} \Omega = \Omega$ by \eqref{eq:StarOpsFixGNSVector}.  
Now, if $A \in \fA_{\loc}$, then there exists $N \in \bbN$ such that for all $n \geq N$, 
\[
\supp (A) \cap (\gamma \setminus \gamma_n) 
= 
\supp(A) \cap \widetilde \gamma_n 
= 
\emptyset.
\]
We then have that for all $n \geq N$, 
\[
\Gamma^X_{\gamma_n} \Gamma^X_{\widetilde \gamma_n}
\rho^X_\gamma(A)
\Gamma^X_{\gamma_n} \Gamma^X_{\widetilde \gamma_n}
=
\Gamma^X_{\gamma_n} \Gamma^X_{\widetilde \gamma_n}
\Gamma^X_{\gamma_n}A\Gamma^X_{\gamma_n}
\Gamma^X_{\gamma_n} \Gamma^X_{\widetilde \gamma_n}
=
\Gamma^X_{\widetilde \gamma_n} A \Gamma^X_{\widetilde \gamma_n}
=
A.
\]
Hence, after applying $\pi_0$ to both sides of the above equation, we have that for all $A, B \in \fA_{\loc}$, 
\[
\lim_{n \to \infty}
\langle 
\Gamma^X_{\gamma_n} \Gamma^X_{\widetilde \gamma_n} 
\pi^X_\gamma(A) \Omega |
\pi^X_\gamma(B) \Omega 
\rangle
=
\lim_{n \to \infty}
\langle 
\pi_0(A) \Gamma^X_{\gamma_n} \Gamma^X_{\widetilde \gamma_n} \Omega|
\pi^X_\gamma(B) \Omega 
\rangle
=
\langle \pi_0(A) \Omega| \pi^X_\gamma(B) \Omega \rangle.
\]
On the other hand, we have that for all $A, B \in \fA_{\loc}$, 
\[
\langle V \pi^X_\gamma(A) \Omega| \pi^X_\gamma(B) \Omega \rangle
=
\langle \pi_0(A) V \Omega| \pi^X_\gamma(B) \Omega \rangle
=
\langle \pi_0(A) \Omega| \pi^X_\gamma(B) \Omega \rangle.
\]
Thus, since the sequence $(\Gamma^X_{\gamma_n} \Gamma^X_{\widetilde \gamma_n})$ is uniformly bounded and $\pi^X_\gamma(\fA_{\loc})\Omega$ is dense in $\cH$ (as $\rho^X_\gamma$ is an automorphism of $\fA$ and $\pi^X_\gamma = \pi_0 \circ \rho^X_\gamma$), we obtain that $V = \lim^{WOT} \Gamma^X_{\gamma_n} \Gamma^X_{\widetilde \gamma_n}$, as desired.
\end{proof}

We will also need an explicit description of the intertwiners between two distinct condensed type $X$ excitations in order to construct a tensor functor from the bulk to the boundary in \textsection\ref{sec:BulkToBoundary}.  
We can obtain such an explicit description using \cite[Lem.~4.1]{MR2804555}.

\begin{prop}
\label{prop:IntertwinersBetweenTwoCondensedXExcitations}
Let $\gamma^1$ and $\gamma^2$ be two infinite paths of type $X$ starting at the boundary.  
Then there is a unique unitary $V$ such that $\Ad V \circ \pi^X_{\gamma^1} = \pi^X_{\gamma^2}$ and $V\Omega = \Omega$ (where $\Omega$ is the GNS vector for $\omega_0$).  
Furthermore, if $\widetilde \gamma_n$ is a sequence of paths of type $X$ from the $n$th face of $\gamma^1$ to the $n$th face of $\gamma^2$ satisfying that the distances from $\widetilde \gamma_n$ to the starting bonds of $\gamma^1$ and $\gamma^2$ go to infinity, then $V = \lim^{WOT} \Gamma^X_{\gamma^1_n} \Gamma^X_{\widetilde \gamma_n} \Gamma^X_{\gamma^2_n}$, where $\gamma^i_n$ is the path consisting of the first $n$ bonds of $\gamma^i$.  
\end{prop}

\begin{proof}
As before, existence and uniqueness of $V$ follow by the argument used in \cite[Lem.~4.1]{MR2804555}.  
We let $\widehat{\gamma}^1$ (respectively $\widehat{\gamma}^2$) be the infinite path consisting of all but the first bond of $\gamma^1$ (respectively $\gamma^2$).  
Then by \cite[Lem.~4.1]{MR2804555}, we have that $\widehat{V} \coloneqq \lim^{WOT} \Gamma^X_{\widehat {\gamma}^1_n} \Gamma^X_{\widetilde \gamma_{n + 1}} \Gamma^X_{\widehat {\gamma}^2_n}$ intertwines $\pi^X_{\widehat {\gamma}^1}$ and $\pi^X_{\widehat {\gamma}^2}$.  
(Here $\widehat{\gamma}^i_n$ is the path consisting of the first $n$ bonds of $\widehat{\gamma}^i$.)
We now let $\Gamma^X_1$ (respectively $\Gamma^X_2$) be the Pauli $X$ operator on the first bond of $\gamma^1$ (respectively $\gamma^2$), and we let $V \coloneqq \Gamma^X_1 \widehat{V} \Gamma^X_2 = \Gamma^X_2 \widehat{V} \Gamma^X_1$.  
We then have that $V = \lim^{WOT} \Gamma^X_{\gamma^1_n} \Gamma^X_{\widetilde \gamma_n} \Gamma^X_{\gamma^2_n}$.  
Also, $\rho^X_{\gamma^i} = \Ad \Gamma^X_i \circ \rho^X_{\widehat{\gamma}^i}$ for $i \in \{1, 2\}$, so after applying $\pi_0$, we have that 
\[
V \pi^X_{\gamma^1}(-)
=
\Gamma^X_2 \widehat{V} \Gamma^X_1 \pi^X_{\gamma^1}(-)
=
\Gamma^X_2 \widehat{V} \pi^X_{\widehat{\gamma}^1}(-) \Gamma^X_1
=
\Gamma^X_2 \pi^X_{\widehat{\gamma}^2}(-) \widehat{V} \Gamma^X_1
=
\pi^X_{\gamma^2} (-) \Gamma^X_2 \widehat{V} \Gamma^X_1
=
\pi^X_{\gamma^2} (-) V.
\]
Furthermore, for all $n \in \bbN$, $\Gamma^X_{\gamma^1_n} \Gamma^X_{\widetilde \gamma_n} \Gamma^X_{\gamma^2_n}$ is the product of the star operators in the region enclosed by $\gamma^1_n \cup \widetilde \gamma_n \cup \gamma^2_n$ and the boundary.  
Hence $\Gamma^X_{\gamma^1_n} \Gamma^X_{\widetilde \gamma_n} \Gamma^X_{\gamma^2_n} \Omega = \Omega$ for all $n \in \bbN$, so we have that for all $\xi \in \cH$, 
\[
\langle V \Omega, \xi \rangle
=
\lim_{n \to \infty} \langle \Gamma^X_{\gamma^1_n} \Gamma^X_{\widetilde \gamma_n} \Gamma^X_{\gamma^2_n} \Omega, \xi \rangle
=
\langle \Omega, \xi \rangle
\]
and thus $V \Omega = \Omega$.  
\end{proof}

Finally, we have canonical intertwiners between two type $Y$ excitations, analogous to the intertwiners described in Proposition \ref{prop:IntertwinersBetweenTypeZExcitations}.

\begin{prop}
\label{prop:IntertwinersBetweenTwoCondensedYExcitations}
Let $\gamma^1$ and $\gamma^2$ be two infinite paths of type $Y$ starting at the boundary, with $\gamma^i_k$ denoting the path of type $k$ comprising $\gamma^i$ for $i = 1, 2$ and $k \in \{X, Z\}$.  
Then there is a unique unitary $V$ such that $\Ad V \circ \rho^Y_{\gamma^1} = \rho^Y_{\gamma^2}$ and $V\Omega = \Gamma^Z_{\gamma'}\Omega$, where $\gamma'$ is any path of type $Z$ from the starting site of $\gamma^1_Z$ to the starting site of $\gamma^2_Z$.  
\end{prop}

\begin{proof}
The proof is analogous to that of Proposition \ref{prop:IntertwinersBetweenTwoCondensedXExcitations}.  
As before, uniqueness follows by Schur's Lemma.  
Existence follows by an argument similar to the one used in \cite[Lem.~4.1]{MR2804555}.  
Let $\gamma'$ be a path of type $Z$ from the starting site of $\gamma^1_Z$ to the starting site of $\gamma^2_Z$.  
Note that by the argument in the proof of Theorem \ref{thm:ToricCodeBoundaryExcitationsCondensation}, $\omega_0 \circ \rho_{\gamma^1}^Y = \omega_0 \circ \Ad \Gamma^Z_{\gamma'} \circ \rho_{\gamma^2}^Y$, and thus $(\pi_0 \circ \rho_{\gamma^1}^Y, \cH, \Omega)$ and $(\pi_0 \circ \Ad \Gamma^Z_{\gamma'} \circ \rho_{\gamma^2}^Y, \cH, \Omega)$ are both GNS representations corresponding to this state.  
Hence by uniqueness of the GNS representation, there exists a unitary $\widetilde{V}$ intertwining $\pi_{\gamma^1}^Y$ and $\Ad \Gamma^Z_{\gamma'} \circ \pi_{\gamma^2}^Y$ such that $\widetilde{V} \Omega = \Omega$.  
Now, let $V = \Gamma^Z_{\gamma'} \widetilde{V}$.  
Then 
\[
V \pi_{\gamma^1}^Y(-)
=
\Gamma^Z_{\gamma'} \widetilde{V} \pi_{\gamma^1}^Y(-)
=
\Gamma^Z_{\gamma'} \Gamma^Z_{\gamma'} \pi_{\gamma^2}^Y(-) \Gamma^Z_{\gamma'} \widetilde{V}
=
\pi_{\gamma^2}^Y(-)V,
\]
and $V\Omega = \Gamma^Z_{\gamma'} \widetilde{V}\Omega = \Gamma^Z_{\gamma'}\Omega$, as desired.  
\end{proof}

We want to build a $\rmC^*$-tensor category with objects superselection sectors $\pi \colon \fA \to \scrB(\cH)$ corresponding to paths starting at the boundary and morphisms being intertwiners.  
To motivate the following discussion, we will momentarily identify $\fA \cong \pi_0(\fA)$.  
If we do this, then the superselection sectors corresponding to paths are endomorphisms of $\fA$, so it would make sense to define the tensor product of objects to be composition.  
In this case, the tensor product of morphisms $S \colon \rho_1 \to \rho_2$ and $T \colon \rho_1' \to \rho_2'$ should be $S \otimes T \coloneqq S\rho_1(T)$.  
However, the intertwiners need not live in $\fA$.  
Furthermore, for a more general superselection sector $\pi \colon \fA \to \scrB(\cH)$, $\pi(\fA)$ need not be $\pi_0(\fA)$.  
We remedy the situation similarly to \cite{MR2804555}.  
We let $\fA^{up}$ be the $\rmC^*$-algebra generated by the algebras $\pi_0(\fA(\Lambda))''$, where $\Lambda$ is an upward-oriented cone along the boundary.  
We similarly have a $\rmC^*$-algebra $\fA^{down}$ corresponding to the downward-oriented cones along the boundary.  
Note that $\fA \cong \pi_0(\fA) \subseteq \fA^{up}$ and $\fA \cong \pi_0(\fA) \subseteq \fA^{down}$.  
Furthermore, for all cones $\Lambda$ along the boundary, $\pi_0(\fA(\Lambda))'' \subseteq \fA^{up}$ or $\pi_0(\fA(\Lambda))'' \subseteq \fA^{down}$.  
As in \cite[Prop.~4.2]{MR2804555}, we have that a superselection sector $\pi \colon \fA \to \scrB(\cH)$ corresponding to an infinite path has a unique extension to an endomorphism $\pi^{up} \colon \fA^{up} \to \fA^{up}$, as well as a unique extension to an endomorphism $\pi^{down} \colon \fA^{down} \to \fA^{down}$.  
(Note that in order to view $\fA$ as a subalgebra of $\fA^{up}$ or $\fA^{down}$, one must first identify $\fA$ with $\pi_0(\fA)$.)

\begin{prop}
\label{prop:AuxiliaryEndomorphisms}
Let $\pi \colon \fA \to \scrB(\cH)$ be any superselection sector.  
Then $\pi$ has a unique extension $\pi^{up}$ to $\fA^{up}$ that is WOT-continuous on $\pi_0(\fA(\Lambda))''$ for all upward-oriented cones $\Lambda$ along the boundary.  
Similarly $\pi$ has a unique extension $\pi^{down}$ to $\fA^{down}$ that is WOT-continuous on $\pi_0(\fA(\Lambda))''$ for all downard-oriented cones $\Lambda$ along the boundary.  
Furthermore, if $\pi$ corresponds to an infinite path, then $\pi^{up}(\fA^{up}) \subseteq \fA^{up}$ and $\pi^{down}(\fA^{down}) \subseteq \fA^{down}$, i.e., $\pi^{up}$ and $\pi^{down}$ are endomorphisms of $\fA^{up}$ and $\fA^{down}$ respectively.  
\end{prop}

\begin{proof}
Our proof follows the proof of \cite[Prop.~4.2]{MR2804555}, with modifications when necessary.  
For clarity, we include the entire argument.  
We consider the case of extending $\pi$ to $\fA^{up}$; the proof for $\fA^{down}$ is analogous.  
Let $\Lambda$ be an upward-oriented cone along the boundary.  
Since $\pi$ is transportable, there exists a unitary $V \in \scrB(\cH)$ such that $\pi = \Ad V \circ \widetilde{\pi}$, where $\widetilde{\pi}$ is localized in a downward-oriented cone along the boundary disjoint from $\Lambda$.  
Note that for all $A \in \fA(\Lambda)$, $\widetilde{\pi}(A) = \pi_0(A)$, so $\pi(A) = V\pi_0(A)V^*$ for all $A \in \fA(\Lambda)$.  
Since multiplication is separately WOT-continuous, there is a unique WOT-continuous extension $\widehat \pi$ of $\pi$ to $\pi_0(\fA(\Lambda))''$ given by $\widehat\pi(B) \coloneqq VBV^*$ for $B \in \pi_0(\fA(\Lambda))''$.  
(Note that for $A \in \fA(\Lambda)$, $\pi(A) = \widehat\pi(\pi_0(A))$.)
Now, $\widehat \pi$ is also norm-continuous, so this uniquely determines $\pi^{up}$.  

Now, suppose $\pi$ corresponds to an infinite path.  
To see that $\pi^{up}(\fA^{up}) \subseteq \fA^{up}$, it suffices to show that $\pi^{up}(\pi_0(\fA(\Lambda))'') \subseteq \pi_0(\fA(\Lambda))''$ for all upward-oriented cones $\Lambda$ along the boundary.  
Let $\Lambda$ be an upward-oriented cone along the boundary.  
Then by WOT-continuity of $\pi^{up}$, we have that 
\[
\pi^{up}(\pi_0(\fA(\Lambda))'')
\subseteq
\pi^{up}(\pi_0(\fA(\Lambda)))''
=
\pi(\fA(\Lambda))''.
\]
Now, for $A \in \fA(\Lambda)_{\loc}$, we have that $\pi(A) \in \pi_0(\fA(\Lambda)_{\loc})$.  
Hence $\pi(\fA(\Lambda)) \subseteq \pi_0(\fA(\Lambda))$, which completes the proof.
\end{proof}

The argument in the last paragraph of the above proof gives the following corollary.  

\begin{cor}
\label{cor:AuxiliaryEndomorphismsPreserveConeAlgebras}
Let $\pi$ be a superselection sector corresponding to a path $\gamma$ of any type.  
Suppose $\Lambda \subseteq \bfB$ is contained in an upward-oriented (resp.~downward-oriented) cone along the boundary and satisfies that $\gamma \subseteq \Lambda$.  
Then $\pi^{up}(\pi_0(\fA(\Lambda))'') \subseteq \pi_0(\fA(\Lambda))''$ (resp.~$\pi^{down}(\pi_0(\fA(\Lambda))'') \subseteq \pi_0(\fA(\Lambda))''$).  
In particular, if $\gamma$ is contained in an upward-oriented (resp.~downward-oriented) cone $\Lambda$ along the boundary, then $\pi^{up}(\pi_0(\fA(\Lambda))'') \subseteq \pi_0(\fA(\Lambda))''$ (resp.~$\pi^{down}(\pi_0(\fA(\Lambda))'') \subseteq \pi_0(\fA(\Lambda))''$).  
\end{cor}

\begin{rem}
Once we have shown that our model satisfies Haag duality, we will have that if $\pi \colon \fA \to \scrB(\cH)$ is a superselection sector localized in the cone $\Lambda$ along the boundary, then $\pi(\fA(\Lambda)) \subseteq \pi_0(\fA(\Lambda))''$ by an argument provided in \cite[\textsection 2]{MR3135456}.  
In this case, we have that $\pi^{up}(\fA^{up}) \subseteq \fA^{up}$ if $\Lambda$ is an upward-oriented cone and $\pi^{down}(\fA^{down}) \subseteq \fA^{down}$ if $\Lambda$ is a downward-oriented cone.  
Indeed, suppose $\Lambda$ is an upward-oriented cone; the downward-oriented case is analogous.  
Let $\widetilde \Lambda$ be any other upward-oriented cone.  
Then there exists an upward-oriented cone $\widehat \Lambda$ such that $\Lambda \subseteq \widehat \Lambda$ and $\widetilde \Lambda \subseteq \widehat \Lambda$.  
Since $\pi$ is localized in $\Lambda$, it is also localized in $\widehat \Lambda$.  
Hence we have that 
\[
\pi^{up}(\pi_0(\fA(\widetilde \Lambda))'')
\subseteq 
\pi^{up}(\pi_0(\fA(\widetilde \Lambda)))''
\subseteq
\pi^{up}(\pi_0(\fA(\widehat \Lambda)))''
=
\pi(\fA(\widehat \Lambda))''
\subseteq
\pi_0(\fA(\widehat \Lambda))''
\subseteq
\fA^{up},
\]
so $\pi^{up}(\fA^{up}) \subseteq \fA^{up}$.
\end{rem}

At this point, we can show that all intertwiners between superselection sectors corresponding to infinite paths localized in a cone $\Lambda$ along the boundary live in $\pi_0(\fA(\Lambda))''$.  
The key to proving this is a slightly more general result, which can be used to show that the intertwiners between two type $Y$ excitations live in more restricted regions of the lattice.  

\begin{lem}
\label{lem:ControlOnIntertwinersForCompositeSelectionSectors}
Suppose $\pi_1, \pi_2, \pi_1', \pi_2'$ are superselection sectors corresponding to infinite paths localized in a region $\Lambda \subseteq \bfB$ that is contained in an upward-oriented (resp.~downward-oriented) cone along the boundary.  
For $i = 1, 2$, let $\rho_i$ and $\rho_i'$ be the automorphisms of $\fA$ corresponding to the paths defining $\pi_i$ and $\pi_i'$, so that $\pi_i = \pi_0 \circ \rho_i$ and $\pi_i' = \pi_0 \circ \rho_i'$.
Suppose $S \colon \pi_1 \to \pi_1'$ and $T \colon \pi_2 \to \pi_2'$ are intertwiners such that $S, T \in \pi_0(\fA(\Lambda))''$.  
Then $S\pi_1^{up}(T)$ (resp.~$S\pi_1^{down}(T)$) is an intertwiner from $\pi_0 \circ \rho_1 \circ \rho_2$ to $\pi_0 \circ \rho_1' \circ \rho_2'$, and $S\pi_1^{up}(T) \in \fA(\Lambda)''$ (resp.~$S\pi_1^{down}(T) \in \fA(\Lambda)''$).  
\end{lem}

\begin{proof}
We consider the case where $\Lambda$ is contained in an upward-oriented cone; the downward-oriented case is analogous.
We have that $S\pi_1^{up}(T)$ is an intertwiner from $\pi_0 \circ \rho_1 \circ \rho_2$ to $\pi_0 \circ \rho_1' \circ \rho_2'$, since for all $A \in \fA$, 
\begin{align*}
S\pi_1^{up}(T)\pi_0(\rho_1(\rho_2(A)))
&=
S\pi_1^{up}(T)\pi_1(\rho_2(A))
=
S\pi_1^{up}(T)\pi_1^{up}(\pi_0(\rho_2(A)))
=
S\pi_1^{up}(T\pi_2(A))
\\&=
S\pi_1^{up}(\pi_2'(A)T)
=
S\pi_1^{up}(\pi_0(\rho_2'(A))T)
=
S\pi_1(\rho_2'(A))\pi_1^{up}(T)
\\&=
\pi_1'(\rho_2'(A))S \pi_1^{up}(T)
=
\pi_0(\rho_1'(\rho_2'(A))) S\pi_1^{up}(T).
\end{align*}
Furthermore, $S\pi_1^{up}(T) \in \fA(\Lambda)''$ by Corollary \ref{cor:AuxiliaryEndomorphismsPreserveConeAlgebras}.
\end{proof}

\begin{thm}
\label{thm:IntertwinersLiveInConeAlgebra}
Let $\Lambda$ be a cone along the boundary, and let $\pi$ and $\pi'$ be superselection sectors corresponding to infinite paths contained in $\Lambda$ or vacuum.  
Then any intertwiner from $\pi$ to $\pi'$ is in $\fA(\Lambda)''$.  
\end{thm}

\begin{proof}
Note that if $\pi \not \cong \pi'$, then the only intertwiner $\pi \to \pi'$ is zero.  
Hence, without loss of generality, we have that either $\pi$ and $\pi'$ correspond to paths of the same type, $\pi$ corresponds to a path of type $X$ and $\pi'$ is the identity, or $\pi$ corresponds to a path of type $Y$ and $\pi'$ corresponds to a path of type $Z$.  
In all but the cases where $\pi$ corresponds to a path of type $Y$, the result follows by Schur's Lemma and Proposition \ref{prop:IntertwinersBetweenTypeZExcitations}, Proposition \ref{prop:IntertwinersBetweenTwoCondensedXExcitations}, or Proposition \ref{prop:IntertwinersBetweenCondensedXExcitationAndIdentity}.  
Thus, it suffices to consider the case where $\pi$ corresponds to a path of type $Y$ and $\pi'$ corresponds to a path of type $Z$ or type $Y$.  
In this case, $\pi = \pi_0 \circ \rho_1 \circ \rho_2$, where $\rho_1$ and $\rho_2$ are automorphisms of $\fA$ corresponding to paths of type $X$ and $Z$ respectively.  
Similarly, $\pi' = \pi_0 \circ \rho_1' \circ \rho_2'$, where $\rho_1'$ is an automorphism corresponding  to a path of type $X$ and $\rho_2'$ is either an automorphism corresponding to a path of type $Z$ or the identity.  
By Propositions \ref{prop:IntertwinersBetweenTypeZExcitations}, \ref{prop:IntertwinersBetweenCondensedXExcitationAndIdentity}, and \ref{prop:IntertwinersBetweenTwoCondensedXExcitations}, there exist unitary intertwiners $U_1 \colon \pi_0 \circ \rho_1 \to \pi_0 \circ \rho_1'$ and $U_2 \colon \pi_0 \circ \rho_2 \to \pi_0 \circ \rho_2'$ that live in $\fA(\Lambda)''$.  
By Lemma \ref{lem:ControlOnIntertwinersForCompositeSelectionSectors}, if $\Lambda$ is an upward-oriented cone, then $U_1 (\pi_0 \circ \rho_1)^{up}(U_2) \in \fA(\Lambda)''$ is a unitary intertwiner from $\pi$ to $\pi'$, and by Schur's Lemma, the result follows.  
If $\Lambda$ is downward-oriented, the result follows by using $U_1(\pi_0 \circ \rho_1)^{down}(U_2)$ instead.
\end{proof}

\begin{rem}
\label{rem:BasicFactsAboutSuperselectionSectorExtensions}
Suppose $\pi_1 \colon \fA \to \scrB(\cH)$ and $\pi_2 \colon \fA \to \scrB(\cH)$ are superselection sectors.  
Then $V \in \scrB(\cH)$ is an intertwiner from $\pi_2$ to $\pi_2$ if and only if $V$ intertwines $\pi_1^{up}$ and $\pi_2^{up}$.  
The reverse direction is clear since $\pi_i^{up}$ extends $\pi_i$ for each $i = 1,2$.  
The forward direction follows since $\pi_i^{up}$ is an extension of $\pi_i$ that is WOT-continuous on $\pi_0(\fA(\Lambda))''$ for each upward-oriented cone $\Lambda$ along the boundary and since multiplication is separately WOT-continuous.  
Furthermore, if $\rho_1$ and $\rho_2$ are automorphisms of $\fA$ corresponding to infinite paths, then $(\pi_0 \circ \rho_1 \circ \rho_2)^{up} = (\pi_0 \circ \rho_1)^{up} \circ (\pi_0 \circ \rho_2)$ since the latter is an extension of $\pi_0 \circ \rho_1 \circ \rho_2$ to $\fA^{up}$ that is WOT-continuous on $\pi_0(\fA(\Lambda))''$ for each upward-oriented cone $\Lambda$ along the boundary.  
Analogous results hold if one replaces up with down in all relevant locations.  
\end{rem}

We now fix a cone $\Lambda$ along the boundary.  
Without loss of generality, we may assume that $\Lambda$ is oriented upward.  
For two superselection sectors $\pi_1, \pi_2$ corresponding to paths localized in $\Lambda$, we define $\pi_1 \otimes \pi_2 \coloneqq \pi_1^{up} \circ \pi_2$.  
Furthermore, given intertwiners $S \colon \pi_1 \to \pi_1'$ and $T \colon \pi_2 \to \pi_2'$, we have that $T \in \pi_0(\fA(\Lambda))''$, so we may define $S \otimes T \coloneqq S\pi_1^{up}(T)$.  
Note that since $S \in \pi_0(\fA(\Lambda))''$, by Lemma \ref{lem:ControlOnIntertwinersForCompositeSelectionSectors} and Remark \ref{rem:BasicFactsAboutSuperselectionSectorExtensions}, $S \otimes T \colon \pi_1 \otimes \pi_2 \to \pi_1' \otimes \pi_2'$ and $S \otimes T \in \pi_0(\fA(\Lambda))''$.  
This gives the structure of a $\rmC^*$-tensor category, with tensor unit $\pi_0$.  

Note that the above construction did not use that the superselection sectors corresponded to paths.  
That is, there is a more general $\rmC^*$-tensor category consisting of all superselection sectors localized in $\Lambda$, with the same tensor product as just defined.  
In \textsection\ref{sec:BoundingExcitations}, we will show that the only simple sectors correspond to paths, so we are justified in restricting ourselves in this way.  

We also have that the $\rmC^*$-tensor category of sectors corresponding to paths is \emph{rigid}, meaning that any object $\pi$ has a dual object.  
A \emph{dual object} for an object $\pi$ consists of an object $\overline{\pi}$ along with morphisms $R \colon \pi_0 \to \overline{\pi} \otimes \pi$ and $\overline{R} \colon \pi_0 \to \pi \otimes \overline{\pi}$ satisfying the following \emph{zig-zag equations}: 
\[
\pi(R^*)\overline{R} = I, 
\qquad\qquad
R^*\overline{\pi}(\overline{R}) = I.
\]
Since every automorphism $\rho$ corresponding to an infinite path satisfies that $\rho \circ \rho = \mathbbm{1}$, we have that $\pi \otimes \pi = \pi_0$ for all superselection sectors $\pi$ corresponding to paths.  
Hence every such sector is self-dual, with $R = I$ and $\overline{R} = I$.  

Lastly, by Schur's Lemma, we have that $\End(\pi) \cong \bbC$ for any superselection $\pi$ corresponding to an infinite path.  
Here, $\End(\pi)$ is the space of self-intertwiners of $\pi$, the endomorphisms of $\pi$ in the category we have just constructed.
Thus, we will have a fusion category of excitations if we can take direct sums of superselection sectors.  
To show we can do this, we must first show that $\pi_0(\fA(\Lambda))''$ is an infinite factor.  


\section{Cone algebras}
\label{sec:ConeAlgebras}


In this section, we show that for a cone $\Lambda$ along the boundary, $\pi_0(\fA(\Lambda))''$ is an infinite factor.  
Following \cite{MR2804555}, for a subset $\Lambda \subseteq \bfB$, we define $\cR_\Lambda \coloneqq \pi_0(\fA(\Lambda))''$.  
We then have the following lemma, which is analogous to \cite[Lem.~5.1]{MR2804555}.  

\begin{lem}
For any subset $\Lambda \subseteq \bfB$, we have that $\cR_\Lambda \vee \cR_{\Lambda^c} = \scrB(\cH)$, where $\cH$ is the GNS Hilbert space for $\omega_0$.  
\end{lem}

\begin{proof}
The proof of \cite[Lem.~5.1]{MR2804555} still holds in this setting.  
In particular, note that if $\Lambda \subseteq \bfB$, then $\cR_\Lambda = \bigvee_{b \in \Lambda} \pi_0(\fA(\{b\}))''$.  
Thus, since $\pi_0$ is an irreducible representation, we have that for all $\Lambda \subseteq \bfB$, $\cR_\Lambda \vee \cR_{\Lambda^c} = \pi_0(\fA)'' = \scrB(\cH)$.  
\end{proof}

The key to showing that $\cR_\Lambda$ is an infinite factor will be showing that $\cR_\Lambda$ being finite implies that $\omega_0$ is tracial.  
We will therefore need the following lemma in order to obtain a contradiction.  

\begin{lem}
\label{lem:GroundStateNotTracial}
The state $\omega_0 \colon \fA \to \bbC$ is not tracial.  
\end{lem}

\begin{proof}
Let $\widetilde \gamma$ be a loop of type $Z$, and let $\gamma_1$ and $\gamma_2$ be paths of type $X$ starting at the boundary such that $\gamma_1$ and $\gamma_2$ each intersect $\widetilde \gamma$ at one bond and such that $\gamma_1$ and $\gamma_2$ share the same non-boundary endpoint (see Figure \ref{fig:PathsIllustratingThatGroundStateIsNotTracial}).  
\begin{figure}[h]
\centering
\includegraphics{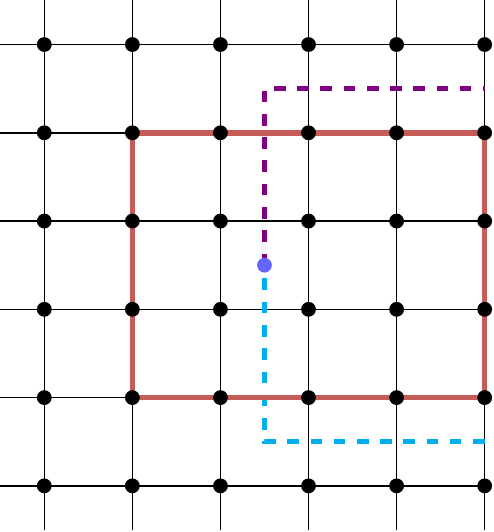}
\caption{A loop $\widetilde \gamma$ of type $Z$ (bold, orange path), and paths $\gamma_1$ and $\gamma_2$ of type $X$ starting at the boundary (dashed, cyan and purple) each intersecting $\widetilde \gamma$ once and sharing the same non-boundary endpoint (blue dot).}
\label{fig:PathsIllustratingThatGroundStateIsNotTracial}
\end{figure}
Then $\Gamma^Z_{\widetilde\gamma}$ is a product of plaquette operators, and $\Gamma^X_{\gamma_1} \Gamma^X_{\gamma_2} = \Gamma^X_{\gamma_2} \Gamma^X_{\gamma_1}$ is a product of star operators since $\gamma_1 \cup \gamma_2$ is a path of type $X$ starting and ending at the boundary.  
Hence $\omega_0(\Gamma^X_{\gamma_1} \Gamma^X_{\gamma_2} \Gamma^Z_{\widetilde\gamma}) = 1$.  
On the other hand, $\Gamma^Z_{\widetilde \gamma}$ and $\Gamma^X_{\gamma_1}$ anti-commute since $\gamma_1$ and $\widetilde\gamma$ intersect at one bond, so we have that 
\[
\omega_0(\Gamma^X_{\gamma_2} \Gamma^Z_{\widetilde\gamma} \Gamma^X_{\gamma_1})
=
-\omega_0(\Gamma^X_{\gamma_2} \Gamma^X_{\gamma_1} \Gamma^Z_{\widetilde\gamma})
=
-1
\neq
\omega_0(\Gamma^X_{\gamma_1} \Gamma^X_{\gamma_2} \Gamma^Z_{\widetilde\gamma}).
\]
Thus $\omega_0$ is not tracial.  
\end{proof}

For $n \in \bbZ$, we let $\tau_n \colon \fA \to \fA$ denote translation by $n$ in the direction parallel to the boundary.  
Recall that $\omega_0$ is translation invariant, so $\omega_0 \circ \tau_n = \omega_0$ for all $n \in \bbZ$.  
Using this, we have the following theorem, analogous to \cite[Thm.~5.1]{MR2804555}.  

\begin{thm}
\label{thm:ConeAlgebraInfiniteFactor}
Let $\Lambda \subseteq \bfB$ be a cone along the boundary.  
Then $\cR_\Lambda$ is an infinite factor.  
\end{thm}

\begin{proof}
We follow the proof of \cite[Thm.~5.1]{MR2804555}, modifying the argument as appropriate for our situation.  
For clarity, we present the argument in full.  
Note that the proof in \cite{MR2804555} is itself an adaptation of an argument from \cite{MR2281418}.  
We first show that $\cR_\Lambda$ is a factor.  
Note that since $\cZ(\cR_\Lambda) = \cR_\Lambda \cap \cR_\Lambda'$, we have that $\cZ(\cR_\Lambda)' \supseteq \cR_\Lambda \vee \cR_\Lambda'$.  
Thus, since $\cR_{\Lambda^c} \subseteq \cR_\Lambda'$, we get that $\cZ(\cR_\Lambda)' = \scrB(\cH)$, from which it follows that $\cR_\Lambda$ is a factor.  

We now show that $\cR_\Lambda$ is an infinite factor.  
We first note that the restriction of $\omega_0$ to $\fA(\Lambda)$ is a factor state.  
Indeed, since $\omega_0$ is a factor state, we have by \cite[Thm.~2.6.10]{MR887100} that for all $\Lambda_1 \subseteq \bfB$ finite and $\varepsilon > 0$, there exists $\Lambda_1' \subseteq \bfB$ finite such that for all $\Lambda_2 \subseteq \bfB$ finite with $\Lambda_2 \cap \Lambda_1' = \emptyset$,
\[
|\omega_0(AB) - \omega_0(A)\omega_0(B)|
\leq
\varepsilon \|A\|\|B\|
\]
for all $A \in \fA(\Lambda_1)$ and $B \in \fA(\Lambda_2)$.
We thus have that the above statement holds replacing $\bfB$ with $\Lambda$, so the restriction of $\omega_0$ to $\fA(\Lambda)$ is a factor state by \cite[Thm.~2.6.10]{MR887100}.

Now suppose, towards contradiction, that $\cR_\Lambda$ is finite.  
Then there exists a normal tracial state $\psi$ on $\cR_\Lambda$, which gives a tracial state $\widetilde \psi \coloneqq \psi \circ \pi_0$ on $\fA(\Lambda)$.  
Since the restriction of $\omega_0$ to $\fA(\Lambda)$ is a factor state, we have by \cite[Prop.~10.3.14]{MR1468230} that $\widetilde \psi$ is quasi-equivalent to the restriction of $\omega_0$ to $\fA(\Lambda)$.  
Furthermore, $\widetilde \psi$ is also a factor state, since quasi-equivalence preserves factoriality.  

We will now show that $\omega_0$ is tracial, contradicting Lemma \ref{lem:GroundStateNotTracial}.  
Let $A, B \in \fA_{\loc}$, and let $\varepsilon > 0$.  
By \cite[Cor.~2.6.11]{MR887100}, there exists a finite subset $\widehat{\Lambda} \subseteq \Lambda$ such that for all $X \in \fA(\Lambda \setminus \widehat{\Lambda})$, $|\omega_0(X) - \widetilde \psi(X)| < \varepsilon \|X\|$.  
Now, since $A$ and $B$ are local operators and $\Lambda$ is a cone along the boundary, there exists $n \in \bbZ$ such that $\tau_n(\supp(A)) \cup \tau_n(\supp(B)) \subseteq \Lambda \setminus \widehat{\Lambda}$.  
We then have that 
\[
|\omega_0(AB) - \widetilde\psi(\tau_n(AB))|
=
|\omega_0(\tau_n(AB)) - \widetilde\psi(\tau_n(AB))|
<
\varepsilon \|A\| \|B\|,
\]
and similarly $|\omega_0(BA) - \widetilde\psi(\tau_n(BA))| < \varepsilon \|A\| \|B\|$.  
Hence we have that 
\[
|\omega_0(AB) - \omega_0(BA)|
\leq
|\omega_0(AB) - \widetilde\psi(\tau_n(AB))|
+
|\omega_0(BA) - \widetilde\psi(\tau_n(BA))|
<
2\varepsilon.
\]
Since $\varepsilon > 0$ was arbitrary, we have that $\omega_0$ is tracial, the desired contradiction.
\end{proof}

Note that Theorem \ref{thm:ConeAlgebraInfiniteFactor} implies that \cite[Cor.~5.1]{MR2804555} holds in this setting.  

\begin{cor}[{\cite[Cor.~5.1]{MR2804555}}]
\label{cor:CuntzAlgebraInConeAlgebra}
Let $\Lambda \subseteq \bfB$ be a cone along the boundary.  
Then there exist isometries $V_1, V_2 \in \cR_\Lambda$ such that $V_1V_1^* + V_2 V_2^* = 1$.  
\end{cor}

We further have that $\cR_\Lambda$ is the hyperfinite $\rmI\rmI_\infty$ factor for a cone $\Lambda$ along the boundary, by the argument in \cite{Ogata_2022}.  

\begin{prop}
Let $\Lambda$ be a cone along the boundary.  
Then $\cR_\Lambda$ is type $\rmI\rmI_\infty$.  
\end{prop}

\begin{proof}
We have that $\cR_\Lambda$ is not type $\rmI$ by adapting the proof of \cite[Prop.~2.2]{MR1828987} to obtain that $\cR_\Lambda$ is type $\rmI$ only if $\omega_0$ is quasiequivalent to $\omega_0|_{\fA(\Lambda)} \otimes \omega_0|_{\fA(\Lambda^c)}$.  The argument in \cite[Thm.~5.1]{MR2804555} showing that this condition is not met holds in this setting as well.  
The fact that $\cR_\Lambda$ is not type $\rmI\rmI\rmI$ follows by adapting the argument in \cite{Ogata_2022} to this setting.  
\end{proof}


\section{The fusion category \(\Delta(\Lambda)\)}
\label{sec:FusionCategoryForBoundaryExcitations}


Since \cite[Cor.~5.1]{MR2804555} holds for cones along the boundary, we have that \cite[Lem.~6.1]{MR2804555} holds for these cones, as the proof of this lemma holds without modification.  

\begin{lem}[{\cite[Lem.~6.1]{MR2804555}}]
Suppose $\pi_1$ and $\pi_2$ are superselection sectors localized in a cone $\Lambda$ along the boundary.  
Then we have a \emph{direct sum} superselection sector $\pi_1 \oplus \pi_2$ that is also localized in $\Lambda$.  
\end{lem}

The direct sum sector $\pi_1 \oplus \pi_2$ is given by $(\pi_1 \oplus \pi_2)(A) \coloneqq V_1 \pi_1(A) V_1^* + V_2 \pi_2(A) V_2^*$ for $A \in \fA$, where $V_1$ and $V_2$ are isometries satisfying the conditions in \cite[Cor.~5.1]{MR2804555}.  
The idea behind this definition is that $\pi_1 \oplus \pi_2$ satisfies the categorical definition of a direct sum.  
Specifically, $V_1$ and $V_2$ witness the inclusion of $\pi_1$ and $\pi_2$ into $\pi_1 \oplus \pi_2$, and $V_1^*$ and $V_2^*$ are the projections from $\pi_1 \oplus \pi_2$ onto $\pi_1$ and $\pi_2$ respectively.  
While $\pi_1 \oplus \pi_2$ is not a direct sum in the Hilbert space sense, one can verify that $V_1$ and $V_2$, along with their adjoints, satisfy the same relations as the inclusion maps do in the case of Hilbert space direct sums.  
This is why we may refer to $\pi_1 \oplus \pi_2$ as a direct sum in our setting.

We fix a cone $\Lambda$ along the boundary.  
Without loss of generality, we assume $\Lambda$ is oriented upward.
We define a fusion category $\Delta(\Lambda)$ as follows.  
The objects of $\Delta(\Lambda)$ are the superselection secotrs given by infinite paths of all types starting at the boundary and contained in $\Lambda$, as well as the direct sums of these sectors.  
(Again, we will show in \textsection\ref{sec:BoundingExcitations} that the superselection sectors corresponding to infinite paths are the only simple sectors, so this restriction is justified.)
The tensor product of objects is defined as it is at the end of \textsection\ref{sec:Intertwiners}, i.e., $\pi_1 \otimes \pi_2 \coloneqq \pi_1^{up} \circ \pi_2$.  
The morphisms in $\Delta(\Lambda)$ are intertwiners.  
Note that because the intertwiners for the simple objects (i.e., the sectors given by infinite paths) live in $\pi_0(\fA(\Lambda))''$ and the isometries witnessing the direct sums also live in $\pi_0(\fA(\Lambda))''$ by Corollary \ref{cor:CuntzAlgebraInConeAlgebra}, we have that all morphisms in $\Delta(\Lambda)$ live in $\pi_0(\fA(\Lambda))''$.  
We can therefore define the tensor product of morphisms as follows: if $S \colon \pi_1 \to \pi_1'$ and $T \colon \pi_2 \to \pi_2'$, then $S \otimes T \coloneqq S\pi_1^{up}(T)$.  
Hence $\Delta(\Lambda)$ is a strict monoidal category.  
Furthermore, since each simple object of $\Delta(\Lambda)$ is self-dual, $\Delta(\Lambda)$ is a unitary fusion category.  
We now show that $\Delta(\Lambda) \cong \mathsf{Hilb_{fd}}(\bbZ/2\bbZ)$, as expected from \cite{MR2942952}.  

\begin{prop}
The fusion category $\Delta(\Lambda)$ is monoidally equivalent to $\mathsf{Hilb_{fd}}(\bbZ/2\bbZ)$.  
\end{prop}

\begin{proof}
We proceed similarly to the proof of \cite[Thm.~6.2]{MR2804555}.  
We view $\mathsf{Hilb_{fd}}(\bbZ/2\bbZ)$ as a skeletal category, meaning that there is exactly one object in each isomorphism class.  
In this case there are two simple objects, $1$ and $g$, in $\mathsf{Hilb_{fd}}(\bbZ/2\bbZ)$, which satisfy that $g \otimes g = 1$.  
In addition, the associators in $\mathsf{Hilb_{fd}}(\bbZ/2\bbZ)$ are trivial.  
We define the functor $F \colon \mathsf{Hilb_{fd}}(\bbZ/2\bbZ) \to \Delta(\Lambda)$ to be the unique linear functor satisfying that $F(1) = \pi_0$ and $F(g) = \pi$, where $\pi$ is a superselection sector corresponding to an infinite path of type $Z$ contained in $\Lambda$.  
Since $\pi \otimes \pi = \pi_0$ and $\mathsf{Hilb_{fd}}(\bbZ/2\bbZ)$ and $\Delta(\Lambda)$ are strict tensor categories, $F$ is a strict tensor functor.  
Furthermore, $F$ is fully faithful by construction, and $F$ is essentially surjective by Proposition \ref{prop:ToricCodeExcitationsBulkBehavior} and Theorem \ref{thm:ToricCodeBoundaryExcitationsCondensation}.  
Thus $F$ is a monoidal equivalence, as desired.  
\end{proof}


\section{Functor from bulk to boundary}
\label{sec:BulkToBoundary}


Let $\Lambda$ be a cone along the boundary.  
We now wish to equip the category $\Delta(\Lambda)$ with the structure of a module tensor category over the category of sectors for bulk toric code, described in \cite[\textsection 6]{MR2804555}.
To do so, we wish to define a braided tensor functor $F$ from the category of bulk toric code sectors to $\cZ(\Delta(\Lambda))$.  
We would like $F$ to be defined as follows: we extend an infinite path $\gamma$ defining a bulk excitation to the boundary, and then we map this boundary excitation to a half-braiding in a way that remembers which type of bulk excitation it came from.  
In what follows, we proceed in the opposite order.  
Namely, given an infinite path $\gamma$ defining a boundary excitation of type $k$, we first define a half-braiding of $\pi^k_\gamma$ that remembers the type $k$ of bulk excitation $\pi^k_\gamma$ corresponds to.  
We will then define a functor that maps bulk excitations to these half-braidings in the way just described and show that this functor is in fact a braided tensor functor.  

\begin{figure}[h]
\centering
\includegraphics{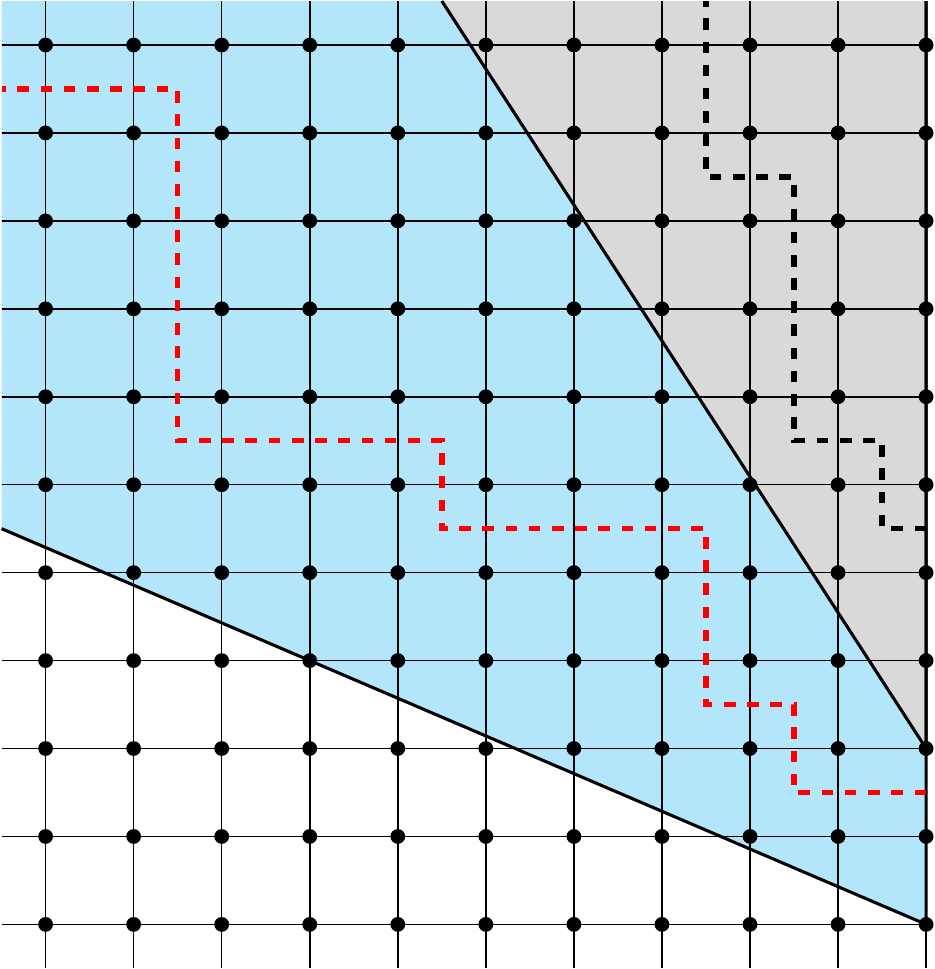}
\caption{A cone $\Lambda$ along the boundary (gray shaded region) with a path $\gamma$ of type $X$ contained in $\Lambda$ (black dashed curve).  
The entire shaded area depicts a cone $\widetilde \Lambda$ containing $\Lambda$ such that there exists a path $\widetilde \gamma$ of type $X$ (red dashed curve) contained in $\widetilde \Lambda \setminus \Lambda$ (cyan shaded region).  
}
\label{fig:HalfBraidingExtraWiggleRoom}
\end{figure}

Let $\Lambda$ be a cone along the boundary.
Without loss of generality, we assume that $\Lambda$ is oriented upward.
Let $\pi$ be a superselection sector corresponding to an infinite path $\gamma$ of type $X$, $Y$, or $Z$, starting at the boundary and contained in $\Lambda$.  
We wish to construct a half-braiding $\sigma_{-, \pi}$, where $\sigma_{\varpi, \pi} \colon \varpi \otimes \pi \to \pi \otimes \varpi$ for $\varpi \in \Delta(\Lambda)$.  
We proceed similarly to the discussion preceding \cite[Lem.~4.2]{MR2804555}.  
Let $\varpi \in \Delta(\Lambda)$. 
Let $\widetilde \Lambda \subseteq \bfB$ be a cone along the boundary such that $\Lambda \subseteq \widetilde \Lambda$ and such that there exists an infinite path $\widetilde \gamma$ of the same type as $\gamma$ starting at the boundary and contained in $\widetilde \Lambda \setminus \Lambda$ (see Figure \ref{fig:HalfBraidingExtraWiggleRoom}).  
Then there exists a unitary $V$ intertwining $\pi$ and $\pi_{\widetilde \gamma}$, and $\pi_{\widetilde \gamma}$ is localized in $\widetilde \Lambda \setminus \Lambda$.  
Furthermore, by Theorem \ref{thm:IntertwinersLiveInConeAlgebra}, we have that $V \in \pi_0(\fA(\widetilde \Lambda))'' \subseteq \fA^{up}$.  
We also have that $\varpi^{up} \circ \pi_{\widetilde \gamma} = (\pi_{\widetilde \gamma})^{up} \circ \varpi$.  
Indeed, $\varpi$ and $\pi_{\widetilde \gamma}$ are localized in disjoint regions, so $\varpi^{up} \circ \pi_{\widetilde \gamma}(A) = (\pi_{\widetilde \gamma})^{up} \circ \varpi(A)$ for all local operators $A$.  
Recalling that $\varpi^{up} \circ \pi_{\widetilde \gamma} = \varpi \otimes \pi_{\widetilde \gamma}$ and $(\pi_{\widetilde \gamma})^{up} \circ \varpi = \pi_{\widetilde \gamma} \otimes \varpi$, we can define an intertwiner 
$\sigma_{\varpi, \pi} \colon \varpi \otimes \pi \to \pi \otimes \varpi$ by 
\begin{equation}
\label{eq:HalfBraidingDefinition}
\sigma_{\varpi, \pi}
\coloneqq
V^* \varpi^{up}(V).
\end{equation}
Note that since $\pi$ and $\pi_{\widetilde \gamma}$ are irreducible representations of $\fA$, we must have that any two unitary intertwiners $V, V'$ between $\pi$ and $\pi_{\widetilde \gamma}$ differ by a scalar.  
Thus $\sigma_{\varpi, \pi}$ does not depend on the choice of unitary $V$.  
We now show that $\sigma_{\varpi, \pi}$ does not depend on the choice of cone $\widetilde \Lambda$ and path $\widetilde \gamma$.  

\begin{prop}
Let $\Lambda$, $\pi$, and $\varpi$ be as in the discussion above.  
Then $\sigma_{\varpi, \pi}$ does not depend on the choice of $\widetilde \Lambda$ and $\widetilde \gamma$ in the discussion above.
\end{prop}

\begin{proof}
We proceed as in the proofs of \cite[Prop.~8.42]{Halvorson_2006} and \cite[Lem.~8.40]{Halvorson_2006}.
Let $\widetilde \Lambda$ and $\widetilde \Lambda'$ be two cones containing $\Lambda$ such that there exist paths $\widetilde \gamma$ and $\widetilde \gamma'$ of the same type as $\gamma$ contained in $\widetilde \Lambda \setminus \Lambda$ and $\widetilde \Lambda' \setminus \Lambda$ respectively.  
We can then take a cone $\widehat{\Lambda}$ containing both $\widetilde \Lambda$ and $\widetilde \Lambda'$, and $\widetilde \gamma$ and $\widetilde \gamma'$ are both contained in $\widehat \Lambda$.
Thus, we may assume without loss of generality that $\widetilde \Lambda = \widetilde \Lambda' = \widehat \Lambda$.

Now, let $V$ be an intertwiner from $\pi$ to $\pi_{\widetilde \gamma}$ and let $V'$ be an intertwiner from $\pi$ to $\pi_{\widetilde \gamma'}$.  
Then $W \coloneqq V'V^*$ is an intertwiner from $\pi_{\widetilde \gamma}$ to $\pi_{\widetilde \gamma'}$.  
Furthermore, by Propositions \ref{prop:IntertwinersBetweenTypeZExcitations} and \ref{prop:IntertwinersBetweenTwoCondensedXExcitations}, Lemma \ref{lem:ControlOnIntertwinersForCompositeSelectionSectors}, and Schur's Lemma, $W \in \pi_0(\fA(\widetilde{\Lambda} \setminus \Lambda))''$.  
Since $\varpi$ is localized in $\Lambda$, $\varpi^{up}$ is the identity on $\pi_0(\fA(\widetilde{\Lambda} \setminus \Lambda))''$.  
Hence $\varpi^{up}(W) = W$, and thus
\[
(V')^*\varpi^{up}(V')
=
V^*W^* \varpi^{up}(WV)
=
V^*W^*W\varpi^{up}(V)
=
V^*\varpi^{up}(V).
\qedhere
\]
\end{proof}

We now show that $\sigma_{-, \pi}$ does in fact define a half-braiding.  

\begin{prop}
Let $\Lambda$ be a cone along the boundary, and let $\pi$ be a superselection sector corresponding to an infinite path $\gamma$ of type $X$, $Y$, or $Z$, starting at the boundary and contained in $\Lambda$.  
Then $\sigma_{-, \pi}$, as defined in \eqref{eq:HalfBraidingDefinition}, is a half-braiding.  
\end{prop}

\begin{proof}
We proceed as in the proof of \cite[Prop.~8.50]{Halvorson_2006}.  
We first show naturality.  
To do so, we must show that if $\varpi, \varpi' \in \Delta(\Lambda)$ and $T$ is an intertwiner from $\varpi$ to $\varpi'$, then 
\[
(I_\rho \otimes T)\sigma_{\varpi, \pi}
=
\sigma_{\varpi', \pi}(T \otimes I_\pi).
\]
Let $\varpi, \varpi' \in \Delta(\Lambda)$, and let $T$ be an intertwiner from $\varpi$ to $\varpi'$.  
Then $T \in \pi_0(\fA(\Lambda))''$.
Let $\widetilde \Lambda \subseteq \bfB$ be a cone along the boundary such that $\Lambda \subseteq \widetilde \Lambda$ and such that there exists an infinite path $\widetilde \gamma$ of the same type as $\gamma$ starting at the boundary and contained in $\widetilde \Lambda \setminus \Lambda$.
Let $V$ be an intertwiner from $\pi$ to $\pi_{\widetilde \gamma}$.  
Since $\pi_{\widetilde \gamma}$ is localized in $\widetilde{\Lambda} \setminus \Lambda$, we have that $(\pi_{\widetilde \gamma})^{up}(T) = T$.
Thus, we have that 
\begin{align*}
(I_\pi \otimes T)\sigma_{\varpi, \pi}
&=
\pi^{up}(T) V^*\varpi^{up}(V)
=
V^*(\pi_{\widetilde \gamma})^{up}(T) \varpi^{up}(V)
\\&=
V^*T\varpi^{up}(V)
=
V^*(\varpi')^{up}(V)T
=
\sigma_{\varpi', \pi}(T \otimes I_\pi).
\end{align*}

We now show that $\sigma_{-, \pi}$ satisfies the braid equation for a half-braiding.  
Let $\varpi, \varpi' \in \Delta(\Lambda)$.  
We must show that 
\[
\sigma_{\varpi \otimes \varpi', \pi}
=
(\sigma_{\varpi, \pi} \otimes I_{\varpi'}) 
(I_\varpi \otimes \sigma_{\varpi', \pi}).
\]
As before, let $\widetilde \Lambda \subseteq \bfB$ be a cone along the boundary such that $\Lambda \subseteq \widetilde \Lambda$ and such that there exists an infinite path $\widetilde \gamma$ of the same type as $\gamma$ starting at the boundary and contained in $\widetilde \Lambda \setminus \Lambda$.
Let $V$ be an intertwiner from $\pi$ to $\pi_{\widetilde \gamma}$.  
We then have that 
\begin{align*}
\sigma_{\varpi \otimes \varpi', \pi}
&=
V^*(\varpi^{up} \circ (\varpi')^{up})(V)
=
V^*\varpi^{up}(VV^*(\varpi')^{up}(V))
\\&=
V^* \varpi^{up}(V) \varpi^{up}(V^*(\varpi')^{up}(V))
=
(\sigma_{\varpi, \pi} \otimes I_{\varpi'}) 
(I_\varpi \otimes \sigma_{\varpi', \pi}).
\qedhere
\end{align*}
\end{proof}

We now wish to construct a functor from the bulk to the boundary that equips the boundary with the structure of a module tensor category.  
Let $\Lambda$ be a cone along the boundary.
On simple objects, we define the functor as follows: we take a superselection sector $\pi_\gamma$ corresponding to a bulk excitation with $\gamma$ an infinite path localized in $\Lambda$, and we extend $\gamma$ to a path $\widetilde \gamma$ starting at the boundary localized in $\Lambda$.  
The functor then maps $\pi_\gamma$ to $F(\pi_\gamma) \coloneqq (\pi_{\widetilde \gamma}, \sigma_{-, \pi_{\widetilde \gamma}})$.  
We also define $F$ to map $\pi_0$ to the trivial half-braiding of $\pi_0$.
On morphisms, we wish to map the canonical intertwiner between $\pi$ and $\pi'$, as described in Propositions \ref{prop:IntertwinersBetweenTypeZExcitations}, \ref{prop:IntertwinersBetweenTwoCondensedXExcitations}, and \ref{prop:IntertwinersBetweenTwoCondensedYExcitations}, to the canonical intertwiner between $F(\pi)$ and $F(\pi')$.  
Since the canonical intertwiner $U \colon \pi \to \pi'$ is characterized by the property that $U\Omega = \Gamma_\gamma \Omega$, if $\gamma$ is a path from the starting site of $\pi$ to the starting site of $\pi'$, this assignment of morphisms is clearly functorial.  
(Note that by ``site" we are only referring to vertex endpoints, not condensed type $X$ excitations.  
If $\rho$ corresponds to a path of type $X$ starting at the boundary, then $\gamma = \emptyset$.)
However, for this assignment to give a well-defined functor, we must show that the canonical intertwiners are half-braiding morphisms.  

\begin{prop}
\label{prop:CanonicalIntertwinersAreHalfBraidingMorphisms}
Let $\Lambda$ be a cone along the boundary, and let $\pi$ and $\pi'$ be superselection sectors corresponding to paths $\gamma$ and $\gamma'$ of the same type contained in $\Lambda$.  
Then the unique unitary intertwiner $U \colon \pi \to \pi'$ satisfying that $U\Omega = \Gamma_{\widehat{\gamma}} \Omega$ for any path $\widehat{\gamma}$ from the starting site of $\gamma$ to the starting site of $\gamma'$ is a morphism from $(\pi, \sigma_{-, \pi})$ to $(\pi', \sigma_{-, \pi'})$.  
\end{prop}

\begin{proof}
Let $\widetilde \Lambda \subseteq \bfB$ be a cone along the boundary such that $\Lambda \subseteq \widetilde \Lambda$ and such that there exists an infinite path $\widetilde \gamma$ of the same type as $\gamma$ starting at the boundary and contained in $\widetilde \Lambda \setminus \Lambda$.
Let $V$ be the canonical intertwiner from $\pi$ to $\pi_{\widetilde \gamma}$, and let $V'$ be the canonical intertwiner from $\pi'$ to $\pi_{\widetilde \gamma}$.  
Then $U = (V')^*V$, since $(V')^*V$ is a unitary intertwiner from $\pi$ to $\pi'$ satisfying that $(V')^*V\Omega = \Gamma_{\widehat{\gamma}} \Omega$ for any path $\widehat{\gamma}$ from the starting site of $\gamma$ to the starting site of $\gamma'$.  
Now, let $\varpi \in \Delta(\Lambda)$.  
Then $\sigma_{\varpi, \pi} = V^*\varpi^{up}(V)$ and $\sigma_{\varpi, \pi'} = (V')^* \varpi^{up}(V')$.  
We thus have that 
\begin{align*}
(U \otimes I_\varpi)\sigma_{\varpi, \pi}
&=
UV^*\varpi^{up}(V)
=
(V')^*\varpi^{up}(V)
=
(V')^*\varpi^{up}(V')\varpi^{up}((V')^*V)
\\&=
\sigma_{\varpi, \pi'} \varpi^{up}(U)
=
\sigma_{\varpi, \pi'} (I_\varpi \otimes U),
\end{align*}
so $U \colon (\pi, \sigma_{-, \pi}) \to (\pi', \sigma_{-, \pi'})$.
\end{proof}

By semisimplicity, defining $F$ on simple objects and morphisms between simple objects, as we have done, uniquely determines $F$.  
We now show that the functor $F$ is a strict tensor functor.  
To prove this, again by semisimplicity, it suffices to show that if $\pi$ and $\pi'$ are superselection sectors corresponding to infinite paths localized in a cone $\Lambda$ along the boundary, then $(\pi \otimes \pi', \sigma_{-, \pi \otimes \pi'}) = (\pi, \sigma_{-, \pi}) \otimes (\pi', \sigma_{-, \pi'})$, where the tensor product on the right is the tensor product in $\cZ(\Delta(\Lambda))$.  
This result follows from the next proposition.  

\begin{prop}
\label{prop:FunctorFromBulkToBoundaryRespectsTensorProduct}
Let $\Lambda$ be a cone along the boundary, and let $\pi$ and $\pi'$ be superselection sectors corresponding to paths $\gamma$ and $\gamma'$ contained in $\Lambda$.  
Then for all $\varpi \in \Delta(\Lambda)$, we have that 
\[
\sigma_{\varpi, \pi \otimes \pi'}
=
(I_\pi \otimes \sigma_{\varpi, \pi'}) (\sigma_{\varpi, \pi} \otimes I_{\pi'}).
\]
\end{prop}

\begin{proof}
Let $\widetilde \Lambda' \subseteq \bfB$ be a cone along the boundary such that $\Lambda \subseteq \widetilde \Lambda'$ and such that there exists an infinite path $\widetilde \gamma'$ of the same type as $\gamma'$ starting at the boundary and contained in $\widetilde \Lambda' \setminus \Lambda$.
Now, let $\widetilde \Lambda \subseteq \bfB$ be a cone along the boundary such that $\widetilde \Lambda' \subseteq \widetilde \Lambda$ and such that there exists an infinite path $\widetilde \gamma$ of the same type as $\gamma$ starting at the boundary and contained in $\widetilde \Lambda \setminus \widetilde \Lambda'$.
Let $V$ be the canonical intertwiner from $\pi$ to $\pi_{\widetilde \gamma}$, and let $V'$ be the canonical intertwiner from $\pi'$ to $\pi_{\widetilde \gamma'}$.  
Note that $\pi_{\widetilde \gamma}$ is localized in $\widetilde \Lambda \setminus \widetilde \Lambda'$, so $(\pi_{\widetilde \gamma})^{up}(V') = V'$ as $V' \in \pi_0(\fA(\widetilde \Lambda'))''$.  
Hence we have that 
\[
V \otimes V'
=
(\pi_{\widetilde{\gamma}})^{up}(V')V
=
V'V.
\]
Since $V \otimes V'$ is a unitary intertwiner from $\pi \otimes \pi'$ to $\pi_{\widetilde\gamma} \otimes \pi_{\widetilde\gamma'}$, we have that for all $\varpi \in \Delta(\Lambda)$, 
\[
\sigma_{\varpi, \pi \otimes \pi'}
=
(V \otimes V')^*\varpi^{up}(V \otimes V')
=
V^*(V')^*\varpi^{up}(V'V).
\]
Finally, for all $\varpi \in \Delta(\Lambda)$, we have that $\varpi^{up}(V') \in \pi_0(\fA(\widetilde \Lambda'))''$ and thus
\begin{align*}
(I_\pi \otimes \sigma_{\varpi, \pi'}) (\sigma_{\varpi, \pi} \otimes I_{\pi'})
&=
\pi^{up}((V')^*\varpi^{up}(V'))V^*\varpi^{up}(V)
=
V^*(\pi_{\widetilde\gamma})^{up}((V')^*\varpi^{up}(V'))\varpi^{up}(V)
\\&=
V^*(V')^*\varpi^{up}(V')\varpi^{up}(V)
=
\sigma_{\varpi, \rho \otimes \rho'}.
\qedhere
\end{align*}
\end{proof}

Using Proposition \ref{prop:FunctorFromBulkToBoundaryRespectsTensorProduct}, we have that the functor $F$ is essentially surjective once we have shown that $(\pi^X, \sigma_{-, \pi^X})$ is not the trivial half-braiding, as in that case $(\pi^X, \sigma_{-, \pi^X})$ and $(\pi^Z, \sigma_{-, \pi^Z})$ tensor generate $\cZ(\Delta(\Lambda))$.  

\begin{prop}
Let $\Lambda$ be a cone along the boundary, and let $\pi$ be a superselection sector corresponding to a path $\gamma$ of type $X$ contained in $\Lambda$ and starting at the boundary.  
Then $(\pi, \sigma_{-, \pi})$ is not isomorphic to $(\pi_0, \sigma_{-, \pi_0})$, where $\sigma_{-, \pi_0}$ is the trivial half-braiding.  
\end{prop}

\begin{proof}
By Proposition \ref{prop:CanonicalIntertwinersAreHalfBraidingMorphisms}, we may assume that $\gamma$ is contained in a cone $\Lambda_0 \subseteq \Lambda$ along the boundary such that there exists a superselection sector $\varpi$ of type $Z$ localized in $\Lambda \setminus \Lambda_0$.  
It suffices to show that a unitary intertwiner $U \colon \pi \to \pi_0$ is not a morphism $(\pi, \sigma_{-, \pi}) \to (\pi_0, \sigma_{-, \pi_0})$ in $\cZ(\Delta(\Lambda))$.  
Note that since $\pi$ is localized in $\Lambda_0$, $U \in \pi_0(\fA(\Lambda_0))''$, so $\varpi^{up}(U) = U$.  
Let $\widetilde \Lambda \subseteq \bfB$ be a cone along the boundary such that $\Lambda \subseteq \widetilde \Lambda$ and such that there exists an infinite path $\widetilde \gamma$ of type $X$ starting at the boundary and contained in $\widetilde \Lambda \setminus \Lambda$.
Let $V$ be the canonical intertwiner from $\pi$ to $\pi_{\widetilde \gamma}$, as described in Proposition \ref{prop:IntertwinersBetweenTwoCondensedXExcitations}.  
Then $\sigma_{\varpi, \pi} = V^*\varpi^{up}(V)$.  
Now, recall from Proposition \ref{prop:IntertwinersBetweenTwoCondensedXExcitations} that $V = \lim^{WOT} \Gamma^X_{\gamma_n} \Gamma^X_{\widehat{\gamma}_n} \Gamma^X_{\widetilde{\gamma}_n}$, where $\gamma_n$ and $\widetilde{\gamma}_n$ are the paths consisting of the first $n$ bonds of $\gamma$ and $\widetilde \gamma$ respectively and $(\widehat{\gamma}_n)$ is a sequence of paths of type $X$ from the $n$th face of $\gamma$ to the $n$th face of $\widetilde \gamma$ satisfying that the distance from $\widehat \gamma_n$ to the starting bonds of $\gamma$ and $\widetilde \gamma$ goes to infinity.  
Note that each path $\widehat\gamma_n$ intersects the path giving rise to $\varpi$ an odd number of times, while $\gamma$ and $\widetilde \gamma$ do not intersect the path giving rise to $\varpi$.  
Hence $\varpi(\Gamma^X_{\gamma_n} \Gamma^X_{\widehat{\gamma}_n} \Gamma^X_{\widetilde{\gamma}_n}) = - \Gamma^X_{\gamma_n} \Gamma^X_{\widehat{\gamma}_n} \Gamma^X_{\widetilde{\gamma}_n}$ for all $n \in \bbN$, so $\varpi^{up}(V) = -V$ by WOT-continuity of $\varpi^{up}$.  
Thus, we have that 
\[
\sigma_{\varpi, \rho}
=
V^*\varpi^{up}(V)
=
-I,
\]
so we have that 
\[
(U \otimes I_\varpi)\sigma_{\varpi, \pi}
=
-U,
\qquad\qquad
\sigma_{\varpi, \pi_0}(I_\varpi \otimes U)
=
\varpi^{up}(U)
=
U.
\]
Hence $U$ is not a morphism $(\pi, \sigma_{-, \pi}) \to (\pi_0, \sigma_{-, \pi_0})$ in $\cZ(\Delta(\Lambda))$.  
\end{proof}

Lastly, we have that $F$ respects the braiding, as we have defined the braiding in the manner of \cite{MR2804555}.  

\begin{thm}
\label{thm:PathSectorCatModuleTensor}
The functor $F$ defined in the preceding paragraphs is a strict braided tensor functor from bulk toric code to $\cZ(\Delta(\Lambda))$.  
Hence, $\Delta(\Lambda)$ is a module tensor category over the category of sectors for bulk toric code.  
\end{thm}

\begin{proof}
It remains to show that $F$ is braided.  
By semisimplicity, it suffices to show that for all superselection sectors $\pi, \pi'$ corresponding to paths in the bulk, we have that $F(\beta_{\pi, \pi'}) = \sigma_{F(\pi), F(\pi')}$, where $\beta_{\pi, \pi'}$ denotes the braiding defined in \cite{MR2804555}.  
Let $\Lambda'$ be a cone (not along the boundary) disjoint from $\Lambda$ such that there exists a cone $\widetilde \Lambda$ along the boundary with $\Lambda \cup \Lambda' \subseteq \widetilde \Lambda$ (see Figure \ref{fig:BraidingDisjointCones}).  
\begin{figure}[h]
\centering
\includegraphics{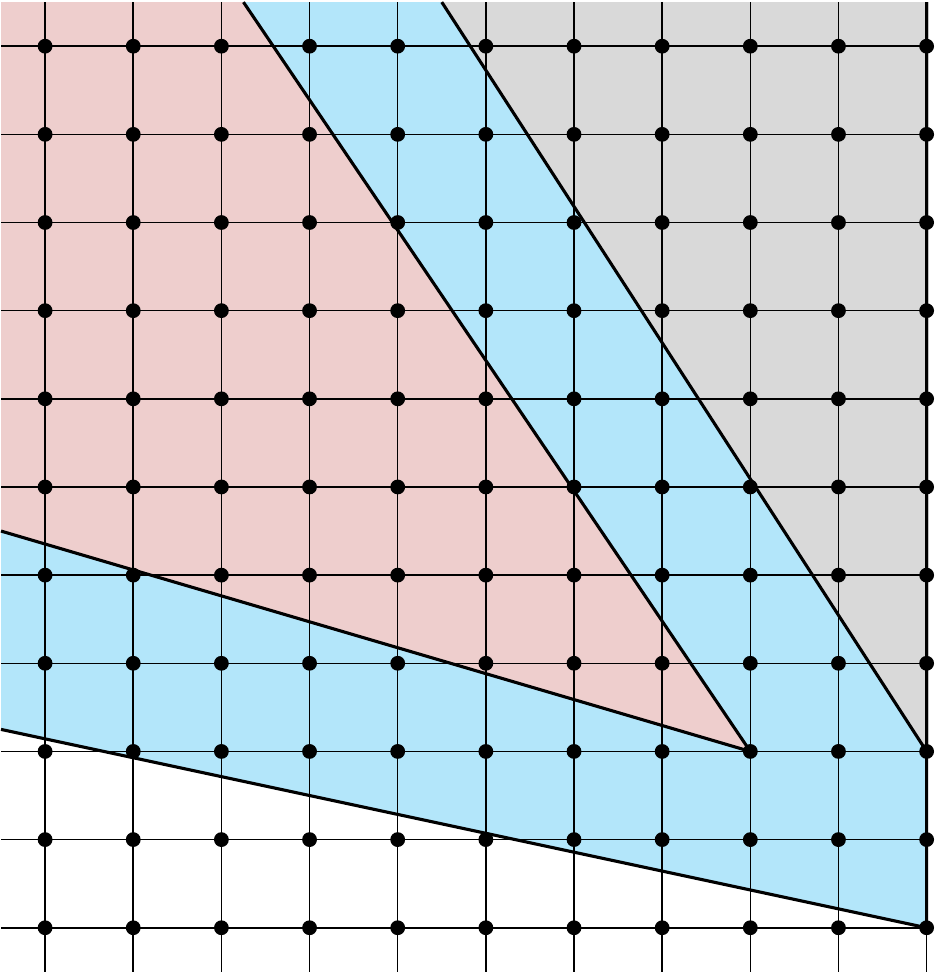}
\caption{A cone $\Lambda$ along the boundary (gray shaded region) along with a cone $\Lambda'$ disjoint from $\Lambda$ (orange shaded region) such that there exists a cone $\widetilde \Lambda$ along the boundary (entire shaded region) with $\Lambda \cup \Lambda' \subseteq \widetilde{\Lambda}$.  
}
\label{fig:BraidingDisjointCones}
\end{figure}
Let $\gamma$ be the path corresponding to $\pi'$, and let $\overline{\gamma}$ be a path in $\Lambda'$ of the same type as $\gamma$.  
Then $\beta_{\pi, \pi'} = V^*\pi^{up}(V)$, where $V$ is a unitary intertwiner from $\pi'$ to $\pi_{\overline{\gamma}}$ (see \cite[p.~365]{MR2804555}).  
Now, we may assume without loss of generality that $\widetilde\Lambda \setminus \Lambda$ contains a path $\widetilde{\gamma}$ starting at the boundary and extending $\overline{\gamma}$.  
Note that we can define a functor $\widetilde{F}$ mapping into $\cZ(\Delta(\widetilde{\Lambda}))$ analogously to how we defined $F$, and $\widetilde{F}$ extends $F$ by construction.  
Furthermore, by how $\widetilde{F}$ is defined, we may assume that $\widetilde{F}(\pi_{\overline{\gamma}}) = \pi_{\widetilde{\gamma}}$.  
Then $\widetilde{F}(V)$ is a unitary intertwiner from $\widetilde{F}(\pi') = F(\pi')$ to $\pi_{\widetilde{\gamma}}$, so $\sigma_{F(\pi), F(\pi')} = \widetilde{F}(V)^*F(\pi)^{up}(\widetilde{F}(V))$.  
Hence, we have that
\[
F(\beta_{\pi, \pi'})
=
\widetilde{F}(\beta_{\pi, \pi'})
=
\widetilde{F}(V^*\pi^{up}(V))
=
\widetilde{F}(V)^*\widetilde{F}(\pi)^{up}(\widetilde{F}(V))
=
\widetilde{F}(V)^*F(\pi)^{up}(\widetilde{F}(V))
=
\sigma_{F(\pi), F(\pi')}.
\qedhere
\]
\end{proof}

Theorem \ref{thm:PathSectorCatModuleTensor} is very close in statement to Theorem \ref{thm:SectorCatModuleTensor}.  
However, we have not yet shown that the only excitations are given by infinite paths.  
In the remainder of the paper, we prove technical results that will allow us to conclude that the only simple superselection sectors are those corresponding to infinite paths.  
This will complete the proof of Theorem \ref{thm:SectorCatModuleTensor}.  


\section{Haag duality for cones along the boundary}
\label{sec:HaagDuality}


In this section, we prove Haag duality for cones along the boundary, i.e., that the following theorem holds.  

\begin{thm}
\label{thm:HaagDuality}
Let $\Lambda$ be a cone along the boundary.  
Then 
\[
\pi_0(\fA(\Lambda))''
=
\pi_0(\fA(\Lambda^c))'.
\]
\end{thm}
Let $\Lambda \subseteq \bfB$ be a cone along the boundary.  
Note that $\pi_0(\fA(\Lambda))'' \subseteq \pi_0(\fA(\Lambda^c))'$ by locality, so we must show that $\pi_0(\fA(\Lambda^c))' \subseteq \pi_0(\fA(\Lambda))''$.  
We also use the notation $\cR_\Lambda \coloneqq \pi_0(\fA(\Lambda))''$ and $\cR_{\Lambda^c} \coloneqq \pi_0(\fA(\Lambda^c))''$, as was done in section \ref{sec:ConeAlgebras}.
Our proof of Haag duality in this setting will follow the proof of Haag duality in \cite[\textsection 3]{MR2956822}, with appropriate modifications.

For any subset $\widetilde \Lambda \subseteq \bfB$, we can consider the set of path operators 
\[
\cF_{\widetilde \Lambda}
\coloneqq
\set{\Gamma_{\gamma}}{\gamma \text{ is a finite path of type $X$ or $Z$ contained in $\widetilde \Lambda$}}.
\]
Note that for paths of type $X$, we include cases where $\gamma$ is a path starting at the boundary.  
In what follows, we specifically consider the cases where $\widetilde \Lambda = \Lambda$ and $\widetilde \Lambda = \Lambda^c$.  
Note that \cite[Lem.~3.3]{MR2956822} holds in this setting, as the proof still holds without modification.  

\begin{lem}[{\cite[Lem.~3.3]{MR2956822}}]
\label{lem:DenseSubspaceOfGNSSpace}
The vector space 
\[
\Span
\set{\Gamma_1 \cdots \Gamma_n 
\widehat{\Gamma}_1 \cdots \widehat{\Gamma}_m\Omega}
{\Gamma_1, \dots,\Gamma_n \in \cF_\Lambda, \widehat{\Gamma}_1, \dots,
\widehat{\Gamma}_m \in \cF_{\Lambda^c}}
\]
is dense in $\cH$, the GNS Hilbert space associated to $\pi_0$.  
\end{lem}

We now consider the vector space 
\[
\cH_\Lambda
\coloneqq
\overline{\Span 
\set{\Gamma_1 \cdots \Gamma_n \Omega}
{\Gamma_1, \dots,\Gamma_n \in \cF_\Lambda}}
\subseteq
\cH,
\]
and we let $P_\Lambda \in \scrB(\cH)$ be the projection onto this subspace.  
We have that $\cH_\Lambda$ is an invariant subspace for $\pi_0(\fA(\Lambda))$ (and hence $\cR_\Lambda$), as the proof of \cite[Lem.~3.5]{MR2956822} holds in this setting.  

\begin{lem}[{\cite[Lem.~3.5]{MR2956822}}]
\label{lem:InvarianceOfLambdaSubspaceUnderLambdaAlgebra}
The subspace $\cH_\Lambda \subseteq \cH$ is invariant for $\pi_0(\fA(\Lambda))$, i.e., $\pi_0(\fA(\Lambda))\cH_\Lambda \subseteq \cH_\Lambda$.  
\end{lem}

Note that \cite[Lem.~3.5]{MR2956822} also includes the statement that the an operator $A \in \cR_\Lambda$ is uniquely determined by its restriction to $\cH_\Lambda$.  
However, the proof given there holds for operators in $\cR_{\Lambda^c}' \supseteq \cR_\Lambda$ as well.  

\begin{lem}
\label{lem:HaagDualityOperatorsDeterminedByRestriction}
An operator $A \in \cR_{\Lambda^c}'$ is uniquely determined by its restriction to $\cH_\Lambda$.  
\end{lem}

\begin{proof}
This result follows by an argument in the proof of \cite[Lem.~3.5]{MR2956822}, but we include the proof here for convenience.
Suppose $A_1, A_2 \in \cR_{\Lambda^c}'$ such that $A_1|_{\cH_\Lambda} = A_2|_{\cH_\Lambda}$.  
By Lemma \ref{lem:DenseSubspaceOfGNSSpace}, it suffices to show that $A_1 \widehat{\Gamma} \Gamma \Omega = A_2 \widehat{\Gamma} \Gamma \Omega$ if $\widehat{\Gamma}$ is a product of operators in $\cF_{\Lambda^c}$ and $\Gamma$ is a product of operators in $\cF_\Lambda$.
Let $\widehat{\Gamma}$ be a product of operators in $\cF_{\Lambda^c}$ and $\Gamma$ be a product of operators in $\cF_\Lambda$.  
Then since $A_1, A_2 \in \cR_{\Lambda^c}'$, $A_1$ and $A_2$ commute with $\widehat{\Gamma}$, so 
\[
A_1 \widehat{\Gamma} \Gamma \Omega
=
\widehat{\Gamma} A_1 \Gamma \Omega
=
\widehat{\Gamma} A_2 \Gamma \Omega
=
A_2 \widehat{\Gamma} \Gamma \Omega.
\qedhere
\]
\end{proof}

Since we wish to show that $\cR_{\Lambda^c}' \subseteq \cR_\Lambda$, we would expect that $\cH_\Lambda$ is also an invariant subspace for $\cR_{\Lambda^c}'$.  
This is true, and it is in fact an important step in proving Haag duality.  
To prove this fact, we need to define the boundary of a cone $\Lambda$ along the boundary.  

\begin{defn}
Let $\Lambda$ be a cone along the boundary.  
We say that a star $s$ or plaquette $p$ is in the \emph{boundary of $\Lambda$} if $s$ (or $p$) contains bonds both in $\Lambda$ and in $\Lambda^c$.  
\end{defn}

We can now prove invariance of $\cH_\Lambda$ under $\cR_{\Lambda^c}'$.  

\begin{lem}
\label{lem:InvarianceOfLambdaSubspaceUnderCommutantOfComplementAlgebra}
The subspace $\cH_\Lambda \subseteq \cH$ is invariant for $\cR_{\Lambda^c}'$, i.e., $\cR_{\Lambda^c}'\cH_\Lambda \subseteq \cH_\Lambda$.  
\end{lem}

\begin{proof}
We follow the proof of \cite[Lem.~3.6]{MR2956822}, modifying the argument as appropriate.  
For clarity, we include the full proof here.
Let $B \in \cR_{\Lambda^c}'$.  
We wish to show that $B\cH_\Lambda \subseteq \cH_\Lambda$.  
Note that by density, it is sufficient to show that $B\Gamma_1 \cdots\Gamma_n \Omega \in \cH_\Lambda$, for all $\Gamma_1, \dots, \Gamma_n \in \cF_\Lambda$.  
Let $\Gamma_1, \dots, \Gamma_n \in \cF_\Lambda$, and let $\xi \coloneqq \Gamma_1 \cdots\Gamma_n \Omega$.  
Again by density, in order to show that $B \xi \in \cH_\Lambda$, it is sufficient to show that $\langle \eta | B \xi \rangle = 0$ for all $\eta \in \cH_\Lambda^\perp$ of the form $\eta = \Gamma \widehat{\Gamma}_1 \cdots \widehat{\Gamma}_m\Omega$, where $\widehat{\Gamma}_1, \dots, \widehat{\Gamma}_m \in \cF_{\Lambda^c}$ and $\Gamma$ is the product of operators in $\cF_\Lambda$.  
We let $\widehat{\Gamma}_1, \dots, \widehat{\Gamma}_m \in \cF_{\Lambda^c}$ and $\Gamma$ be the product of operators in $\cF_\Lambda$, and we set $\eta \coloneqq \Gamma \widehat{\Gamma}_1 \cdots \widehat{\Gamma}_m\Omega$, not necessarily in $\cH_\Lambda^\perp$.  

First, suppose $\eta \in \cH_\Lambda^\perp$ and suppose there exists a star operator $A_s$ or a plaquette operator $B_p$ in $\cR_{\Lambda^c}$ that anti-commutes with $\widehat{\Gamma}_1 \cdots \widehat{\Gamma}_m$.  
We consider the case of a star operator $A_s \in \cR_{\Lambda^c}$ that anti-commutes with $\widehat{\Gamma}_1 \cdots \widehat{\Gamma}_m$; the case of a plaquette operator is treated analagously.  
In this case, since $\cF_\Lambda \subseteq \cR_\Lambda \subseteq \cR_{\Lambda^c}'$, we have that 
\[
\langle \eta|B\xi\rangle
=
\langle \eta|B \Gamma_1 \dots \Gamma_n A_s \Omega\rangle
=
\langle \eta|A_s B\xi \rangle
=
\langle A_s \eta| B\xi \rangle
=
-\langle \Gamma \widehat{\Gamma}_1 \cdots \widehat{\Gamma}_m A_s \Omega|B\xi\rangle
=
-\langle \eta|B\xi\rangle,
\]
and thus $\langle \eta | B\xi \rangle = 0$.  

Now, suppose every star and plaquette operator $A_s, B_p \in \cR_{\Lambda^c}$ commutes with $\widehat{\Gamma}_1 \cdots \widehat{\Gamma}_m$.  
(Note that a star or plaquette operator must either commute or anti-commute with $\widehat{\Gamma}_1 \cdots \widehat{\Gamma}_m$.)
We claim that in this case, $\eta \in \cH_\Lambda$, so $\eta \notin \cH_\Lambda^\perp$ unless $\eta = 0$.  
For all $i = 1, \dots, m$, we let $\gamma_i \subseteq \Lambda^c$ denote the path giving $\widehat{\Gamma}_i$.  
Note that if two paths $\gamma_i$ and $\gamma_j$ share an endpoint, which is a face in the bulk or a vertex in the bulk or along the boundary, we can concatenate $\gamma_i$ and $\gamma_j$ to form a new path.  
We can thus combine the operators $\widehat{\Gamma}_i$ and $\widehat{\Gamma}_j$ in the product $\widehat{\Gamma}_1 \cdots \widehat{\Gamma}_m$, possibly at the expense of a minus sign.  
Proceeding in this way, we can assume that no two paths $\gamma_i$ and $\gamma_j$ share an endpoint, not including the ``endpoint" of the boundary for paths of type $X$.  
Furthermore, if $\gamma_i$ is a path that is a loop or a path of type $X$ starting and ending at the boundary, then $\widehat \Gamma_i$ is a product of star or plaquette operators, and hence $\widehat \Gamma_i \Omega = \Omega$.  
We can therefore remove from the product $\widehat{\Gamma}_1 \cdots \widehat{\Gamma}_m$ any $\widehat \Gamma_i$ corresponding to loops or paths of type $X$ starting and ending at the boundary, again possibly at the expense of a minus sign.

With these simplifications, we have that the endpoints of the paths $\gamma_1, \dots, \gamma_m$ are all disjoint, no path $\gamma_i$ is a loop, and no path $\gamma_i$ of type $X$ both starts and ends on the boundary.  
Note that a star or plaquette operator acting on an endpoint site of a path $\gamma_i$ must anti-commute with $\widehat \Gamma_i$.  
Thus, by assumption, any such star or plaquette must be in the boundary of $\Lambda$.  
If $\gamma_i$ is a path with both endpoints sites in the lattice (i.e., $\gamma_i$ is not a path of type $X$ starting at the boundary), then there exists a path $\gamma_i' \subseteq \Lambda$ with the same endpoints as $\gamma_i$.  
On the other hand, if $\gamma_i$ is a path of type $X$ starting at the boundary, then there exists a path $\gamma_i' \subseteq \Lambda$ of type $X$ starting at the boundary, with $\gamma_i'$ and $\gamma_i$ sharing the same non-boundary endpoint.  
For $i = 1, \dots, m$, we let $\Gamma_i'$ be the string operator associated to the path $\gamma_i'$.  
Then for all $i$, $\Gamma_i' \widehat{\Gamma}_i$ is a product of star or plaquette operators since $\gamma_i \cup \gamma_i'$ is a loop or a path of type $X$ starting and ending at the boundary.  
Hence for all $i$, $\Gamma_i' \Omega = \widehat{\Gamma}_i \Omega$, so we have that 
\[
\eta
=
\Gamma \widehat{\Gamma}_1 \cdots \widehat{\Gamma}_m \Omega
=
\pm\Gamma \Gamma'_1 \cdots \Gamma'_m \Omega
\in
\cH_\Lambda,
\]
as desired.
\end{proof}

Note that Lemma \ref{lem:DenseSubspaceOfGNSSpace} and Lemma \ref{lem:InvarianceOfLambdaSubspaceUnderCommutantOfComplementAlgebra}, coupled with a standard result of von Neumann algebra theory, give the following corollary.  

\begin{cor}
If $P_\Lambda \in \scrB(\cH)$ is the projection onto $\cH_\Lambda$, then $P_\Lambda \in \cR_\Lambda'$ and $P_\Lambda \in \cR_{\Lambda^c}$.  
\end{cor}

We now let $\cA_\Lambda \coloneqq \cR_\Lambda P_\Lambda \subseteq \scrB(\cH_\Lambda)$ and $\cB_\Lambda \coloneqq P_\Lambda \cR_{\Lambda^c} P_\Lambda \subseteq \scrB(\cH_\Lambda)$.  
Note that $\cA_\Lambda$ and $\cB_\Lambda$ are von Neumann algebras by a standard result of von Neumann algebra theory.  
Furthermore, $\Omega \in \cH_\Lambda$ is a cyclic vector for $\cA_\Lambda$, by how $\cH_\Lambda$ was defined.  
We let $\cA_{\Lambda, \sa}$ and $\cB_{\Lambda, \sa}$ denote the self-adjoint elements of $\cA_\Lambda$ and $\cB_\Lambda$ respectively.  
The key step in proving Theorem \ref{thm:HaagDuality} is the following lemma, which is analogous to \cite[Lem.~3.8]{MR2956822}.  

\begin{lem}
\label{lem:HaagRieffelVanDaeleLemma}
The real vector space $\cA_{\Lambda, \sa}\Omega + i \cB_{\Lambda, \sa}\Omega$ is dense in $\cH_\Lambda$.  
\end{lem}

\begin{proof}
We follow the proof of \cite[Lem.~3.8]{MR2956822}, modifying it as appropriate.  
For clarity, we include the proof in full.
Since $\cA_{\Lambda, \sa}\Omega + i \cB_{\Lambda, \sa}\Omega$ is a real vector space, it is sufficent to show that $\Gamma \Omega \in \cA_{\Lambda, \sa}\Omega + i \cB_{\Lambda, \sa}\Omega$ and $i \Gamma \Omega \in \cA_{\Lambda, \sa}\Omega + i \cB_{\Lambda, \sa}\Omega$ for all products $\Gamma$ of operators in $\cF_\Lambda$.  
Let $\Gamma \coloneqq \Gamma_1 \cdots \Gamma_n$, where $\Gamma_i \in \cF_\Lambda$ for all $i$.  
Note that each of the $\Gamma_i$ are self-adjoint, and two operators $\Gamma_i$ and $\Gamma_j$ either commute or anti-commute.  
Hence either $\Gamma^* = \Gamma$ or $\Gamma^* = -\Gamma$.  
Now, we have that $\Gamma P_\Lambda \in \cA_\Lambda$ by how $\cA_\Lambda$ was defined.  
Thus, if $\Gamma^* = \Gamma$, then $\Gamma \Omega \in \cA_{\Lambda, \sa} \Omega$, while if $\Gamma^* = - \Gamma$, then $i\Gamma \Omega \in \cA_{\Lambda, \sa} \Omega$.  

Now, suppose there exists a star operator $A_s$ or a plaquette operator $B_p$ in $\cR_\Lambda$ that anti-commutes with $\Gamma$.  
We consider the case of a star operator $A_s \in \cR_\Lambda$ that anti-commutes with $\Gamma$; the case of a plaquette operator is treated analogously.  
If $\Gamma^* = \Gamma$, then $iA_s\Gamma P_\Lambda \in \cA_{\Lambda}$ is self-adjoint, and thus
\[
i \Gamma \Omega
=
i\Gamma A_s \Omega
=
-iA_s \Gamma \Omega
\in
\cA_{\Lambda, \sa} \Omega.
\]
On the other hand, if $\Gamma^* = - \Gamma$, then $A_s \Gamma P_\Lambda \in \cA_{\Lambda}$ is self-adjoint, and thus 
\[
\Gamma \Omega
=
\Gamma A_s \Omega
=
-A_s \Gamma \Omega
\in
\cA_{\Lambda, \sa} \Omega.
\]
We thus have that $\Gamma \Omega \in \cA_{\Lambda, \sa}\Omega + i \cB_{\Lambda, \sa}\Omega$ and $i\Gamma \Omega \in \cA_{\Lambda, \sa}\Omega + i \cB_{\Lambda, \sa}\Omega$ if there exists a star operator or plaquette operator in $\cR_\Lambda$ that anti-commutes with $\Gamma$.  

Now, suppose every star and plaquette operator in $\cR_\Lambda$ commutes with $\Gamma$.  
(Recall that a star or plaquette operator must either commute or anti-commute with $\Gamma$.)
For all $i = 1, \dots, n$, we let $\gamma_i \subseteq \Lambda$ be the path corresponding to the string operator $\Gamma_i$.  
By the argument in the proof of Lemma \ref{lem:InvarianceOfLambdaSubspaceUnderCommutantOfComplementAlgebra}, we may assume that the endpoints of the paths $\gamma_1, \dots, \gamma_n$ are all disjoint, no path $\gamma_i$ is a loop, and no path $\gamma_i$ of type $X$ both starts and ends on the boundary.  
By the same argument as in that proof, any star or plaquette at an ending site of a path $\gamma_i$ must be in the boundary of $\Lambda$.  
Thus, if $\gamma_i$ is a path with both endpoints sites in the lattice (i.e., $\gamma_i$ is not a path of type $X$ starting at the boundary), then there exists a path $\gamma_i' \subseteq \Lambda^c$ with the same endpoints as $\gamma_i$.  
On the other hand, if $\gamma_i$ is a path of type $X$ starting at the boundary, then there exists a path $\gamma_i' \subseteq \Lambda^c$ of type $X$ starting at the boundary, with $\gamma_i'$ and $\gamma_i$ sharing the same non-boundary endpoint.
For $i = 1, \dots, n$, we let $\widehat{\Gamma}_i$ be the string operator associated with $\gamma_i'$, and we let $\widehat{\Gamma} \coloneqq \widehat{\Gamma}_1 \cdots \widehat{\Gamma}_n$.  
Then $\widehat{\Gamma} \in \cR_{\Lambda^c}$, and by the same argument as in the proof of Lemma \ref{lem:InvarianceOfLambdaSubspaceUnderCommutantOfComplementAlgebra}, we have that $\widehat{\Gamma}\Omega = \pm\Gamma \Omega$.  

We now claim that $\Gamma^* = \Gamma$ if and only if $\widehat{\Gamma}^* = \widehat{\Gamma}$.  
Let $i, j \in \{1, \dots, n\}$, $i \neq j$.  
We claim that $\Gamma_i$ and $\Gamma_j$ commute if and only if $\widehat{\Gamma}_i$ and $\widehat{\Gamma}_j$ commute, which is sufficient to prove the desired claim.  
Indeed, we first note that $\gamma_i$ and $\gamma_j$ are paths of the same type if and only if $\gamma_i'$ and $\gamma_j'$ are paths of the same type, so we may restrict ourselves to the case where $\gamma_i$ and $\gamma_j$ are paths of different type.  
Without loss of generality, we may assume that $\gamma_i$ and $\gamma_i'$ are paths of type $Z$ (i.e., paths on the lattice, not the dual lattice).  
In this case, $\gamma_i \cup \gamma_i'$ is a loop on the lattice.  
Thus, $\gamma_i \cup \gamma_i'$ intersects $\gamma_j \cup \gamma_j'$ in an even number of bonds.  
Hence, $\gamma_i$ and $\gamma_j$ intersect in an even number of bonds if and only if $\gamma_i'$ and $\gamma_j'$ intersect in an even number of bonds, so $\Gamma_i$ and $\Gamma_j$ commute if and only if $\widehat \Gamma_i$ and $\widehat \Gamma_j$ do.  

We can now complete the proof that $\Gamma \Omega \in \cA_{\Lambda, \sa}\Omega + i \cB_{\Lambda, \sa}\Omega$ and $i\Gamma \Omega \in \cA_{\Lambda, \sa}\Omega + i \cB_{\Lambda, \sa}\Omega$.  
Note that we have already shown that if $\Gamma^* = \Gamma$, then $\Gamma \Omega \in \cA_{\Lambda, \sa}\Omega + i \cB_{\Lambda, \sa}\Omega$, and if $\Gamma^* = - \Gamma$, then $i \Gamma \Omega \in \cA_{\Lambda, \sa}\Omega + i \cB_{\Lambda, \sa}\Omega$.  
If $\Gamma^* = \Gamma$, then $\widehat{\Gamma}^* = \widehat{\Gamma}$, so $P_\Lambda \widehat{\Gamma} P_\Lambda \in \cB_{\Lambda, \sa}$.  
We thus have in this case that 
\[
i\Gamma \Omega
=
i P_\Lambda \Gamma \Omega
=
\pm i P_\Lambda \widehat{\Gamma} \Omega
=
\pm i P_\Lambda \widehat{\Gamma} P_\Lambda \Omega
\in
i \cB_{\Lambda, \sa} \Omega.
\]
On the other hand, if $\Gamma^* = -\Gamma$, then $\widehat{\Gamma}^* = - \widehat{\Gamma}$, and hence $i P_\Lambda \widehat{\Gamma} P_\Lambda \in \cB_{\Lambda, \sa}$.  
Thus, in this case, we have that 
\[
\Gamma \Omega
=
P_\Lambda \Gamma \Omega
=
\pm P_\Lambda \widehat{\Gamma} \Omega
=
\pm P_\Lambda \widehat{\Gamma} P_\Lambda \Omega
\in
i \cB_{\Lambda, \sa} \Omega.
\]
We have thus shown that $\Gamma \Omega \in \cA_{\Lambda, \sa}\Omega + i \cB_{\Lambda, \sa}\Omega$ and $i\Gamma \Omega \in \cA_{\Lambda, \sa}\Omega + i \cB_{\Lambda, \sa}\Omega$ in all possible cases, as desired.
\end{proof}

Theorem \ref{thm:HaagDuality} now follows by the proof of \cite[Thm.~3.1]{MR2956822}, which we repeat here for convenience.  

\begin{proof}[Proof of Theorem \ref{thm:HaagDuality}]
It remains to show that $\cR_{\Lambda^c}' \subseteq \cR_\Lambda$.  
Note that by Lemma \ref{lem:HaagRieffelVanDaeleLemma} and \cite[Thm.~2]{MR383096}, we have that $\cA_\Lambda = \cB_\Lambda'$.  
The result then follows by Lemma \ref{lem:HaagDualityOperatorsDeterminedByRestriction}, since $\cA_\Lambda = \cR_\Lambda P_\Lambda$ and $\cB_\Lambda' = \cR_{\Lambda^c}' P_\Lambda$.  
\end{proof}


\section{Distal split property}
\label{sec:DistalSplit}


We wish to show that there are only two nonequivalent simple superselection sectors, namely the vacuum $\pi_0$ and the type $Z$ excitation $\pi_Z$.  
We will show this by using the machinery developed in \cite[\textsection{3}]{MR3135456} with an argument analogous to the one in \cite[\textsection{4}]{MR3135456}.  
To do so, we will need to show that there exists a relation $\Lambda_1 \ll \Lambda_2$ on cones $\Lambda_1, \Lambda_2$ along the boundary such that if this relation is satisfied, then there exists a type I factor $\cN$ such that $\cR_{\Lambda_1} \subseteq \cN \subseteq \cR_{\Lambda_2}$.  
This property is called the \emph{distal split property} \cite[Def.~5.1]{MR2804555}.  

\begin{defn}[{\cite{MR2804555}}]
Let $\Lambda_1, \Lambda_2 \subseteq \bfB$ be cones along the boundary.  
We say $\Lambda_1 \ll \Lambda_2$ if $\Lambda_1 \subseteq \Lambda_2$ and if a star or plaquette is contained in $\Lambda_1 \cup \Lambda_2^c$, then this star or plaquette is either contained in $\Lambda_1$ or $\Lambda_2^c$.  
\end{defn}

Using Haag duality, there is a quick argument to show that if $\Lambda_1 \ll \Lambda_2$, then there exists a type I factor $\cN$ such that $\cR_{\Lambda_1} \subseteq \cN \subseteq \cR_{\Lambda_2}$, which proves that $\omega_0$ satisfies the distal split property (see \cite[Thm.~5.2]{MR2804555}).  
However, in order to apply the argument in \cite[\textsection{4}]{MR3135456}, we will need to adapt the more direct proof found in \cite[\textsection{4}]{MR2956822} to the case of toric code with boundary.  
The following proof closely mirrors the one in \cite[\textsection{4}]{MR2956822} but is modified as appropriate for our setting.  

We fix $\Lambda_1, \Lambda_2$ cones along the boundary such that $\Lambda_1 \ll \Lambda_2$.  
We let $\Lambda_0 \coloneqq \Lambda_1^c \cup \Lambda_2$.  
We fix two vertex sites $v_1$ on the boundary of $\Lambda_1$ and $v_2$ on the boundary of $\Lambda_2$, and we fix a path $\gamma^b$ from $v_1$ to $v_2$ contained in $\Lambda_0$.  
In addition, we let $\cS$ denote the collection of vertices and faces whose stars and plaquettes are contained in $\Lambda_0$.  
If $\cS$ is nonempty, we fix a vertex $\widehat{v} \in \cS$ and a face $\widehat{p} \in \cS$.  
We fix paths $\gamma_{\widehat{v}}$ and $\gamma_{\widehat{p}}$ contained in $\Lambda_0$ from $\widehat{v}$ and $\widehat{p}$ respectively to the boundary of $\Lambda_1$.  
Furthermore, for any $s \in \cS \setminus \{ \widehat{v}, \widehat{p}\}$, we fix a path $\gamma_s$ of the appropriate type from $s$ to $\widehat{v}$ or $\widehat{p}$ contained in $\Lambda_0$.  
We let $\cF_0 \coloneqq \left\{\Gamma^Z_{\gamma^b}\right\} \cup \set{\Gamma_{\gamma_s}}{s \in \cS}$.  
Furthermore, we define $\fF_0 \coloneqq \set{\Gamma_1 \cdots \Gamma_n}{\Gamma_i \in \cF_0}$, $\fF_{\Lambda_1} \coloneqq \set{\Gamma_1 \cdots \Gamma_n}{\Gamma_i \in \cF_{\Lambda_1}}$, and $\fF_{\Lambda_2^c} \coloneqq \set{\Gamma_1 \cdots \Gamma_n}{\Gamma_i \in \cF_{\Lambda_2^c}}$.  
We now define
\[
\cH_0 
\coloneqq 
\overline{\Span \fF_0 \Omega}
\subseteq
\cH.
\]
We have the following lemma, which is analogous to \cite[Lem.~4.3]{MR2956822}.  

\begin{lem}
\label{lem:DenseSubspaceDistalSplit}
We have that $\Span \fF_{\Lambda_1} \fF_0 \fF_{\Lambda_2^c} \Omega$ is dense in $\cH$.  
\end{lem}

\begin{proof}
We follow the proof of \cite[Lem.~4.3]{MR2956822}, modifying it to fit our setting.  
For clarity, we include the full argument.  
Note that by Lemma \ref{lem:DenseSubspaceOfGNSSpace}, it suffices to show that if $\Gamma$ is a product of (finite) path operators, then there exists $\Gamma_1 \in \fF_{\Lambda_1}$, $\widehat{\Gamma} \in \fF_0$, and $\Gamma_2 \in \fF_{\Lambda_2^c}$ such that $\Gamma \Omega = \Gamma_1 \widehat{\Gamma} \Gamma_2 \Omega$.  
Furthermore, since any two path operators either commute or anti-commute, it suffices to assume that $\Gamma$ is actually a single path operator $\Gamma_\gamma$ of type $X$ or $Z$.  
Note that if $\gamma$ is a closed loop or a path of type $X$ starting and ending on the boundary, then $\Gamma_\gamma$ is a product of star or plaquette operators and thus $\Gamma_\gamma \Omega = \Omega$.  
Hence, we may assume that this is not the case, i.e., that $\gamma$ generates an excitation at one or both endpoints.  
We first assume that $\gamma$ generates excitations at both endpoints, i.e., $\gamma$ is not a path of type $X$ starting on the boundary.  
If both endpoints of $\gamma$ lie in $\Lambda_1$ and its boundary or in $\Lambda_2^c$ and its boundary, then these endpoints can be joined by a path $\gamma'$ contained in $\Lambda_1$ or $\Lambda_2^c$, and $\Gamma_\gamma \Omega = \Gamma_{\gamma'} \Omega \in \fF_{\Lambda_1} \fF_0 \fF_{\Lambda^c} \Omega$.  
On the other hand, if both these endpoints lie in $\cS$, then letting $s_1$ and $s_2$ denote the endpoints of $\gamma$, we have that $\Gamma_{\gamma} \Omega = \Gamma_{\gamma_{s_1}} \Gamma_{\gamma_{s_2}} \Omega$ if $s_1, s_2 \notin \{\widehat{v}, \widehat{p}\}$ and $\Gamma_{\gamma} \Omega = \Gamma_{\gamma_{s_1}} \Omega$ if $s_2 \in \{\widehat{v}, \widehat{p}\}$.  
Hence $\Gamma_{\gamma} \Omega \in \fF_0 \Omega$ if both endpoints of $\gamma$ are in $\Lambda_0$.  
If one endpoint of $\gamma$ lies in $\cS$ and one lies in $\Lambda_1$ or its boundary, we let $s_1$ denote the endpoint in $\Lambda_0$ and $s_2$ denote the endpoint in $\Lambda_1$.  
In this case, we can get from $s_1$ to $s_2$ by taking $\gamma_{s_1}$, followed by $\gamma_{\widehat{v}}$ or $\gamma_{\widehat{p}}$ depending on the type of $\gamma$, followed by a path in $\Lambda_1$ from the $\Lambda_1$-boundary endpoint of $\gamma_{\widehat{v}}$ or $\gamma_{\widehat{p}}$ to $s_2$.  
The product $\Gamma'$ of these path operators then lives in $\fF_0 \fF_{\Lambda_1}$ and $\Gamma'\Omega = \Gamma_\gamma \Omega$.  

To handle the remaining cases where $\gamma$ generates excitations at both endpoints, we must handle the cases where $\gamma$ is type $Z$ and where $\gamma$ is type $X$ differently.  
First, suppose one endpoint of $\gamma$ is in $\Lambda_1$ or its boundary and one is in $\Lambda_2^c$ or its boundary.  
If $\gamma$ is type $Z$, we can get from one endpoint to the other by taking $\gamma^b$ along with paths in $\Lambda_1$ and $\Lambda_2^c$ from the endpoints of $\gamma^b$ to the endpoints of $\gamma$.  
The product $\Gamma'$ of these path operators then lives in $\fF_{\Lambda_1} \fF_0 \fF_{\Lambda_2^c}$ and $\Gamma'\Omega = \Gamma_\gamma \Omega$.  
On the other hand, if $\gamma$ is type $X$, we can take paths in $\Lambda_1$ and in $\Lambda_2^c$ from the endpoints of $\gamma$ to the boundary.  
The product $\Gamma'$ of these path operators then lives in $\fF_{\Lambda_1} \fF_{\Lambda_2^c}$ and $\Gamma'\Omega = \Gamma_\gamma \Omega$ since $\Gamma' \Gamma_\gamma$ is the path operator for a path of type $X$ starting and ending at the boundary.  
Lastly, suppose one endpoint (denoted $s_1$) of $\gamma$ lies in $\cS$ and the other (denoted $s_2$) lies in $\Lambda_2^c$ or its boundary.  
If $\gamma$ is type $Z$, we can connect $s_1$ and $s_2$ as follows: we take $\gamma_{s_1}$, followed by $\gamma_{\widehat{v}}$, followed by a path in $\Lambda_1$ from the $\Lambda_1$-boundary endpoint of $\gamma_{\widehat{v}}$ to the $\Lambda_1$-boundary endpoint of $\gamma^b$, followed by $\gamma^b$, followed by a path in $\Lambda_2^c$ from the $\Lambda_2$-boundary endpoint of $\gamma^b$ to $s_2$.  
The product $\Gamma'$ of these path operators then lives in $\fF_{\Lambda_1} \fF_0 \fF_{\Lambda_2^c}$ and $\Gamma'\Omega = \Gamma_\gamma \Omega$.  
If $\gamma$ is type $X$, we can take a path from $s_2$ to the boundary that is contained in $\Lambda_2^c$, and we can construct a path from $s_1$ to the boundary by taking $\gamma_{s_1}$, followed by $\gamma_{\widehat{p}}$, followed by a path from the $\Lambda_1$-boundary endpoint of $\gamma_{\widehat{p}}$ to the actual boundary.  
The product $\Gamma'$ of these path operators then lives in $\fF_{\Lambda_1} \fF_0 \fF_{\Lambda_2^c}$ and $\Gamma'\Omega = \Gamma_\gamma \Omega$.  
This takes care of all possible cases where $\gamma$ generates excitations at both endpoints.  

Finally, we suppose that $\gamma$ is a path of type $X$ starting at the boundary.  
If the non-boundary endpoint of $\gamma$ lies in $\Lambda_1$ or $\Lambda_2^c$ or the boundary of $\Lambda_1$ or $\Lambda_2^c$, we can take a path $\gamma'$ from the non-boundary endpoint of $\gamma$ to the boundary that is entirely contained in $\Lambda_1$ or in $\Lambda_2^c$, and $\Gamma_\gamma \Omega = \Gamma_{\gamma'} \Omega \in \fF_{\Lambda_1} \fF_0 \fF_{\Lambda^c} \Omega$.  
On the other hand, if the non-boundary endpoint $s$ of $\gamma$ lies in $\cS$, we can get from $s$ to the boundary as follows: we take $\gamma_s$, followed by $\gamma_{\widehat{p}}$, followed by a path in $\Lambda_1$ from the $\Lambda_1$-boundary endpoint of $\gamma_{\widehat{p}}$ to the actual boundary.  
The product $\Gamma'$ of these path operators then lives in $\fF_{\Lambda_1} \fF_0$ and $\Gamma'\Omega = \Gamma_\gamma \Omega$.  
\end{proof}

We now wish to construct a unitary map $U \colon \cH \to \cH_{\Lambda_1} \otimes \cH_{\Lambda_2^c} \otimes \cH_0$.  
Note that by Lemma \ref{lem:DenseSubspaceDistalSplit}, it will suffice to define $U$ for vectors in $\Span \fF_{\Lambda_1} \fF_0 \fF_{\Lambda_2^c} \Omega$.  
If $\Gamma_1 \in \fF_{\Lambda_1}$, $\Gamma_2 \in \fF_{\Lambda_2^c}$, and $\widehat{\Gamma} \in \fF_0$, we say that $\Gamma_1 \widehat{\Gamma} \Gamma_2 \Omega$ is in \emph{canonical form}, as in \cite{MR2956822}.  

\begin{lem}
\label{lem:UnitaryDecompositionForGNSSpace}
We have a well-defined unitary map $U \colon \cH \to \cH_{\Lambda_1} \otimes \cH_{\Lambda_2^c} \otimes \cH_0$, given on vectors in canonical form by 
\begin{equation}
\label{eq:UnitaryDecompositionForGNSSpace}
U\Gamma_1 \widehat{\Gamma} \Gamma_2 \Omega
\coloneqq
\Gamma_1\Omega \otimes \Gamma_2 \Omega \otimes \widehat{\Gamma} \Omega.
\end{equation}
\end{lem}

\begin{proof}
We follow the proof of \cite[Lem.~4.4]{MR2956822}, modifying it as appropriate for our setting.  
For clarity, we present the full argument.  
First, observe that \eqref{eq:UnitaryDecompositionForGNSSpace} uniquely determines $U$ by Lemma \ref{lem:DenseSubspaceDistalSplit}.  
We show that $U$ is an isometry, which will imply that $U$ is well-defined.  
To show that $U$ is an isometry, it suffices by Lemma \ref{lem:DenseSubspaceDistalSplit} to show that for all $\eta_1 \coloneqq \Gamma_1 \widehat{\Gamma} \Gamma_2 \Omega$ and $\eta_2 \coloneqq \Gamma_1' \widehat{\Gamma}' \Gamma_2' \Omega$ in canonical form, then $\langle \eta_1 | \eta_2 \rangle = \langle U \eta_1 | U\eta_2 \rangle$.  
Let $\eta_1 \coloneqq \Gamma_1 \widehat{\Gamma} \Gamma_2 \Omega$ and $\eta_2 \coloneqq \Gamma_1' \widehat{\Gamma}' \Gamma_2' \Omega$ be in canonical form.  
First, suppose that $\widehat{\Gamma} \neq \pm \widehat{\Gamma}'$.  
In that case there exists a star or plaquette operator that anti-commutes with one of $\widehat{\Gamma}$ and $\widehat{\Gamma}'$ and commutes with the other, so $\omega_0(\widehat{\Gamma}^* \widehat{\Gamma}') = 0$ by Lemma \ref{lem:AFHStateLemma} and hence $\langle U \eta_1 | U\eta_2 \rangle = 0$.  
We show that $\langle \eta_1 | \eta_2 \rangle = 0$.  
If there exists a star or plaquette contained in $\Lambda_0$ that anti-commutes with exactly one of $\widehat{\Gamma}$ and $\widehat{\Gamma}'$, then $\langle \eta_1 | \eta_2 \rangle = 0$ by Lemma \ref{lem:AFHStateLemma}, since this star or plaquette operator commutes with all operators in $\fF_{\Lambda_1}$ and $\fF_{\Lambda_2^c}$.  
Now, suppose the only star and plaquette operators that anti-commute with exactly one of $\widehat{\Gamma}$ and $\widehat{\Gamma}'$ live in $\fF_{\Lambda_1}$ or $\fF_{\Lambda_2^c}$.  
In that case, the path operator $\Gamma^Z_{\xi^b}$ is a factor in exactly one of $\widehat{\Gamma}$ and $\widehat{\Gamma}'$, say $\widehat{\Gamma}$, since $\Gamma_1 \widehat{\Gamma} \Gamma_2 \Omega$ and $\Gamma_1' \widehat{\Gamma}' \Gamma_2' \Omega$ are in canonical form.  
We then have that the operator $\Gamma_1 \widehat{\Gamma} \Gamma_2$ gives an odd number of type $Z$ excitations in $\Lambda_1$ (and in $\Lambda_2^c$ as well).  
However, $\Gamma_1' \widehat{\Gamma}' \Gamma_2'$ gives an even number of type $Z$ excitations in each of these regions.  
Hence there exists a star operator that anti-commutes with exactly one of $\Gamma_1 \widehat{\Gamma} \Gamma_2$ and $\Gamma_1' \widehat{\Gamma}' \Gamma_2'$, so $\langle \eta_1 | \eta_2 \rangle = 0$.  

We now consider the case where $\widehat{\Gamma} = \pm \widehat{\Gamma}'$.  
Without loss of generality, we may assume that $\widehat{\Gamma} = \widehat{\Gamma}'$.  
In that case, we must show that 
\[
\omega_0(\Gamma_1^*\Gamma_1' \Gamma_2^* \Gamma_2') 
= 
\omega_0(\Gamma_1^* \Gamma_1') \omega_0(\Gamma_2^* \Gamma_2').
\]
Note that since $\Lambda_1 \ll \Lambda_2$, a star or plaquette operator cannot anti-commute both with a path operator in $\cF_{\Lambda_1}$ and with a path operator in $\cF_{\Lambda_2^c}$.  
Thus, if there is a star or plaquette operator that anti-commutes with exactly one of $\Gamma_1$ and $\Gamma_1'$, then both sides of the above equation are zero as this path operator must commute with $\Gamma_2$ and $\Gamma_2'$.  
Similarly, both sides of the above equation are zero if there is a star or plaquette operator that anti-commutes with exactly one of $\Gamma_2$ and $\Gamma_2'$.  
If there are no such star and plaquette operators, then $\Gamma_1^* \Gamma_1'$ and $\Gamma_2^* \Gamma_2'$ commute with all star and plaquette operators, so they are, up to sign, the product of star and plaquette operators.  
Hence in this case, $\omega_0(\Gamma_1^*\Gamma_1' \Gamma_2^* \Gamma_2') = \pm 1$ and $\omega_0(\Gamma_1^* \Gamma_1') \omega_0(\Gamma_2^* \Gamma_2') = \pm 1$, with the signs being the same.  

Finally, the image of $U$ is clearly dense in $\cH_{\Lambda_1} \otimes \cH_{\Lambda_2^c} \otimes \cH_0$, so $U$ is a unitary.  
\end{proof}

We will now show that the distal split property holds for $\omega_0$ by explicitly constructing a type I factor $\cN$ such that $\cR_{\Lambda_1} \subseteq \cN \subseteq \cR_{\Lambda_2}$.  
To do so, we will show that $U\cR_{\Lambda_1}U^*$ acts solely on the first factor of $\cH_{\Lambda_1} \otimes \cH_{\Lambda_2^c} \otimes \cH_0$ and $U\cR_{\Lambda_2^c}U^*$ acts solely on the second factor, a result that will be important for obtaining a bound on the number of nonequivalent simple superselection sectors.  

\begin{prop}
\label{prop:ActionOfDistalSplitFactorsOnDecomposedGNSSpace}
Let $U \colon \cH \to \cH_{\Lambda_1} \otimes \cH_{\Lambda_2^c} \otimes \cH_0$ be as in Lemma \ref{lem:UnitaryDecompositionForGNSSpace}.  
Then we have that $U\cR_{\Lambda_1}U^* = \cR_{\Lambda_1} P_{\Lambda_1} \otimes I \otimes I$ and $U\cR_{\Lambda_2^c}U^* = I \otimes \cR_{\Lambda_2^c} P_{\Lambda_2^c} \otimes I$, where $P_{\Lambda_1}$ and $P_{\Lambda_2^c}$ are the projections onto $\cH_{\Lambda_1}$ and $\cH_{\Lambda_2^c}$ respectively.  
\end{prop}

\begin{proof}
This proof is identical to an argument in the proof of \cite[Thm.~4.5]{MR2956822}, but we repeat it here for convience.  
First note that $\cR_{\Lambda_1} \cH_{\Lambda_1} \subseteq \cH_{\Lambda_1}$ by Lemma \ref{lem:InvarianceOfLambdaSubspaceUnderLambdaAlgebra}, and similarly $\cR_{\Lambda_2^c} \cH_{\Lambda_2^c} \subseteq \cH_{\Lambda_2^c}$.  
Hence $P_{\Lambda_1} \in \cR_{\Lambda_1}'$ and $P_{\Lambda_2^c} \in \cR_{\Lambda_2^c}'$.  
We show that $U\cR_{\Lambda_1}U^* = \cR_{\Lambda_1} P_{\Lambda_1} \otimes I \otimes I$; the other result follows by an analogous argument.  
To show this, it suffices to show that $UAU^* = AP_{\Lambda_1} \otimes I \otimes I$ for all $A \in \cR_{\Lambda_1}$.  
Let $A \in \cR_{\Lambda_1}$.  
Then by a density argument, it suffices to show that 
\[
UAU^*(\eta \otimes \Gamma \Omega \otimes \widehat{\Gamma}\Omega) 
= 
A \eta \otimes \Gamma \Omega \otimes \widehat{\Gamma}\Omega
\] 
for all $\eta \in \cH_{\Lambda_1}$, $\Gamma \in \fF_{\Lambda_2^c}$, and $\widehat{\Gamma} \in \fF_0$.  
Let $\eta \in \cH_{\Lambda_1}$, $\Gamma \in \fF_{\Lambda_2^c}$, and $\widehat{\Gamma} \in \fF_0$.  
Then $U^*(\eta \otimes \Gamma \Omega \otimes \widehat{\Gamma}\Omega) = \widehat{\Gamma}\Gamma \eta$ by the definition of $U$.  
Furthermore, since $A \in \cR_{\Lambda_1}$, we have by locality that $A$ commutes with $\Gamma$ and $\widehat{\Gamma}$, and since $\cR_{\Lambda_1} \cH_{\Lambda_1} \subseteq \cH_{\Lambda_1}$, we have that $U\widehat{\Gamma}\Gamma A\eta = A \eta \otimes \Gamma \Omega \otimes \widehat{\Gamma}\Omega$.  
Thus, we have that 
\[
UAU^*(\eta \otimes \Gamma \Omega \otimes \widehat{\Gamma}\Omega) 
=
UA\widehat{\Gamma}\Gamma \eta
=
U\widehat{\Gamma}\Gamma A\eta
=
A \eta \otimes \Gamma \Omega \otimes \widehat{\Gamma}\Omega.
\qedhere
\]
\end{proof}

At this point, the distal split property follows from Proposition \ref{prop:ActionOfDistalSplitFactorsOnDecomposedGNSSpace} by an argument in the proof of \cite[Thm.~4.5]{MR2956822}.  

\begin{thm}
Let $U \colon \cH \to \cH_{\Lambda_1} \otimes \cH_{\Lambda_2^c} \otimes \cH_0$ be as in Lemma \ref{lem:UnitaryDecompositionForGNSSpace}.  
Then the type I factor $\cN \coloneqq U^*(\scrB(\cH_0) \otimes I \otimes I) U$ satisfies that $\cR_{\Lambda_1} \subseteq \cN \subseteq \cR_{\Lambda_2}$.  
\end{thm}


\section{Bounding the number of excitations}
\label{sec:BoundingExcitations}

\begin{figure}[h]
\centering
\includegraphics{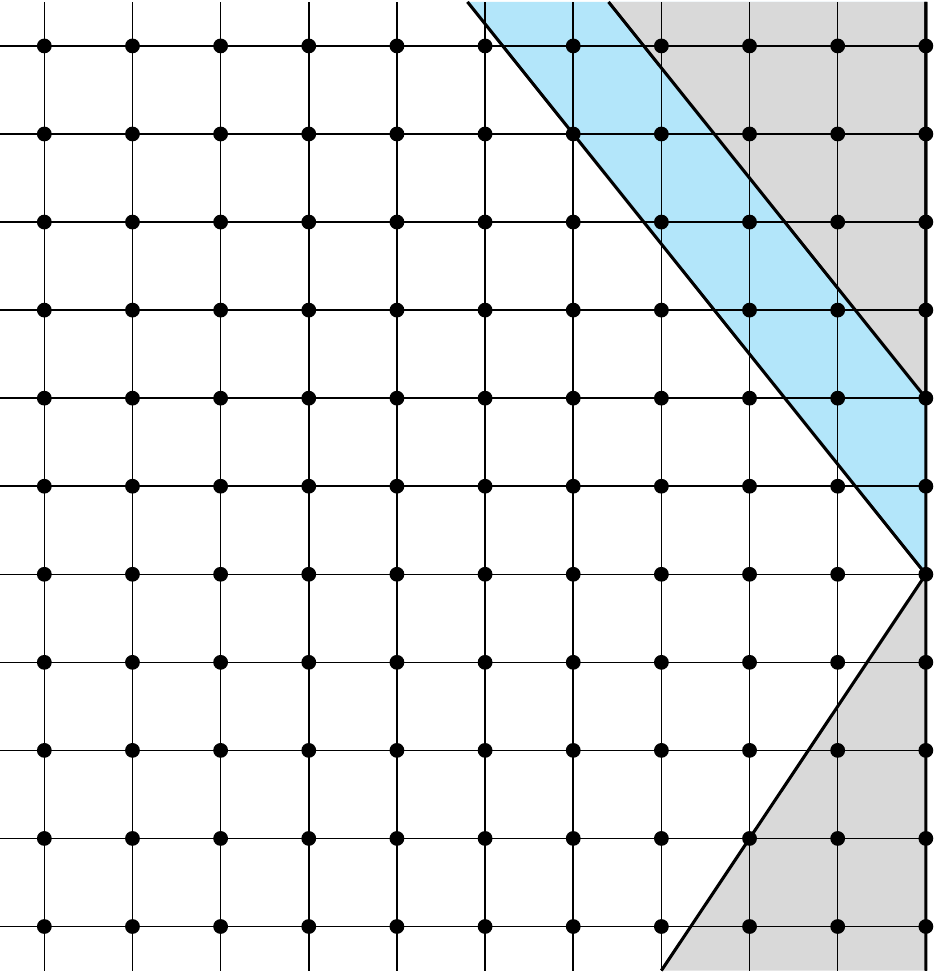}\caption{The gray shaded regions depict a region $\Xi \in \cC^2$.  The top gray shaded region is a cone $\Lambda_1$ along the boundary, and the cyan shaded region depicts the set $\Lambda \cap \Lambda_1^c$, where $\Lambda$ is a cone along the boundary with $\Lambda_1 \ll \Lambda$.  
The bottom shaded region depicts a cone $\Lambda_2 \subseteq \Lambda^c$ along the boundary.}
\label{fig:SplitConeRegions}
\end{figure}

In this section, we show that the number of nonisomorphic simple superselection sectors is at most 2, and hence is equal to 2 since we have already constructed two distinct such sectors (the vacuum and $\pi^Z$).  
By simple, we mean that $\End(\pi) \cong \bbC$, where again $\End(\pi)$ denotes the space self-intertwiners of $\pi$ (the endomorphisms of $\pi$ in the category $\Delta(\Lambda)$).  
We use the approach in \cite{MR3135456}.  
Following \cite{MR3135456}, we let $\cC^2$ be the collection of subsets of $\bfB$ of the form $\Lambda_1 \cup \Lambda_2$, where $\Lambda_1$ and $\Lambda_2$ are (disjoint) cones along the boundary such that there exists a cone  $\Lambda$ along the boundary with $\Lambda_1 \ll \Lambda$ and $\Lambda_2 \subseteq \Lambda^c$ (see Figure \ref{fig:SplitConeRegions}).
Note that by the distal split property, if $\Xi = \Lambda_1 \cup \Lambda_2 \in \cC^2$, with $\Lambda_1$ and $\Lambda_2$ disjoint cones as just described, then $\cR_\Xi = \cR_{\Lambda_1} \vee \cR_{\Lambda_2} \cong \cR_{\Lambda_1} \otimes \cR_{\Lambda_2}$.  
For $\Xi \in \cC^2$, we define $\widehat{\cR}_\Xi \coloneqq \cR_{\Xi^c}'$.  
Since $\pi_0$ is an irreducible representation, the proof of \cite[Lem.~3.2]{MR3135456} still applies here to give that $\cR_{\Xi} \subseteq \widehat{\cR}_{\Xi}$ is an irreducible subfactor.  
Note that for any subfactor $\fN \subseteq \fM$ with a normal conditional expectation $\cE \colon \fM \to \fN$, we can define the \emph{Kosaki-Longo index} \cite[Thm.~4.1]{MR1027496} by $[\fM : \fN]_\cE \coloneqq \lambda^{-1}$, where 
\[
\lambda
\coloneqq
\sup\set{r \geq 0}{\cE(X) \geq rX \text{ for all } X \in \fM_+}.  
\]
(Note that if $\lambda = 0$, then $[\fM : \fN]_\cE = \infty$.)
We let $[\fM : \fN] = \inf_{\cE} [\fM : \fN]_\cE$; if $[\fM : \fN] < \infty$, there is a unique normal conditional expectation such that $[\fM : \fN]_\cE = [\fM : \fN]$ \cite[Thm.~5.5]{MR1027496}.  
We now define $\mu \coloneqq \inf_{\Xi \in \cC^2} [\widehat{\cR}_\Xi : \cR_\Xi]$.  
By the argument used in the proof of \cite[Lem.~13]{MR1838752}, we have the following result.  

\begin{prop}
\label{prop:MuIndexBoundsExcitations}
The number of nonisomorphic simple superselection sectors for toric code with boundary is bounded above by $\mu$.  
\end{prop}

We now wish to show that for all $\Xi \in \cC^2$, $[\widehat{\cR}_\Xi : \cR_\Xi] = 2$.  
This will show that $\mu = 2$ and hence that there are at most 2 nonisomorphic simple superselection sectors.  
To show this, we follow the proof in \cite[\textsection 4]{MR3135456}, modifying it to fit our setting.  
Let $\Xi \in \cC^2$.  
In order to show that $[\widehat{\cR}_\Xi : \cR_\Xi] = 2$, we will show that $\widehat{\cR}_\Xi$ is given by a crossed product of $\cR_\Xi$ with $\bbZ/2\bbZ$.  
Proving this fact requires several steps.  
Since $\Xi \in \cC^2$, $\Xi = \Lambda_1 \cup \Lambda_2$, where $\Lambda_1$ and $\Lambda_2$ are (disjoint) cones along the boundary such that there exists a cone  $\Lambda$ along the boundary with $\Lambda_1 \ll \Lambda$ and $\Lambda_2 \subseteq \Lambda^c$.
We let $V$ be a unitary intertwiner between two type $Z$ superselection sectors, one in $\Lambda_1$ and one in $\Lambda_2$, satisfying the conditions of Proposition \ref{prop:IntertwinersBetweenTypeZExcitations}.  
Recall that  $V = \lim^{WOT} \Gamma^Z_{\gamma^1_n} \Gamma^Z_{\widetilde \gamma_n} \Gamma^Z_{\gamma^2_n}$, where $\gamma^1$ and $\gamma^2$ are infinite paths of type $Z$ with $\gamma^1 \subseteq \Lambda_1$ and $\gamma^2 \subseteq \Lambda_2$, $\gamma^1_n$ and $\gamma^2_n$ are the first $n$ bonds of $\gamma^1$ and $\gamma^2$ respectively, and $(\widetilde \gamma_n)$ is a sequence of paths from the $n$th vertex of $\gamma^1$ to the $n$th vertex of $\gamma^2$ such that the distance from $\widetilde{\gamma}_n$ to the starting sites of $\gamma^1$ and $\gamma^2$ goes to infinity.  
The main step in proving that $\widehat{\cR}_\Xi$ is given by a crossed product of $\cR_\Xi$ with $\bbZ/2\bbZ$ is the following lemma.  

\begin{lem}
\label{lem:AlgebraObtainedByAdjoiningIntertwiner}
We have that $\widehat{\cR}_\Xi = \cR_\Xi \vee \{V\}$.  
\end{lem}

For ease of notation, we write $\cA \coloneqq \cR_\Xi \vee \{V\}$ for the remainder of this section.  
We also write $\cB \coloneqq \cR_{\Xi^c}$.  
Note that $\cA \subseteq \cB'$ by locality and by the fact that every local operator in $\fA(\Xi^c)$ eventually commutes with $\Gamma^Z_{\gamma^1_n} \Gamma^Z_{\widetilde \gamma_n} \Gamma^Z_{\gamma^2_n}$.  
Hence in order to prove Lemma \ref{lem:AlgebraObtainedByAdjoiningIntertwiner}, it remains to show that $\cB' \subseteq \cA$.  
The proof of this will be similar to the proof of Haag duality in \textsection \ref{sec:HaagDuality}.  
Like with the proof of Haag duality, we will proceed by restricting to a subspace of $\cH$.  
In particular, we consider the subspace $\cH_{\overline{\Xi}} \coloneqq \overline{\cA \Omega}$.  
We now fix a path $\gamma_Z$ from a vertex site entirely contained in $\Lambda_1$ to a vertex site entirely contained in $\Lambda_2$.  
(By ``entirely contained in" we mean that the star at the vertex is contained in the cone.)
As before, we let $\cF_\Xi$ be the collection of all path operators corresponding to paths in $\Xi$.  
We also define $\cF_{\overline{\Xi}} \coloneqq \cF_\Xi \cup \{\Gamma^Z_{\gamma_Z}\}$.  
Similarly, we have 
\[
\fF_\Xi
\coloneqq
\set{\Gamma_1 \cdots \Gamma_n}{\Gamma_i \in \cF_\Xi}, 
\qquad
\fF_{\Xi^c}
\coloneqq
\set{\Gamma_1 \cdots \Gamma_n}{\Gamma_i \in \cF_{\Xi^c}}, 
\qquad
\fF_{\overline{\Xi}}
\coloneqq
\set{\Gamma_1 \cdots \Gamma_n}{\Gamma_i \in \cF_{\overline{\Xi}}}.
\]
We now describe a useful dense subspace of $\cH_{\overline{\Xi}}$, which is an analogue of \cite[Lem.~4.4]{MR3135456}.  

\begin{lem}
\label{lem:DenseSubspaceForAuxiliarySubspaceForBoundingExcitations}
The space $\Span \fF_{\overline{\Xi}} \Omega$ is a dense subspace of $\cH_{\overline{\Xi}}$.  
In fact, we have that 
\begin{equation}
\label{eq:GeneratingSetForAuxiliarySubspaceForBoundingExcitations}
\fF_{\overline{\Xi}} \Omega = \set{\Gamma V^i \Omega}{\Gamma \in \fF_{\Xi}, i \in \{0, 1\}}.
\end{equation}
\end{lem}

\begin{proof}
We follow the proof of \cite[Lem.~4.4]{MR3135456}, modifying it as appropriate for our setting.  
For clarity, we include the full argument.  
We first claim that $V$ commutes or anti-commutes with each $\Gamma \in \fF_\Xi$.  
Since $\fF_\Xi \cup \{V\}$ generates $\cA = \cR_\Xi \vee \{V\}$ as a von Neumann algebra, this will imply that $\Span \set{\Gamma V^i \Omega}{\Gamma \in \fF_{\Xi}, i \in \{0, 1\}}$ is a dense subspace of $\cH_{\overline{\Xi}}$ since $V^2 = I$.  
Let $\Gamma \in \fF_{\Xi}$.  
Recall that $V = \lim^{WOT} \Gamma^Z_{\gamma^1_n} \Gamma^Z_{\widetilde \gamma_n} \Gamma^Z_{\gamma^2_n}$ as described above.  
For each $n \in \bbN$, $\Gamma \Gamma^Z_{\gamma^1_n} \Gamma^Z_{\widetilde \gamma_n} \Gamma^Z_{\gamma^2_n} = \pm \Gamma^Z_{\gamma^1_n} \Gamma^Z_{\widetilde \gamma_n} \Gamma^Z_{\gamma^2_n} \Gamma$, and since $\Gamma$ is local this sign eventually becomes constant.  
Thus, since multiplication is separately WOT-continuous, $\Gamma V =\pm V \Gamma$.  

We now show that \eqref{eq:GeneratingSetForAuxiliarySubspaceForBoundingExcitations} holds.  
For $i \in \{1, 2\}$, we let $\gamma_i$ be a path from the starting site of $\gamma^i$ to the endpoint of $\gamma_Z$ in $\Lambda_i$.  
Then $\gamma_1 \cup \gamma_Z \cup \gamma_2$ is a path from the starting site of $\gamma^1$ to the starting site of $\gamma^2$.  
By \cite[Lem.~4.1]{MR2804555}, we have that $V\Omega = \Gamma^Z_{\gamma_1} \Gamma^Z_{\gamma_Z} \Gamma^Z_{\gamma_2} \Omega$, from which \eqref{eq:GeneratingSetForAuxiliarySubspaceForBoundingExcitations} follows.  
\end{proof}

In order to restrict to $\cH_{\overline{\Xi}}$, we need $\cH_{\overline{\Xi}}$ to be an invariant subspace for $\cA$ and $\cB'$.  
This is true, as detailed in the following lemma.  

\begin{lem}
\label{lem:BoundaryExcitationsInvariantSubspaces}
We have that $\cA \cH_{\overline{\Xi}} \subseteq \cH_{\overline{\Xi}}$ and $\cB' \cH_{\overline{\Xi}} \subseteq \cH_{\overline{\Xi}}$.  
Furthermore, elements of $\cB'$ are uniquely determined by their restriction to $\cH_{\overline{\Xi}}$.  
\end{lem}

\begin{proof}
We follow the proof of \cite[Lem.~4.5]{MR3135456}, modifying it to fit our setting. 
For clarity, we present the argument in full.  
The fact that $\cA \cH_{\overline{\Xi}} \subseteq \cH_{\overline{\Xi}}$ is clear from the definition of $\cH_{\overline{\Xi}}$.  
We now show that $\cB' \cH_{\overline{\Xi}} \subseteq \cH_{\overline{\Xi}}$.  
Let $B \in \cB'$.  
Note that by Lemma \ref{lem:DenseSubspaceForAuxiliarySubspaceForBoundingExcitations}, it suffices to show that $B\Gamma V^i \Omega \in \cH_{\overline{\Xi}}$ for all $\Gamma \in \fF_\Xi$ and $i \in \{0, 1\}$.  
Let $\Gamma \in \fF_\Xi$ and $i \in \{0, 1\}$, and let $\xi \coloneqq \Gamma V^i \Omega$.  
Note that by an argument similar to the proof of Lemma \ref{lem:DenseSubspaceDistalSplit}, the space $\Span \fF_{\Xi^c} \fF_{\overline{\Xi}} \Omega$ is dense in $\cH$.  
Thus, in order to show that $B \xi \in \cH_{\overline{\Xi}}$, it is sufficient to show that $\langle \eta | B \xi \rangle = 0$ for all $\eta \in \cH_{\overline{\Xi}}^\perp$ of the form $\eta = \widehat{\Gamma} \widetilde{\Gamma} \Omega$, where $\widehat{\Gamma} \in \fF_{\Xi^c}$ and $\widetilde{\Gamma} \in \fF_{\overline{\Xi}}$.  
We let $\widehat{\Gamma} \in \fF_{\Xi^c}$ and $\widetilde{\Gamma} \in \fF_{\overline{\Xi}}$, and we set $\eta \coloneqq \widehat{\Gamma} \widetilde{\Gamma} \Omega$, not necessarily in $\cH_{\overline{\Xi}}^\perp$.  

First, suppose there exists a star or plaquette operator in $\fA(\Xi^c)$ that anti-commutes with $\widehat{\Gamma}$.  
We consider the case of an anti-commuting star operator $A_s$; the case of a plaquette operator is handled analogously.  
Note that $A_s$ commutes with $\Gamma V^i$ since $A_s \in \cB$ and $\Gamma V^i \in \cA \subseteq \cB'$.  
Furthermore, $A_s$ commutes with $\widetilde \Gamma$ since any excitations from $\widetilde \Gamma$ lie in $\Xi$.  
Thus, since $B \in \cB'$, we have that 
\[
\langle \eta | B\xi\rangle
=
\langle \eta | B \Gamma V^i A_s \Omega \rangle
=
\langle \eta| A_s B \Gamma V^i \Omega \rangle
=
\langle A_s \widehat{\Gamma} \widetilde{\Gamma} \Omega | \xi \rangle
=
- \langle \widehat{\Gamma} \widetilde{\Gamma} A_s \Omega| \xi \rangle
=
-\langle \eta | B\xi \rangle,
\]
so $\langle \eta | B\xi\rangle = 0$ as desired.  

Now, suppose that there are no star and plaquette operators in $\fA(\Xi^c)$ that anti-commute with $\widehat{\Gamma}$.  
We claim that in this case $\eta \in \cH_{\overline{\Xi}}$, so $\eta \notin \cH_{\overline{\Xi}}^\perp$ unless $\eta = 0$.  
Proceeding as in the proof of Lemma \ref{lem:InvarianceOfLambdaSubspaceUnderCommutantOfComplementAlgebra}, we may assume that the excitations at the end of each path operator in the product $\widehat{\Gamma} \in \fF_{\Xi^c}$ live on the boundary of $\Xi = \Lambda_1 \cup \Lambda_2$, and that there are no paths giving no excitations (i.e., no closed loops or paths of type $X$ starting and ending at the boundary).  
A path operator giving a pair of excitations on the boundary of $\Lambda_1$ or on the boundary of $\Lambda_2$ acts identically on $\Omega$ to a path operator in $\fA(\Lambda_1)$ or $\fA(\Lambda_2)$ giving the same excitations.  
Similarly, a path operator giving a single type $X$ excitation on the boundary of $\Xi$ (i.e., an operator given by a path starting at the actual boundary) acts identically on $\Omega$ to a path operator in $\fA(\Xi)$ giving the same excitation.  
A path operator of type $Z$ giving one excitation on the boundary of $\Lambda_1$ and one excitation on the boundary of $\Lambda_2$ acts identically on $\Omega$ to an operator given by a path with the same endpoints consisting of $\gamma_Z$ and segments in $\Xi$.  
Finally, a path of type $X$ giving one excitation on the boundary of $\Lambda_1$ and one excitation on the boundary of $\Lambda_2$ acts identically on $\Omega$ to the product of two path operators, one of which corresponds to a path in $\Lambda_1$ from the $\Lambda_1$-endpoint of the original path to the boundary and the other corresponds to a path in $\Lambda_2$ from the $\Lambda_2$-endpoint of the original path to the boundary.  
Proceeding in this way, we have that $\widehat{\Gamma} \Omega = \pm\widehat{\Gamma}' \Omega$ for some $\widehat{\Gamma}' \in \fF_{\Xi}$, so 
\[
\eta
=
\widehat{\Gamma} \widetilde{\Gamma} \Omega
=
\pm\widehat{\Gamma}' \widetilde{\Gamma} \Omega
\in
\cH_{\overline{\Xi}}.  
\]

Finally, we show that elements of $\cB'$ are uniquely determined by their restriction to $\cH_{\overline{\Xi}}$.  
Let $A_1, A_2 \in \cB'$ such that $A_1|_{\cH_{\overline{\Xi}}} = A_2|_{\cH_{\overline{\Xi}}}$.  
Since $\Span \fF_{\Xi^c} \fF_{\overline{\Xi}} \Omega$ is dense in $\cH$, it suffices to show that $A_1 \widehat{\Gamma} \widetilde{\Gamma} \Omega = A_2 \widehat{\Gamma} \widetilde{\Gamma} \Omega$ for all $\widehat{\Gamma} \in \fF_{\Xi^c}$ and $\widetilde{\Gamma} \in \fF_{\overline{\Xi}}$.  
But this holds since $A_1$ and $A_2$ commute with all $\widehat{\Gamma} \in \fF_{\Xi^c}$ and $\widetilde{\Gamma} \Omega \in \cH_{\overline{\Xi}}$ if $\widetilde{\Gamma} \in \fF_{\overline{\Xi}}$.  
\end{proof}

We wish to prove Lemma \ref{lem:AlgebraObtainedByAdjoiningIntertwiner} using the argument in the proof of Theorem \ref{thm:HaagDuality}.  
To do so, we need an analogue of Lemma \ref{lem:HaagRieffelVanDaeleLemma}.  
We let $\cA_{\overline{\Xi}} \coloneqq \cA P_{\overline{\Xi}} \subseteq \scrB(\cH_{\overline{\Xi}})$ and $\cB_{\overline{\Xi}} \coloneqq P_{\overline{\Xi}} \cB P_{\overline{\Xi}} \subseteq \scrB(\cH_{\overline{\Xi}})$, where $P_{\overline{\Xi}}$ is the projection onto $\cH_{\overline{\Xi}}$.  
Note that by Lemma \ref{lem:BoundaryExcitationsInvariantSubspaces}, $P_{\overline{\Xi}} \in \cA'$ and $P_{\overline{\Xi}} \in \cB$. 
We also let $\cA_{\overline{\Xi}, \sa}$ and $\cB_{\overline{\Xi}, \sa}$ be the self-adjoint elements of $\cA_{\overline{\Xi}}$ and $\cB_{\overline{\Xi}}$ respectively.  
Lemma \ref{lem:AlgebraObtainedByAdjoiningIntertwiner} will then follow by the proof of Theorem \ref{thm:HaagDuality}, once we have proven the following lemma.  

\begin{lem}
The real vector space $\cA_{\overline{\Xi}, \sa}\Omega + i \cB_{\overline{\Xi}, \sa} \Omega$ is dense in $\cH_{\overline{\Xi}}$.  
\end{lem}

\begin{proof}
We follow the proof of \cite[Lem.~4.6]{MR3135456}, modifying it to fit our setting.  
For clarity, we include the full argument.  
Note that by Lemma \ref{lem:DenseSubspaceForAuxiliarySubspaceForBoundingExcitations}, it suffices to show that $\widehat{\Gamma} \Omega \in \cA_{\overline{\Xi}, \sa}\Omega + i \cB_{\overline{\Xi}, \sa} \Omega$ and $i \widehat{\Gamma} \Omega \in \cA_{\overline{\Xi}, \sa}\Omega + i \cB_{\overline{\Xi}, \sa} \Omega$ for all $\widehat{\Gamma} \in \fF_{\overline{\Xi}}$.  
Let $\widehat{\Gamma} \in \fF_{\overline{\Xi}}$.  
Note that by Lemma \ref{lem:DenseSubspaceForAuxiliarySubspaceForBoundingExcitations}, we have that $\widehat{\Gamma} \Omega = \Gamma V^i \Omega$ for some $\Gamma \in \fF_\Xi$ and $i \in \{0, 1\}$.  
We let $A \coloneqq \Gamma V^i$.  
By the proof of Lemma \ref{lem:DenseSubspaceForAuxiliarySubspaceForBoundingExcitations}, $A^* = A$ or $A^* = -A$.  
If $A^* = A$, then $AP_\Lambda \in \cA_{\overline{\Xi}, \sa}$ and hence $\widehat{\Gamma}\Omega = A \Omega \in \cA_{\overline{\Xi}, \sa}\Omega$.  
On the other hand, if $A^* = -A$, then $iAP_\Lambda \in \cA_{\overline{\Xi}, \sa}$ and hence $i\widehat{\Gamma}\Omega = iA \Omega \in \cA_{\overline{\Xi}, \sa}\Omega$.  

Now, suppose there exists a star or plaquette operator in $\fA(\Xi)$ that anti-commutes with $A$.  
We consider the case of an anti-commuting star operator $A_s$; the case of a plaquette operator is handled similarly.  
We then have that $iA_s A$ is self-adjoint if $A$ is self-adjoint and $A_sA$ is self-adjoint if $iA$ is self-adjoint.  
In the first case, we have that 
\[
i\widehat{\Gamma}\Omega
=
iA\Omega
=
iAA_s\Omega
=
-iA_sA\Omega
\in
\cA_{\overline{\Xi}, \sa},
\]
and similarly in the second case we have that $\widehat{\Gamma} \Omega \in \cA_{\overline{\Xi}, \sa}$.  

It remains to consider the case where there are no star and plaquette operators in $\fA(\Xi)$ that anti-commute with $A$.  
In this case, we have that any excitations generated by $A$ must live on the boundary of $\Xi$, where we view $V$ as generating excitations at the endpoints of the paths whose operators converge weakly to $V$.  
For any pair of type $Z$ excitations generated by $A$, we can find a path of type $Z$ in $\Xi^c$ that generate the same excitations.  
In addition, for any type $X$ excitation generated by $A$, we can find a path in $\Xi^c$ of type $X$ starting at the boundary and ending at this excitation.  
The product $B \in \cR_{\Xi^c}$ of all of these path operators then generates the same excitations as $A$.  
We claim that $B$ is self-adjoint if and only if $A$ is (and thus that $B^* = -B$ if and only if $A^* = -A$).  
Since $A$ and $B$ commute, we have that $(AB)^* = A^*B^*$.  
Hence the desired claim will follow if we can show that $AB$ is self-adjoint.  
Note that the operator $AB$ does not generate any excitations.  
Thus, rearranging the factors of $AB$ if necessary, we have that all of the path operators in $AB$ correspond to closed loops or paths of type $X$ starting and ending on the boundary, where we view a path operator including a factor of $V$ as corresponding to an infinite loop.  
Note that any finite closed loop or path of type $X$ starting and ending on the boundary is the product of star and plaquette operators.  
Hence if $AB$ does not contain a factor of $V$, we have that $AB$ is self-adjoint.  
On the other hand, if $AB$ does contain a factor of $V$, then $AB$ is the weak limit of operators that are the product of star and plaquette operators (and hence are self-adjoint).  
The result then follows since the adjoint is WOT-continuous.  

Note that the operator $B \in \cR_{\Xi^c}$ satisfies that $B\Omega = \pm A\Omega = \pm\widehat{\Gamma}\Omega$.  
Hence if $A^* = A$, then $P_\Lambda B P_\Lambda \in \cB_{\overline{\Xi}, \sa}$ and thus 
\[
i\widehat{\Gamma}\Omega
=
\pm iP_\Lambda B P_\Lambda \Omega
\in
i\cB_{\overline{\Xi}, \sa} \Omega.
\]
Similarly, $\widehat{\Gamma}\Omega \in i\cB_{\overline{\Xi}, \sa} \Omega$ if $A^* = -A$, which completes the proof.  
\end{proof}

We now wish to show that $\cR_\Xi \vee \{V\}$ is isomorphic to the crossed product of $\cR_\Xi$ under a $\bbZ/2\bbZ$-action implemented by $V$.  
We first claim that $V$ implements an action of $\bbZ/2\bbZ$ on $\cR_\Xi$.  
Since $V^2 = I$, it suffices to show that $V\cR_\Xi V \subseteq \cR_\Xi$.  
To see this, observe that $\Span \fF_\Xi$ forms a WOT-dense $*$-subalgebra of $\cR_\Xi$.  
As explained in the proof of Lemma \ref{lem:DenseSubspaceForAuxiliarySubspaceForBoundingExcitations}, $V$ either commutes or anti-commutes with each operator in $\fF_\Xi$, so $V(\Span \fF_\Xi)V \subseteq \Span \fF_\Xi$.  
The result then follows since multiplication is separately WOT-continuous.  

We now show that $\cR_\Xi \vee \{V\}$ is isomorphic to the crossed product $\cR_\Xi \rtimes_\alpha \bbZ/2\bbZ$, where $\alpha$ is the action of $\bbZ/2\bbZ$ on $\cR_\Xi$ implemented by $V$.  

\begin{prop}
\label{prop:BoundingExcitationsCrossedProduct}
We have that $\widehat{\cR}_\Xi$ is isomorphic to $\cR_\Xi \rtimes_\alpha \bbZ/2\bbZ$, where $\alpha$ is the action of $\bbZ/2\bbZ$ on $\cR_\Xi$ implemented by $V$.  
In particular, there exists a $*$-isomorphism $\Phi \colon \cR_\Xi \rtimes_\alpha \bbZ/2\bbZ \to \widehat{\cR}_\Xi$ such that $\Phi(\cR_\Xi) = \cR_\Xi$.  
Furthermore, $\alpha$ is an outer action on $\cR_\Xi$.  
\end{prop}

\begin{proof}
We follow the proof of \cite[Lem.~4.7]{MR3135456}, with modifications to fit our setting.  
For clarity, we include the full argument.  
Note that viewing $\cH \otimes \ell^2(\bbZ/2\bbZ) \cong \cH \oplus \cH$, we can view $\cR_\Xi \rtimes_\alpha \bbZ/2\bbZ \subseteq \scrB(\cH \otimes \ell^2(\bbZ/2\bbZ))$ as consisting of operators of the following form \cite[Def.~13.1.3]{MR1468230}: 
\begin{equation}
\label{eq:CanonicalCrossedProductElement}
X
=
\begin{pmatrix}
R_I & R_Z V
\\
R_Z V & R_I
\end{pmatrix}.
\end{equation}
We define a map $\Phi \colon \cR_\Xi \rtimes_\alpha \bbZ/2\bbZ \to \widehat{\cR}_\Xi$ by $\Phi(X) = R_I + R_ZV$.  
Note that $\Phi$ is a $*$-homomorphism since $V$ implements the action $\alpha \colon \bbZ/2\bbZ \to \cR_\Xi$.  
Furthermore, $\Phi(\cR_\Xi) = \cR_\Xi$, when we view $\cR_\Xi \subseteq \cR_\Xi \rtimes_\alpha \bbZ/2\bbZ$ in the canonical way.  
We also have that $\Phi$ is normal.  
Indeed, suppose $\varphi \colon \widehat{\cR}_\Xi \to \bbC$ is a normal state on $\widehat{\cR}_\Xi$.  
Then there exists a sequence $(\xi_n)$ of vectors in $\cH$ such that $\|\xi_n\|^2 = 1$ and $\varphi = \sum_n \langle \xi_n| \cdot \xi_n\rangle$.  
We let $\eta_n \coloneqq \xi_n \oplus 0$ and $\widetilde \eta_n \coloneqq \xi_n \oplus \xi_n$, and we consider the normal state $\psi$ on $\cR_\Xi \rtimes_\alpha \bbZ/2\bbZ$ given by $\psi \coloneqq \sum_n \langle \eta_n | \cdot \widetilde \eta_n \rangle$.  
Then we have that for all $X \in \cR_\Xi \rtimes_\alpha \bbZ/2\bbZ$ as in \eqref{eq:CanonicalCrossedProductElement},
\[
\psi(X)
=
\sum_n \langle \eta_n | X \widetilde \eta_n \rangle
=
\sum_n \langle \xi_n | (R_I + R_ZV) \xi_n \rangle
=
\varphi(\Phi(X)).
\]
Hence $\varphi \circ \Phi$ is a normal state on $\cR_\Xi \rtimes_\alpha \bbZ/2\bbZ$, so $\Phi$ is normal.  

Since $\Phi$ is a normal $*$-homomorphism, $\Phi(\cR_\Xi \rtimes_\alpha \bbZ/2\bbZ)$ is a von Neumann algebra, which contains both $\cR_\Xi$ and $V$.  
Thus, since $\widehat{\cR}_\Xi = \cR_\Xi \vee \{V\}$ by Lemma \ref{lem:AlgebraObtainedByAdjoiningIntertwiner}, we have that $\Phi(\cR_\Xi \rtimes_\alpha \bbZ/2\bbZ) = \widehat{\cR}_\Xi$.  
In order to show that $\Phi \colon \cR_\Xi \rtimes_\alpha \bbZ/2\bbZ \to \widehat \cR_\Xi$ is an isomorphism, therefore, it remains to show that $\Phi$ is injective.  

To show injectivity of $\Phi$, we use the unitary decomposition of $\cH$ described in Lemma \ref{lem:UnitaryDecompositionForGNSSpace}.  
Recall that $\Xi = \Lambda_1 \cup \Lambda_2$, where there exists $\Lambda$ a cone along the boundary with $\Lambda_1 \ll \Lambda$ and $\Lambda_2 \subseteq \Lambda^c$.  
By Lemma \ref{lem:UnitaryDecompositionForGNSSpace}, we have a well-defined unitary $U \colon \cH \to \cH_{\Lambda_1} \otimes \cH_{\Lambda^c} \otimes \cH_0$ given by 
\[
U\Gamma_1 \widehat{\Gamma} \Gamma_2 \Omega
\coloneqq
\Gamma_1\Omega \otimes \Gamma_2 \Omega \otimes \widehat{\Gamma} \Omega
\]
for $\Gamma_1 \in \fF_{\Lambda_1}$, $\Gamma_2 \in  \fF_{\Lambda^c}$, and $\widehat{\Gamma} \in \fF_0$.  
Here $\cH_0$ and $\fF_0$ are as defined in \textsection\ref{sec:DistalSplit}.  
Furthermore, since $\cR_{\Lambda_2} \subseteq \cR_{\Lambda^c}$, we have by Proposition \ref{prop:ActionOfDistalSplitFactorsOnDecomposedGNSSpace} that $U\cR_{\Lambda_1}U^*$ acts only on the $\cH_{\Lambda_1}$ tensor factor and $U\cR_{\Lambda_2}U^*$ acts only on the $\cH_{\Lambda_2}$ tensor factor.  
Note that since $\cR_\Xi = \cR_{\Lambda_1} \vee \cR_{\Lambda_2}$, we have that $U\cR_\Xi U^* = U\cR_{\Lambda_1}U^* \otimes U\cR_{\Lambda_2}U^*$.  
Thus, $U\cR_\Xi U^*$ acts only on $\cH_{\Lambda_1} \otimes \cH_{\Lambda^c}$.  
For ease of notation, we let $\cK \coloneqq \cH_{\Lambda_1} \otimes \cH_{\Lambda^c}$.  

We recall from \textsection\ref{sec:DistalSplit} that one of the path operators generating $\fF_0$ corresponds to a path $\gamma^b$ of type $Z$ in $\Lambda \cap \Lambda_1^c$ from a site on the boundary of $\Lambda_1$ to a site on the boundary of $\Lambda$.  
Without loss of generality, we may assume that $\gamma^b = \gamma_Z \cap (\Lambda \cap \Lambda_1^c)$, where $\gamma_Z$ is the path defined earlier in this section.  
Note that $\xi_I \coloneqq \Omega$ and $\xi_Z \coloneqq \Gamma^Z_{\gamma^b} \Omega$ are orthogonal vectors in $\cH_0$.  
For $k \in \{I, Z\}$, we define $P_k \coloneqq I \otimes |\xi_k\rangle \langle \xi_k|$.  
Observe that $P_k$ commutes with $U\cR_{\Lambda_1}U^*$ and $U \cR_{\Lambda_2}U^*$ and thus commutes with $U\cR_\Xi U^* = U\cR_{\Lambda_1}U^* \otimes U\cR_{\Lambda_2}U^*$.  

Now, suppose $X \in \cR_\Xi \rtimes_\alpha \bbZ/2\bbZ$ is as described in \eqref{eq:CanonicalCrossedProductElement}.  
We then have that for $k \in \{I, Z\}$, 
\[
P_kU\Phi(X)U^*
=
P_kU(R_I + R_Z V)U^*
=
UR_IU^*P_k + UR_ZU^*P_kUVU^*.
\]
We claim that for all $\eta \in \cK$, $P_kUVU^*(\eta \otimes \Omega) = \delta_{k, Z} UVU^*(\eta \otimes \Omega)$.  
By continuity, it suffices to consider the case where $\eta = \Gamma_1\Omega \otimes \Gamma_2 \Omega$ for some $\Gamma_1 \in \fF_{\Lambda_1}$ and $\Gamma_2 \in \fF_{\Lambda_2^c}$.  
Let $\Gamma_1 \in \fF_{\Lambda_1}$ and $\Gamma_2 \in \fF_{\Lambda_2^c}$, and let $\eta \coloneqq \Gamma_1\Omega \otimes \Gamma_2 \Omega$.  
Recall from the proof of Lemma \ref{lem:DenseSubspaceForAuxiliarySubspaceForBoundingExcitations} that $V\Omega = \Gamma^Z_{\gamma_1'} \Gamma^Z_{\gamma_Z} \Gamma^Z_{\gamma_2'}\Omega$ for some paths $\gamma_1'$ in $\Lambda_1$ and $\gamma_2'$ in $\Lambda_2$.  
Using that $\gamma^b = \gamma_Z \cap (\Lambda \cap \Lambda_1^c)$, we have that $V\Omega = \Gamma^Z_{\gamma_1} \Gamma^Z_{\gamma^b} \Gamma^Z_{\gamma_2} \Omega$ for some paths $\gamma_1$ in $\Lambda_1$ and $\gamma_2$ in $\Lambda^c$.  
Thus, since $V$ either commutes or anti-commutes with each path operator, we have that 
\begin{align*}
UVU^*(\eta \otimes \Omega)
&=
UV\Gamma_1\Gamma_2\Omega
=
\pm U\Gamma_1 \Gamma_2 V\Omega
=
\pm U\Gamma_1 \Gamma_2 \Gamma^Z_{\gamma_1} \Gamma^Z_{\gamma^b} \Gamma^Z_{\gamma_2} \Omega
\\&=
\pm U\Gamma_1 \Gamma^Z_{\gamma_1} \Gamma^Z_{\gamma^b} \Gamma_2 \Gamma^Z_{\gamma_2} \Omega
=
\pm \Gamma_1 \Gamma^Z_{\gamma_1} \Omega \otimes \Gamma_2 \Gamma^Z_{\gamma_2} \Omega \otimes \Gamma^Z_{\gamma^b} \Omega.
\end{align*}
We thus have that $P_kUVU^*(\eta \otimes \Omega) = \delta_{k, Z} UVU^*(\eta \otimes \Omega)$, as desired.  

Now, suppose $\Phi(X) = 0$.  
Then for each $k \in \{I, Z\}$, we have that $P_k U\Phi(X)U^* = 0$.  
In particular, we have that for all $\eta \in \cK$, 
\[
0
=
P_I U\Phi(X)U^*(\eta \otimes \Omega)
=
UR_IU^*P_I(\eta \otimes \Omega) + UR_ZU^*P_IUVU^*(\eta \otimes \Omega)
=
UR_IU^*(\eta \otimes \Omega).
\]
Since $UR_IU^*$ only acts on $\cK$ (and not on the factor of $\cH_0$), we have that $UR_IU^* = 0$ and hence $R_I = 0$.  
Similarly, we have that for all $\eta \in \cK$, 
\[
0
=
P_Z U\Phi(X)U^*(\eta \otimes \Omega)
=
UR_IU^*P_Z(\eta \otimes \Omega) + UR_ZU^*P_ZUVU^*(\eta \otimes \Omega)
=
UR_ZVU^*(\eta \otimes \Omega).
\]
We wish to conclude that $UR_ZU^* = 0$, which will complete the proof.  
To see this, it suffices to show that $UR_ZU^*(\Gamma_1 \Omega \otimes \Gamma_2 \Omega \otimes \Gamma^Z_{\gamma^b} \Omega) = 0$ for all $\Gamma_1 \in \fF_{\Lambda_1}$ and $\Gamma_2 \in \fF_{\Lambda^c}$, by density and the fact that $UR_ZU^*$ only acts on $\cK$.  
Let $\Gamma_1 \in \fF_{\Lambda_1}$ and $\Gamma_2 \in \fF_{\Lambda^c}$.  
We let $\gamma_1$ and $\gamma_2$ be paths in $\Lambda_1$ and $\Lambda^c$ respectively so that $V\Omega = \Gamma^Z_{\gamma_1} \Gamma^Z_{\gamma^b} \Gamma^Z_{\gamma_2} \Omega$.  
Furthermore, we let $\eta \coloneqq \Gamma_1 \Gamma^Z_{\gamma_1} \Omega \otimes \Gamma_2 \Gamma^Z_{\gamma_2} \Omega \in \cK$.  
Then by the argument in the preceding paragraph, we have that 
\[
UVU^*(\eta \otimes \Omega)
=
\pm \Gamma_1 \Omega \otimes \Gamma_2 \Omega \otimes \Gamma^Z_{\gamma^b} \Omega.  
\]
Thus, we have that 
\[
0
=
UR_ZVU^*(\eta \otimes \Omega)
=
\pm UR_ZU^*(\Gamma_1 \Omega \otimes \Gamma_2 \Omega \otimes \Gamma^Z_{\gamma^b} \Omega),
\]
as desired.  

To show that $\alpha$ is an outer action on $\cR_\Xi$, it suffices to show that $V \notin \cR_\Xi$, since $\alpha$ is implemented by $V$.  
We show this by showing that there exists an operator in $\scrB(\cH)$ that commutes with $\cR_\Xi$ but not $V$.  
Recall that $U^*P_IU$ commutes with $\cR_\Xi$.  
However, $U^*P_IU$ does not commute with $V$. 
Indeed, for nonzero $\eta \in \cK$, $P_IUVU^*(\eta \otimes \Omega) = 0$, but $P_I(\eta \otimes \Omega) = \eta \otimes \Omega$ and hence $UVU^*P_I (\eta \otimes \Omega) \neq 0$ since $V$ is a unitary.  
Thus $V \notin \cR_\Xi$, so $\alpha$ is outer.
\end{proof}

It is a well-known fact from subfactor theory that $[\cR_\Xi \rtimes_\alpha \bbZ/2\bbZ : \cR_\Xi] = 2$ for an outer action $\alpha$.  
Thus, Proposition \ref{prop:BoundingExcitationsCrossedProduct} along with Proposition \ref{prop:MuIndexBoundsExcitations} gives the following result.  

\begin{thm}
There are exactly two nonisomorphic simple superselection sectors for toric code with boundary.
\end{thm}

\begin{proof}
By Proposition \ref{prop:BoundingExcitationsCrossedProduct}, $\mu = 2$, since $[\widehat{\cR}_\Xi : \cR_\Xi] = 2$ for all $\Xi \in \cC^2$.  
Hence, by Proposition \ref{prop:MuIndexBoundsExcitations}, there are at most two nonisomorphic simple superselection sectors for toric code with boundary.  
We have that there are exactly two such sectors, since by Theorem \ref{thm:ToricCodeBoundaryExcitationsCondensation}, there exist two such sectors (namely the vacuum and the type $Z$ sector $\pi^Z$).  
\end{proof}


\subsection*{Acknowledgments}
We would like to thank David Penneys for his support and invaluable comments on this work.  
We would also like to thank Pieter Naaijkens for his comments on this paper and for answering questions about his prior work on the subject.  
We would like to thank Corey Jones for suggesting this avenue of research as well as for his helpful exposition of the basics of algebraic quantum field theory.  
We would like to thank Peter Huston and Kyle Kawagoe for discussions about topological order, and we would like to thank Sean Sanford for discussions about module tensor categories.
We would like to thank Eric Roon for discussions on time evolution in the thermodynamic limit and ground states.
Finally, we would like to thank Yoshiko Ogata for noticing an error in an earlier version of this paper.
The author was partially supported by NSF DMS 1654159 and 2154389.


\bibliographystyle{plain}
\bibliography{References.bib}

\end{document}